\pdfoutput=1
\documentclass[prd,showpacs,letterpaper,aps,10pt,notitlepage]{revtex4-1}

\usepackage{amsfonts,amsmath,graphicx,bm}

\newcommand{\ua}{\uparrow}
\newcommand{\da}{\downarrow}
\newcommand{\nb}{\phantom{0}}
\newcommand{\wm}{\phantom{-}}
\newcommand{\wdt}{\phantom{.}}
\newcommand{\bs}[1]{\ensuremath{{\boldsymbol{#1}}}}

\def\sfrac#1#2{{\textstyle\frac{#1}{#2}}}

\def\mathbi#1{\textbf{\em #1}}

\begin{document}

\title{Excited-state spectroscopy of triply-bottom baryons from lattice QCD}

\author{Stefan Meinel}
\affiliation{Department of Physics, College of William \&  Mary, Williamsburg,
  VA 23187-8795, USA}

\date{May 22, 2012}


\begin{abstract}
The spectrum of baryons containing three $b$ quarks is calculated in nonperturbative QCD,
using the lattice regularization. The energies of ten excited $bbb$ states with
$J^P=\frac12^+$, $\frac32^+$, $\frac52^+$, $\frac72^+$, $\frac12^-$, and $\frac32^-$ are
determined with high precision. A domain-wall action is used for the up-, down- and strange quarks,
and the bottom quarks are implemented with NRQCD. The computations are done at lattice spacings
of $a\approx0.11$ fm and $a\approx 0.08$ fm, and the results demonstrate the improvement of
rotational symmetry as $a$ is reduced. A large lattice volume of $(2.7\:\:{\rm fm})^3$ is used,
and extrapolations of the $bbb$ spectrum to realistic values of the light sea-quark masses are performed.
All spin-dependent energy splittings are resolved with total uncertainties of order 1 MeV, and the
dependence of these splittings on the couplings in the NRQCD action is analyzed.
\end{abstract}

\maketitle

\section{Introduction}

Heavy quarkonium has been studied in great detail both experimentally
and theoretically. Because its valence quark masses are large compared to $\Lambda_{QCD}$, heavy quarkonium
is an excellent system for probing the strong force on multiple scales \cite{Brambilla:2010cs}.
In addition to these familiar heavy quark-antiquark bound states, QCD also predicts the existence of
an analogous system in the baryonic sector: the bound states of three heavy quarks. Given the huge importance of quarkonium,
it is desirable to investigate triply-heavy baryons in similar depth.

Several continuum-based calculations of triply-heavy baryon spectra can be found in the literature. The methods used there include
quark models \cite{Ponce:1978gk, Hasenfratz:1980ka, Bjorken:1985ei, Tsuge:1985ei, Basdevant:1985ux, SchaffnerBielich:1998ci, 
Martin:1995vk, SilvestreBrac:1996bg, Vijande:2004at, Migura:2006ep, Faessler:2006ft, Martynenko:2007je, Gerasyuta:2007un, Roberts:2007ni,
Bernotas:2008bu, Flynn:2011gf}, QCD sum rules \cite{Zhang:2009re, Wang:2011ae}, and potential NRQCD (pNRQCD) with static potentials
from perturbation theory, at leading order \cite{Jia:2006gw} and next-to-next-to-leading-order \cite{LlanesEstrada:2011kc, Brambilla:2009cd}.
No experimental results are available so far for triply heavy baryons (see Ref.~\cite{Chen:2011mb}
for a recent calculation of production cross sections at the LHC). This means that first-principles
nonperturbative lattice QCD calculations are essential to test the model-dependent or perturbative approaches. For the
$\Omega_{bbb}$, the ground-state mass was already calculated with high precision using lattice QCD in Ref.~\cite{Meinel:2010pw}.
However, much more information about the interactions between three heavy quarks can be gained by computing
the spectrum of the corresponding excited states, including in particular the spin-dependent energy splittings.
The first such calculation of $bbb$ excited states in lattice QCD is reported here. Lattice calculations of light-baryon
excited states can be found for example in Refs.~\cite{Melnitchouk:2002eg, Mathur:2003zf,
Guadagnoli:2004wm, Burch:2006cc, Basak:2007kj, Bulava:2009jb, Bulava:2010yg, Mahbub:2010rm, Edwards:2011jj}.

To fully accommodate the physics of the light sea quarks in lattice QCD, the spatial box size $L$ has to be chosen such that
$L \gg 1/m_\pi$. With the presently available computing resources, this requirement means that the lattice spacing
is too coarse to treat the $b$ quarks in the same way as the light quarks. Therefore, as in Ref.~\cite{Meinel:2010pw},
the $b$ quarks are implemented here with improved lattice NRQCD \cite{Thacker:1990bm, Lepage:1992tx}. NRQCD is an
effective field theory for heavy quarks that retains all the gluon and light-quark degrees of freedom without change.
For the heavy quark Lagrangian, a nonrelativistic expansion is performed in powers of the heavy-quark velocity $v$, and the coefficients
of the NRQCD effective operators are determined by matching to QCD. Thereby, the results of QCD can be reproduced
in principle to an arbitrary order in $v$. For $b \bar{b}$ and $bbb$ hadrons, one has $\langle v^2 \rangle \approx 0.1$. The lattice NRQCD
action used in Ref.~\cite{Meinel:2010pw} was complete through order $v^4$. Because the present work aims to accurately
compute also spin-dependent $bbb$ energy splittings (fine and hyperfine structure), here the spin-dependent order-$v^6$
terms are included in the NRQCD action, as already done in the calculation of the bottomonium spectrum of Ref.~\cite{Meinel:2010pv}.
Furthermore, the coefficients of the leading spin-dependent operators, which are of order $v^4$, are tuned nonperturbatively.

As usual in lattice QCD, the Euclidean path integral is performed by averaging over importance-sampled gauge field configurations.
The ensembles of gauge fields used here match those used in Refs.~\cite{Meinel:2010pw} and \cite{Meinel:2010pv}, and have
been generated by the RBC/UKQCD collaboration \cite{Aoki:2010dy}. These ensembles include the effects of dynamical $u$-, $d$- and $s$- quarks,
which were implemented using a domain-wall action \cite{Kaplan:1992bt,Shamir:1993zy,Furman:1994ky}. Seven different
ensembles with a range of light-quark masses and lattice spacings of $a\approx 0.11$ fm and $a \approx 0.08$ fm are included in the analysis.

The $bbb$ energy levels are extracted from the time-dependence of Euclidean two-point functions of interpolating operators with
the desired quantum numbers. The construction of these interpolating operators, which takes into account
the reduction of the continuum rotational symmetries to the lattice rotational symmetries, follows the highly successful method
originally developed for light baryons in Ref.~\cite{Edwards:2011jj}. This method, as well as the computation of the $bbb$ two-point functions,
is explained in Sec.~\ref{sec:ops}. The details of the lattice actions and parameters are given in Sec.~\ref{sec:parameters}. Next, Sec.~\ref{sec:fitting}
describes the fitting of the two-point functions and the angular momentum identification. The spectrum results are extrapolated
in the light-quark masses to obtain the final results in Sec.~\ref{sec:chiral}. An additional section (Sec.~\ref{sec:coeff_dep})
is devoted to investigating the dependence of the $bbb$ energy splittings on the various operators in the NRQCD action.

\section{\label{sec:ops}Construction of baryon interpolating operators}

In this section we construct interpolating operators, $\Omega$, that give access to $bbb$ states up to $J=\frac72$. The method
is taken from Ref.~\cite{Edwards:2011jj}, but is described again in the following specifically
for the case needed here, where all three quark flavors are equal and only two-component Pauli spinors are used.
Going through the derivation of the interpolating operators also gives some insight into the structure of the $bbb$ states extracted
in the numerical part of the calculation. However, it is important to remember that the spectrum calculated here is
that of the (lattice) QCD+NRQCD Hamiltonian: $H_\mathrm{QCD}|n\rangle = E_n |n\rangle$. The interpolating operators
determine only the overlap factors $\langle n | \Omega | 0 \rangle$,
not the energies $E_n$. For the numerical calculation it is nevertheless advantageous to construct operators
that have large overlaps only with selected $bbb$ states, to get good statistical precision for the energy levels and identify
their angular momentum $J$.

A key feature of the approach from Ref.~\cite{Edwards:2011jj} is the
initial construction of operators with definite quantum numbers $J$ and $m$
according to the \emph{continuum} rotational symmetry (Sec.~\ref{sec:continuumops}).
This is then followed by the \emph{subduction}, where linear combinations of the different $m$-components
at a given $J$ are formed such that these transform irreducibly under the lattice rotational symmetries (Sec.~\ref{sec:subduction}).
The numerical calculations demonstrate that the rotational symmetry breaking is very weak, and operators
subduced from continuum operators with different values of $J$ retain an approximate orthogonality even if they fall in the same irreducible
representation of the octahedral group. This feature dramatically simplifies the angular momentum identification for the extracted energy levels.

Following the group-theoretical operator construction, Sec.~\ref{sec:2ptfns} then describes
the initial smearing of the quark fields and the calculation of the baryon two-point functions on the lattice.

\subsection{\label{sec:continuumops}Operators with definite continuum $J$}

In all baryon operators, the colors of the three quarks are combined into a singlet using the totally antisymmetric color wave function $\epsilon_{abc}$.
In the case considered here, the three quarks have equal flavor. Therefore, to satisfy the Pauli principle, the product of the spin and spatial wave functions must be totally
symmetric. The spatial structure is obtained by applying up to two derivative operators to Gaussian-smeared quark fields. The derivatives are combined to
a definite total orbital angular momentum $L$ and a definite permutation symmetry. Similarly, the spins of the three quarks are combined to
a definite total spin $S$ and definite permutation symmetry. Finally the derivative and spin wave functions obtained in these two separate steps
are combined to obtain baryon operators with a definite total angular momentum $J$ and the desired total symmetry of the
product of the spin and spatial wave functions. Note that $L$ and $S$ are not conserved quantum numbers, and are only used to label
the structure of the interpolating operators.

We begin by combining the three quark fields to definite total spin $S$. Because NRQCD is used for
the heavy quarks, there are only two spin components, denoted by $\tilde{\psi}_\ua$ and $\tilde{\psi}_\da$. The color indices are omitted here,
but remain uncontracted at this stage (the contraction with $\epsilon_{abc}$ is only performed after the gauge-covariant derivatives have been applied).
The $S=\frac32$ combinations are given by
\begin{eqnarray}
\nonumber  O_{\mathsf{S}}(\sfrac32, +\sfrac32) &=& \tilde{\psi}_\ua \tilde{\psi}_\ua \tilde{\psi}_\ua, \\
\nonumber  O_\mathsf{S}(\sfrac32, +\sfrac12) &=& \frac{1}{\sqrt{3}}\Big( \tilde{\psi}_\ua \tilde{\psi}_\ua \tilde{\psi}_\da  + \tilde{\psi}_\ua \tilde{\psi}_\da \tilde{\psi}_\ua  + \tilde{\psi}_\da \tilde{\psi}_\ua  \tilde{\psi}_\ua  \Big), \\
\nonumber  O_\mathsf{S}(\sfrac32, -\sfrac12) &=& \frac{1}{\sqrt{3}}\Big( \tilde{\psi}_\da \tilde{\psi}_\da  \tilde{\psi}_\ua  + \tilde{\psi}_\da \tilde{\psi}_\ua \tilde{\psi}_\da  + \tilde{\psi}_\ua \tilde{\psi}_\da \tilde{\psi}_\da  \Big), \\
 O_\mathsf{S}(\sfrac32, -\sfrac32) &=& \tilde{\psi}_\da \tilde{\psi}_\da \tilde{\psi}_\da , \label{eq:spin32S}
\end{eqnarray}
where the subscript $\mathsf{S}$ indicates the total symmetry under permutations. For $S=\frac12$, one can construct both mixed-symmetric ($\mathsf{MS}$)
or mixed-antisymmetric ($\mathsf{MA}$) combinations:
\begin{eqnarray}
\nonumber  O_\mathsf{MS}(\sfrac12, +\sfrac12) &=& \wm\frac{1}{\sqrt{6}}\Big(\tilde{\psi}_\ua \tilde{\psi}_\da \tilde{\psi}_\ua  + \tilde{\psi}_\da \tilde{\psi}_\ua  \tilde{\psi}_\ua   - 2 \tilde{\psi}_\ua \tilde{\psi}_\ua \tilde{\psi}_\da \Big), \\
 O_\mathsf{MS}(\sfrac12, -\sfrac12) &=& -\frac{1}{\sqrt{6}}\Big( \tilde{\psi}_\da \tilde{\psi}_\ua \tilde{\psi}_\da  + \tilde{\psi}_\ua \tilde{\psi}_\da \tilde{\psi}_\da   - 2 \tilde{\psi}_\da \tilde{\psi}_\da  \tilde{\psi}_\ua  \Big), \label{eq:spin12MS}
\end{eqnarray}
\begin{eqnarray}
\nonumber  O_\mathsf{MA}(\sfrac12, +\sfrac12) &=& \wm\frac{1}{\sqrt{2}}\Big(\tilde{\psi}_\ua \tilde{\psi}_\da \tilde{\psi}_\ua  - \tilde{\psi}_\da \tilde{\psi}_\ua  \tilde{\psi}_\ua  \Big), \\
 O_\mathsf{MA}(\sfrac12, -\sfrac12) &=& -\frac{1}{\sqrt{2}}\Big(\tilde{\psi}_\da \tilde{\psi}_\ua \tilde{\psi}_\da  - \tilde{\psi}_\ua \tilde{\psi}_\da \tilde{\psi}_\da  \Big). \label{eq:spin12MA}
\end{eqnarray}
Next, we come to the derivatives. A single derivative is an $L=1$ object, with $m$-components given by
\begin{eqnarray}
\nonumber  D_{\pm 1} &=& \pm \sfrac i2 (D_x \pm i D_y), \\
 D_{0} &=& -\sfrac {i}{\sqrt{2}} D_z.
\end{eqnarray}
Recall that in this section we work in continuous space; lattice derivatives will be defined in Sec.~\ref{sec:2ptfns}.
In the following, we use the notation $D_m^{(k)}$ for a derivative acting on the $k$-th quark in the baryon operator. As in Ref.~\cite{Edwards:2011jj},
we define the following combinations with definite permutation symmetry:
\begin{eqnarray}
\nonumber  D_{\mathsf{MS}}^{[1]}(1, m) &=& \frac{1}{\sqrt{6}} \Big( 2 D^{(3)}_m - D^{(1)}_m - D^{(2)}_m \Big),  \\
 D_{\mathsf{MA}}^{[1]}(1, m) &=& \frac{1}{\sqrt{2}} \Big( D^{(1)}_m - D^{(2)}_m \Big). \label{eq:singleder}
\end{eqnarray}
(No totally antisymmetric combination exists, and the totally symmetric combination vanishes at zero-momentum.)
Using the Clebsch-Gordan coefficients $\langle L, m | 1, m_1; 1, m_2 \rangle$, we can combine two single-derivative operators
of the form (\ref{eq:singleder}) into double-derivative operators with definite total $L$ and definite permutation symmetry as follows \cite{Edwards:2011jj}:
\begin{eqnarray}
\nonumber  D_{\mathsf{S}}^{[2]}(L, m) &=& \frac{1}{\sqrt{2}} \sum_{m_1, m_2} \langle L, m | 1, m_1; 1, m_2 \rangle \Big( +D_{\mathsf{MS}}^{[1]}(1, m_1)D_{\mathsf{MS}}^{[1]}(1, m_2) +D_{\mathsf{MA}}^{[1]}(1, m_1)D_{\mathsf{MA}}^{[1]}(1, m_2) \Big),  \\
\nonumber  D_{\mathsf{MS}}^{[2]}(L, m) &=& \frac{1}{\sqrt{2}} \sum_{m_1, m_2} \langle L, m | 1, m_1; 1, m_2 \rangle \Big( -D_{\mathsf{MS}}^{[1]}(1, m_1)D_{\mathsf{MS}}^{[1]}(1, m_2) +D_{\mathsf{MA}}^{[1]}(1, m_1)D_{\mathsf{MA}}^{[1]}(1, m_2) \Big),  \\
\nonumber  D_{\mathsf{MA}}^{[2]}(L, m) &=& \frac{1}{\sqrt{2}} \sum_{m_1, m_2} \langle L, m | 1, m_1; 1, m_2 \rangle \Big( +D_{\mathsf{MS}}^{[1]}(1, m_1)D_{\mathsf{MA}}^{[1]}(1, m_2) +D_{\mathsf{MA}}^{[1]}(1, m_1)D_{\mathsf{MS}}^{[1]}(1, m_2) \Big),  \\
 D_{\mathsf{A}}^{[2]}(1, m) &=& \frac{1}{\sqrt{2}} \sum_{m_1, m_2} \langle 1, m | 1, m_1; 1, m_2 \rangle \Big( +D_{\mathsf{MS}}^{[1]}(1, m_1)D_{\mathsf{MA}}^{[1]}(1, m_2) -D_{\mathsf{MA}}^{[1]}(1, m_1)D_{\mathsf{MS}}^{[1]}(1, m_2) \Big). \label{eq:doubleder}
\end{eqnarray}
The first three of the above combinations can give either $L=0$ or $L=2$, while the last combination is restricted to $L=1$.

Now we can combine the spin- and spatial wave functions, distinguishing the cases of zero, one, and two derivatives. Without derivatives,
the requirement of total symmetry restricts the spin to $S=\frac32$. Since $L=0$, we only get $J=\frac32$ in this case:
\begin{eqnarray}
\big[O_{\mathsf{S}}(\sfrac32)\big]^{J=\sfrac32}_m &=& O_{\mathsf{S}}(\sfrac32, m). \label{eq:zeroderivop}
\end{eqnarray}
In one-derivative baryon operators, the derivative part, Eq.~(\ref{eq:singleder}), always has mixed symmetry. Therefore, to get a totally symmetric combination,
the spin part must also have mixed symmetry, and hence $S=\frac12$.
Because the derivative has $L=1$, we can combine $L$ and $S$ to the total angular momenta $J=\frac12$ and $J=\frac32$:
\begin{eqnarray}
\big[D^{[1]}_{\mathsf{M}}(1) \:\: O_{\mathsf{M}}(\sfrac12) \big]^{J=\sfrac12,\: \sfrac32}_m &=& \frac{1}{\sqrt{2}}  \sum_{m_1, m_2} \langle J, m | 1, m_1; \sfrac12, m_2 \rangle \Big( D_{\mathsf{MS}}^{[1]}(1, m_1) \: O_\mathsf{MS}(\sfrac12, m_2) + D_{\mathsf{MA}}^{[1]}(1, m_1) \: O_\mathsf{MA}(\sfrac12, m_2) \Big).  \label{eq:onederivop}
\end{eqnarray}
Finally, we consider the double-derivative operators. Because no totally antisymmetric spin combinations exist, the totally antisymmetric derivative combination
in the last line of Eq.~(\ref{eq:doubleder}) is excluded, and the two derivatives can only combine to $L=0$ or $L=2$. In each case,
one can have $S=\frac12$ with mixed symmetry or $S=\frac32$ with total symmetry. Thus, one obtains the following combinations:
\begin{eqnarray}
\nonumber \big[D^{[2]}_{\mathsf{S}}(0) \:\: O_{\mathsf{S}}(\sfrac32) \big]^{J=\sfrac32}_m &=& D_{\mathsf{S}}^{[2]}(0, 0) \: O_\mathsf{S}(\sfrac32, m), \\
\nonumber \big[D^{[2]}_{\mathsf{M}}(0) \:\: O_{\mathsf{M}}(\sfrac12) \big]^{J=\sfrac12}_m &=& \frac{1}{\sqrt{2}} \Big( D_{\mathsf{MS}}^{[2]}(0, 0) \: O_\mathsf{MS}(\sfrac12, m) + D_{\mathsf{MA}}^{[2]}(0, 0) \: O_\mathsf{MA}(\sfrac12, m) \Big), \\
\nonumber \big[D^{[2]}_{\mathsf{M}}(2) \:\: O_{\mathsf{M}}(\sfrac12) \big]^{J=\sfrac32,\: \sfrac52}_m &=& \frac{1}{\sqrt{2}}  \sum_{m_1, m_2} \langle J, m | 2, m_1; \sfrac12, m_2 \rangle \Big( D_{\mathsf{MS}}^{[2]}(2, m_1) \: O_\mathsf{MS}(\sfrac12, m_2) + D_{\mathsf{MA}}^{[2]}(2, m_1) \: O_\mathsf{MA}(\sfrac12, m_2) \Big), \\
\big[D^{[2]}_{\mathsf{S}}(2) \:\: O_{\mathsf{S}}(\sfrac32) \big]^{J=\sfrac12,\: \sfrac32,\: \sfrac52,\: \sfrac72}_m &=& \sum_{m_1, m_2} \langle J, m | 2, m_1; \sfrac32, m_2 \rangle \:  D_{\mathsf{S}}^{[2]}(2, m_1) \: O_\mathsf{S}(\sfrac32, m_2).  \label{eq:twoderivop}
\end{eqnarray}
Note that the combination with $D_{\mathsf{S}}^{[2]}(0, 0)$, which corresponds to the spatial Laplacian, was excluded in Ref.~\cite{Edwards:2011jj} with the argument that it vanishes
at zero momentum. However, this is not the case for the method of smearing the quark fields and constructing the two-point functions described in Sec.~\ref{sec:2ptfns}.
In fact, the operator $\big[D^{[2]}_{\mathsf{S}}(0) \:\: O_{\mathsf{S}}(\sfrac32) \big]^{J=\sfrac32}_m$ has a good overlap with the first radially excited $J=\frac32$ state,
and including this operator in the basis significantly improves the extraction of this energy level.

\subsection{\label{sec:subduction}Subduction to irreducible representations of the double cover of the octahedral group}

In the previous section, we constructed operators $\big[\Omega \big]^J_m$ that transform under rotations like the basis vectors $|J,m\rangle$ of irreducible representations of $SU(2)$.
The group $SU(2)$ is the double cover of the continuum three-dimensional rotation group $SO(3)$.
On a cubic lattice, the rotational symmetry is reduced to the discrete group $^2{\mathrm{O}}$, the double cover of the octahedral group $\mathrm{O}$. The group $^2{\mathrm{O}}$,
which is obtained from $\mathrm{O}$ by adding a negative identity for $\pm 2\pi$ rotations, has 48 elements in 8 conjugacy classes. Correspondingly, $^2{\mathrm{O}}$
has 8 irreducible representations denoted as $A_1$, $A_2$, $E$, $T_1$, $T_2$, $G_1$, $G_2$, $H$ (see, for example, Ref.~\cite{Johnson:1982yq}). Their dimensions are
1, 1, 2, 3, 3, 2, 2, 4, respectively. Starting from an operator $\big[\Omega \big]^J_m$, it is possible to form suitable linear combinations of its different $m$-components,
so that these linear combinations transform in irreducible representations, $\Lambda$, of the double-cover octahedral group:
\begin{equation}
\big[\Omega \big]^J_{^n\!\Lambda, r} = \sum_m \mathcal{S}^{J,m}_{^n\!\Lambda, r}\: \big[\Omega \big]^J_m.
\end{equation}
This process is referred to as \emph{reduction} or \emph{subduction} \cite{Johnson:1982yq, Edwards:2011jj}, and the coefficients $\mathcal{S}^{J,m}_{^n\!\Lambda, r}$ form the subduction
matrices. Here, $^n\!\Lambda$ denotes the $n$-th occurrence of an irrep $\Lambda$ of $^2{\mathrm{O}}$,
and $r=1,\:...,\:\mathrm{dim}(\Lambda)$ denotes its row (like $m$ denotes the row for the $SU(2)$ irrep).
For each value of $J$, only selected irreps of $^2{\mathrm{O}}$ appear in the subduction, such that the sum of their dimensions equals $2J+1$ (the dimension of the original $SU(2)$ irrep $J$).
This is indicated in Table \ref{tab:subduction}. For integer values of $J$, only the irreps $A_1$, $A_2$, $E$, $T_1$, and $T_2$ appear. Conversely, for half-integer $J$, only the irreps
$G_1$, $G_2$, and $H$ occur. Since we are considering baryons, we will only be concerned with these three irreps in the remainder of the paper.
The subduction matrices for $(J=\frac12) \rightarrow G_1$ and $(J=\frac32) \rightarrow H$ are simply the $2\times2$ and $4\times4$ identity matrices, so that, for example, $\big[\Omega \big]^{\frac12}_{G_1, 1} = \big[\Omega \big]^{\frac12}_{+\frac12}$
and $\big[\Omega \big]^{\frac12}_{G_1, 2} = \big[\Omega \big]^{\frac12}_{-\frac12}$. The subduction matrices for $J=\frac52$ and $J=\frac72$ can be found in Ref.~\cite{Edwards:2011jj}.

\begin{table}[ht!]
\begin{tabular}{lll}
\hline\hline
$J$ & \hspace{3ex} &  Subduction \\
\hline
$0$   && $A_1$ \\
$1/2$ && $G_1$ \\
$1$   && $T_1$ \\
$3/2$ && $H$ \\
$2$   && $E + T_2$ \\
$5/2$ && $G_2 + H$ \\
$3$   && $A_2+T_1+T_2$ \\
$7/2$ && $G_1 + G_2 + H$ \\
$4$   && $A_1+E+T_1+T_2$ \\
$9/2$ && $G_1 + {}^1\!H + {}^2\!H$ \\
\hline\hline
\end{tabular}
\caption{\label{tab:subduction}Subduction of $SU(2)$ irreps to $^2{\mathrm{O}}$ irreps, up to $J=\frac92$ (from Ref.~\cite{Johnson:1982yq}).}
\end{table}

So far we have only discussed the rotational symmetry. Additionally, we can classify the operators according to their transformation properties under space inversion, which remains an exact symmetry on the lattice.
Then all of the irreducible representations come in parity-even and parity-odd versions, as indicated by subscripts $g$ (gerade) and $u$ (ungerade):  $A_{1g}$, ..., $T_{2g}$, $G_{1g}$, $G_{2g}$, $H_g$, and $A_{1u}$, ..., $T_{2u}$, $G_{1u}$, $G_{2u}$, $H_u$.
In this work, the baryon operators are constructed from two-component NRQCD spinors, and therefore the parity of an operator is determined entirely by the number of derivatives it contains: an even number of derivatives
corresponds to even parity and an odd number of derivatives corresponds to odd parity.

The 11 different baryon operators constructed in Eqs.~(\ref{eq:zeroderivop}-\ref{eq:twoderivop}) subduce to 7 operators in the $H_g$ irrep, 3 operators each in the $G_{1g}$
and $G_{2g}$ irreps, and 1 operator each in the $G_{1u}$ and $H_u$ irreps. This set of operators is summarized in Table \ref{tab:operators}.

\begin{table}[ht!]
\begin{tabular}{lll}
\hline\hline
\\[-2.7ex]
Operator(s)                   & \hspace{3ex} &   Structure $\sim\big[D(L) \:\: O(S) \big]^J$   \\
\\[-2.7ex]
\hline
\\[-2ex]
$H_g^{(1)}$                                  &&  $\big[O_{\mathsf{S}}(\sfrac32)\big]^{J=\sfrac32}$ \\
\\[-2ex]
$G_{1u}^{(1)}$                               &&  $\big[D^{[1]}_{\mathsf{M}}(1) \:\: O_{\mathsf{M}}(\sfrac12) \big]^{J=\sfrac12}$ \\
\\[-2ex]
$H_u^{(1)}$                                  &&  $\big[D^{[1]}_{\mathsf{M}}(1) \:\: O_{\mathsf{M}}(\sfrac12) \big]^{J=\sfrac32}$ \\
\\[-2ex]
$H_g^{(2)}$                                  &&  $\big[D^{[2]}_{\mathsf{S}}(0) \:\: O_{\mathsf{S}}(\sfrac32) \big]^{J=\sfrac32}$ \\
\\[-2ex]
$G_{1g}^{(1)}$                               &&  $\big[D^{[2]}_{\mathsf{M}}(0) \:\: O_{\mathsf{M}}(\sfrac12) \big]^{J=\sfrac12}$ \\
\\[-2ex]
$G_{1g}^{(2)}$                               &&  $\big[D^{[2]}_{\mathsf{S}}(2) \:\: O_{\mathsf{S}}(\sfrac32) \big]^{J=\sfrac12}$ \\
\\[-2ex]
$H_g^{(3)}$                                  &&  $\big[D^{[2]}_{\mathsf{S}}(2) \:\: O_{\mathsf{S}}(\sfrac32) \big]^{J=\sfrac32}$ \\
\\[-2ex]
$H_g^{(4)}$, $G_{2g}^{(1)}$                  &&  $\big[D^{[2]}_{\mathsf{S}}(2) \:\: O_{\mathsf{S}}(\sfrac32) \big]^{J=\sfrac52}$ \\
\\[-2ex]
$H_g^{(5)}$, $G_{1g}^{(3)}$, $G_{2g}^{(2)}$  &&  $\big[D^{[2]}_{\mathsf{S}}(2) \:\: O_{\mathsf{S}}(\sfrac32) \big]^{J=\sfrac72}$ \\
\\[-2ex]
$H_g^{(6)}$                                  &&  $\big[D^{[2]}_{\mathsf{M}}(2) \:\: O_{\mathsf{M}}(\sfrac12) \big]^{J=\sfrac32}$ \\
\\[-2ex]
$H_g^{(7)}$, $G_{2g}^{(3)}$                  &&  $\big[D^{[2]}_{\mathsf{M}}(2) \:\: O_{\mathsf{M}}(\sfrac12) \big]^{J=\sfrac52}$ \\
\\[-2ex]
\hline\hline
\end{tabular}
\caption{\label{tab:operators}Interpolating operators, named according to their parity ($g$: $+$, $u$: $-$) and irreducible representation of $^2{\mathrm{O}}$.
The superscript labels the different operators within a given irrep and parity.}
\end{table}

\subsection{\label{sec:2ptfns}Computation of two-point functions on the lattice}

The group-theoretical construction of baryon operators through subduction was performed here in the same way as done for light baryons in
Ref.~\cite{Edwards:2011jj}. However, the method for smearing the quark fields and computing the two-point functions in terms of quark propagators
differs from that used in Ref.~\cite{Edwards:2011jj}. Instead of \emph{distillation} \cite{Peardon:2009gh}, here the more traditional approach
starting from Gaussian-smeared point sources, as in Ref.~\cite{Lichtl:2006dt}, is chosen. This has the advantage over distillation
that the quark smearing width can be made very narrow without increasing the computational cost.
A narrow smearing width is needed to get a good overlap with the physical $bbb$ states, which are expected to be very small objects
as a consequence of the large $b$-quark mass.

In the approach used here, the smeared $b$-quark fields $\tilde{\psi}$ entering in Eqs.~(\ref{eq:spin32S}-\ref{eq:spin12MA}) are defined in terms of the unsmeared quark fields $\psi$
through
\begin{equation}
 \tilde{\psi} = \left(1 + \frac{r_S^2}{2 n_S}\Delta^{(2)}\right)^{n_S} \psi, \label{eq:smear_op}
\end{equation}
where $\Delta^{(2)}$ is a three-dimensional gauge-covariant lattice Laplace operator,
\begin{equation}
 \Delta^{(2)} \psi(\bs{x}, t) = -\frac{1}{a^2} \sum_{j=1}^3\left(  \tilde{U}_j(\bs{x}, t) \psi(\bs{x}+a\bs{\hat{j}},t) - 2\psi(\bs{x}, t) + \tilde{U}_{-j}(\bs{x}, t) \psi(\bs{x}-a\bs{\hat{j}},t) \right).  \label{eq:Laplace}
\end{equation}
In this work, a smearing radius of $r_S\approx 0.14$ fm is used in Eq.~(\ref{eq:smear_op}).
The gauge-covariant derivatives in the baryon operators then act on these smeared quark fields. The continuous derivatives $D_j$ used
in Sec.~\ref{sec:continuumops} are replaced by lattice versions $\nabla_j$, which are defined as
\begin{equation}
 \nabla_j \tilde{\psi}(\bs{x}, t) = \frac{1}{2a} \left(  \tilde{U}_j(\bs{x}, t) \tilde{\psi}(\bs{x}+a\bs{\hat{j}},t) - \tilde{U}_{-j}(\bs{x}, t)\tilde{\psi}(\bs{x}-a\bs{\hat{j}},t) \right). \label{eq:der}
\end{equation}
The tilde on the gauge links in Eqs.~(\ref{eq:Laplace}) and (\ref{eq:der}) indicates that these are also smeared, using the procedure of Ref.~\cite{Morningstar:2003gk}.
The gauge link smearing in the hadron interpolating fields is performed to reduce statistical noise \cite{Lichtl:2006dt}. The baryon operators constructed in the previous two sections
contain quark fields with up to two derivatives. It is convenient to introduce new objects $\tilde{\psi}_i$, where $i$ labels all the required thirteen derivative combinations:
\begin{eqnarray}
\nonumber \tilde{\psi}_1 &=& \tilde{\psi}, \\
\nonumber \tilde{\psi}_2 &=& \nabla_x\: \tilde{\psi}, \\
\nonumber \tilde{\psi}_3 &=& \nabla_y\: \tilde{\psi}, \\
\nonumber \tilde{\psi}_4 &=& \nabla_z\: \tilde{\psi}, \\
\nonumber \tilde{\psi}_5 &=& \nabla_x\: \nabla_x\: \tilde{\psi}, \\
\nonumber \tilde{\psi}_6 &=& \nabla_y\: \nabla_x\: \tilde{\psi}, \\
\nonumber &\vdots& \\
\tilde{\psi}_{13} &=& \nabla_z\: \nabla_z\: \tilde{\psi}. \label{eq:derivs}
\end{eqnarray}
Additionally to the derivative index $i=1,...13$, these fields $\tilde{\psi}_{i}=(\tilde{\psi}_{a \alpha i})$ have a color index $a=1,2,3$
and a spinor index \mbox{$\alpha=1,2$ ($=\:\ua, \da$)}. Then, all baryon interpolating operators used here
have the form
\begin{equation}
\Omega_\Gamma(\bs{x}, t) = \Gamma_{\alpha i\, \beta j \, \gamma k}\:\: \epsilon_{abc}\:\: \tilde{\psi}_{a \alpha i}(\bs{x}, t)\: \tilde{\psi}_{b \beta j}(\bs{x}, t)\: \tilde{\psi}_{c \gamma k}(\bs{x}, t),
\end{equation}
where $\Gamma_{\alpha i\, \beta j \, \gamma k}$ is the set of complex-valued coefficients from Sec.~\ref{sec:subduction} for each operator.
The two-point function at zero momentum, allowing different operators $\Omega_\Gamma$ and $\Omega_{\Gamma'}$ at sink and source, is then defined as
\begin{eqnarray}
\nonumber C_{\Gamma,\: \Gamma'}(t-t') &=& \sum_\bs{x} \left\langle \Omega_{\Gamma}(\bs{x}, t) \: \Omega^\dag_{\Gamma'}(\bs{x'}, t') \right\rangle \\
\nonumber &=& \sum_\bs{x} \Gamma_{\alpha i\, \beta j \, \gamma k}\:\: \epsilon_{abc} \:\:
\Gamma'^*_{\bar{\alpha} \bar{i}\, \bar{\beta} \bar{j} \, \bar{\gamma} \bar{k}} \epsilon_{\bar{a}\bar{b}\bar{c}}  \\
&&\times\:\:\left\langle  \tilde{\psi}_{a \alpha i}(\bs{x}, t)\:
 \tilde{\psi}_{b \beta j}(\bs{x}, t)\: \tilde{\psi}_{c \gamma k}(\bs{x}, t) \:
\tilde{\psi}^\dag_{\bar{a} \bar{\alpha} \bar{i}}(\bs{x'}, t')\: \tilde{\psi}^\dag_{\bar{b} \bar{\beta} \bar{j}}(\bs{x'}, t')\: \tilde{\psi}^\dag_{\bar{c} \bar{\gamma} \bar{k}}(\bs{x'}, t')
  \right\rangle, \label{eq:2ptA}
\end{eqnarray}
where the brackets denote the Euclidean path integral over the gauge fields and fermions, weighted with $e^{-S}$.
The path integral over the fermions can be performed explicitly, giving heavy-quark propagators and
determinants of the Dirac operators for all quark flavors.
Following Ref.~\cite{Lichtl:2006dt}, we define three-quark propagators (for a given gauge field $U$) that have been color-contracted and summed over $\bs{x}$:
\begin{equation}
 \widetilde{G}^{(3)}_{\alpha i\:\bar{\alpha}\bar{i}\:\beta j\:\bar{\beta}\bar{j}\:\gamma k \:\bar{\gamma}\bar{k}}(t,t',\bs{x'}) = \sum_\bs{x} \epsilon_{abc} \: \epsilon_{\bar{a}\bar{b}\bar{c}}
\: \widetilde{G}_{a\alpha i\, \bar{a}\bar{\alpha}\bar{i}}(\bs{x}, t, \bs{x'}, t')
\: \widetilde{G}_{b\beta j\, \bar{b}\bar{\beta}\bar{j}}(\bs{x}, t, \bs{x'}, t')
\: \widetilde{G}_{c\gamma k\, \bar{c}\bar{\gamma}\bar{k}}(\bs{x}, t, \bs{x'}, t'), \label{eq:3qprop}
\end{equation}
where $\widetilde{G}_{a\alpha i\, \bar{a}\bar{\alpha}\bar{i}}(\bs{x}, t, \bs{x'}, t')$ denotes a heavy-quark propagator with smearing and,
depending on $i$ and $\bar{i}$, derivatives at the source and sink.
Performing the fermionic path integral in Eq.~(\ref{eq:2ptA}) gives six contractions because all three heavy-quark flavors are equal.
Using the antisymmetry of the epsilon tensor, one obtains
\begin{eqnarray}
 \nonumber C_{\Gamma,\: \Gamma'}(t-t') &=&  \Gamma_{\alpha i\, \beta j \, \gamma k}\:\:\Gamma'^*_{\bar{\alpha} \bar{i}\, \bar{\beta} \bar{j} \, \bar{\gamma} \bar{k}}\:
\Big\langle\phantom{+}\:\: \widetilde{G}^{(3)}_{\alpha i\:\bar{\alpha}\bar{i}\:\beta j\:\bar{\beta}\bar{j}\:\gamma k \:\bar{\gamma}\bar{k}}(t,t',\bs{x'}) \\
\nonumber &&\phantom{\Gamma_{\alpha i\, \beta j \, \gamma k}\:\:\Gamma'_{\bar{\alpha} \bar{i}\, \bar{\beta} \bar{j} \, \bar{\gamma} \bar{k}}\:
\Big\langle} + \widetilde{G}^{(3)}_{\alpha i\:\bar{\beta}\bar{j}\:\beta j\:\bar{\gamma}\bar{k}\:\gamma k \:\bar{\alpha}\bar{i}}(t,t',\bs{x'}) \\
\nonumber &&\phantom{\Gamma_{\alpha i\, \beta j \, \gamma k}\:\:\Gamma'_{\bar{\alpha} \bar{i}\, \bar{\beta} \bar{j} \, \bar{\gamma} \bar{k}}\:
\Big\langle} + \widetilde{G}^{(3)}_{\alpha i\:\bar{\gamma}\bar{k}\:\beta j\:\bar{\alpha}\bar{i}\:\gamma k \:\bar{\beta}\bar{j}}(t,t',\bs{x'}) \\
\nonumber &&\phantom{\Gamma_{\alpha i\, \beta j \, \gamma k}\:\:\Gamma'_{\bar{\alpha} \bar{i}\, \bar{\beta} \bar{j} \, \bar{\gamma} \bar{k}}\:
\Big\langle} + \widetilde{G}^{(3)}_{\alpha i\:\bar{\beta}\bar{j}\:\beta j\:\bar{\alpha}\bar{i}\:\gamma k \:\bar{\gamma}\bar{k}}(t,t',\bs{x'}) \\
\nonumber &&\phantom{\Gamma_{\alpha i\, \beta j \, \gamma k}\:\:\Gamma'_{\bar{\alpha} \bar{i}\, \bar{\beta} \bar{j} \, \bar{\gamma} \bar{k}}\:
\Big\langle} + \widetilde{G}^{(3)}_{\alpha i\:\bar{\gamma}\bar{k}\:\beta j\:\bar{\beta}\bar{j}\:\gamma k \:\bar{\alpha}\bar{i}}(t,t',\bs{x'}) \\
 &&\phantom{\Gamma_{\alpha i\, \beta j \, \gamma k}\:\:\Gamma'_{\bar{\alpha} \bar{i}\, \bar{\beta} \bar{j} \, \bar{\gamma} \bar{k}}\:
\Big\langle} + \widetilde{G}^{(3)}_{\alpha i\:\bar{\alpha}\bar{i}\:\beta j\:\bar{\gamma}\bar{k}\:\gamma k \:\bar{\beta}\bar{j}}(t,t',\bs{x'}) \:\:\Big \rangle_U. \label{eq:2ptB}
\end{eqnarray}
Here, $\langle \: ... \: \rangle_U$ denotes the path integral over the gauge fields $U$ only, where the weighting factor is given by $e^{-S_\mathrm{gauge}}\times(\mathrm{ fermion\:\:determinants})$.

In the numerical calculations, performing all the multiplications in the three-quark propagator (\ref{eq:3qprop}) is expensive, and it is important to use symmetries to reduce the number of
operations needed. Defining multi-indices $I=(\alpha i\:\bar{\alpha}\bar{i})$, $J=(\beta j\:\bar{\beta}\bar{j})$, and $K=(\gamma k \:\bar{\gamma}\bar{k})$, one finds that
$\widetilde{G}^{(3)}_{I\:J\:K}$ is totally symmetric in $I$, $J$, $K$. Furthermore,
since the baryon operators constructed in the previous two sections contain at most two derivatives total, only those components of $\widetilde{G}^{(3)}_{\alpha i\:\bar{\alpha}\bar{i}\:\beta j\:\bar{\beta}\bar{j}\:\gamma k \:\bar{\gamma}\bar{k}}$
with
\begin{equation}
 n_D(i) + n_D(j) + n_D(k) \leq 2,\:\:\:\:n_D(\bar{i}) + n_D(\bar{j}) + n_D(\bar{k}) \leq 2
\end{equation}
are needed [$n_D(i)$ denotes the number of derivatives associated with the index $i$, see Eq.~(\ref{eq:derivs})].

\section{\label{sec:parameters}Lattice actions and parameters}

The path integral over the gauge fields $U$ in Eq.~(\ref{eq:2ptB}) is performed by averaging over samples of lattice gauge field configurations.
The configurations used here have been generated by the RBC/UKQCD collaboration \cite{Aoki:2010dy}, and include
dynamical $u$-, $d$-, and $s$- quarks, with $m_u=m_d$. These quarks were implemented with a domain-wall action \cite{Kaplan:1992bt,Shamir:1993zy,Furman:1994ky},
which is a five-dimensional action that leads to an approximate lattice chiral symmetry for the four-dimensional theory. This chiral symmetry
becomes exact when the length of the auxiliary fifth dimension is taken to infinity. For the gauge action,
the Iwasaki discretization \cite{Iwasaki:1983ck, Iwasaki:1984cj} is used (the gauge fields are four-dimensional, i.e.~constant in the 5-direction).
The domain-wall formalism requires additional Pauli-Villars fields to cancel bulk modes \cite{Shamir:1993zy,Antonio:2006px}, so that the gauge fields $U$ are distributed with
probability density proportional to
\begin{equation}
\frac{\det [K^{\rm DW}(U; aM_5, a m_{u,d})]^2 \det [K^{\rm DW}(U; aM_5, a m_s)]}{\det [K^{\rm DW}(U; aM_5, 1)]^3}\:\:e^{-S_\mathrm{gauge}[U]},
\end{equation}
where $K^{\rm DW}(U; aM_5, a m)$ is the five-dimensional domain-wall operator with domain-wall height $M_5$ and quark-mass $m$.
Seven ensembles of gauge fields with different parameters are included in the analysis, as shown in Table \ref{tab:lattices}. There
are ensembles with two different values of the gauge coupling $\beta=6/g^2$,
leading to lattice spacings of $a\approx 0.11$ fm and $a\approx 0.085$ fm. The number of lattice points is chosen to be
$24^3\times64$ and $32^3\times64$, respectively, so that the spatial volume in physical units is equal to about $(2.7\:\:{\rm fm})^3$ in both cases.

\begin{table*}[t]
\begin{ruledtabular}
\begin{tabular}{cclllccccccl}
$L^3\times T$    & $\beta$ & $a m_{u,d}$ & $a m_s$   & $a m_b$ & $u_{0L}$ & $c_3$ & $c_4$ & MD range, step  & $n_{\rm src}$ & $a$ (fm) & $m_\pi$ (GeV) \\
\hline
$24^3\times 64$  & $2.13$  & $0.005$ & $0.04$    & $2.487$ & $0.8439$ & 1.196 & 1.168 & $900$  - $8660$, $10$ & 32 & $0.1119(17)$ & $0.3377(54)$  \\
$24^3\times 64$  & $2.13$  & $0.01$  & $0.04$    & $2.522$ & $0.8439$ & 1.196 & 1.168 & $1480$ - $8520$, $10$ & 32 & $0.1139(19)$ & $0.4194(70)$  \\
$24^3\times 64$  & $2.13$  & $0.02$  & $0.04$    & $2.622$ & $0.8433$ & 1.196 & 1.168 & $1800$ - $3600$, $10$ & 32 & $0.1177(29)$ & $0.541(14)$  \\
$24^3\times 64$  & $2.13$  & $0.03$  & $0.04$    & $2.691$ & $0.8428$ & 1.196 & 1.168 & $1280$ - $3060$, $10$ & 32 & $0.1196(29)$ & $0.641(15)$  \\
\\[-1ex]
$32^3\times 64$  & $2.25$  & $0.004$ & $0.03$    & $1.831$ & $0.8609$ & 1.175 & 1.113 & $580$ - $6840$,  $10$ & 24 & $0.0849(12)$ & $0.2950(40)$  \\
$32^3\times 64$  & $2.25$  & $0.006$ & $0.03$    & $1.829$ & $0.8608$ & 1.175 & 1.113 & $552$ - $7632$,  $16$ & 24 & $0.0848(17)$ & $0.3529(69)$  \\
$32^3\times 64$  & $2.25$  & $0.008$ & $0.03$    & $1.864$ & $0.8608$ & 1.175 & 1.113 & $540$ - $5920$,  $10$ & 24 & $0.0864(12)$ & $0.3950(55)$  \\
\end{tabular}
\end{ruledtabular}
\caption{\label{tab:lattices}Summary of lattice parameters. The coupling in the Iwasaki gauge action is given as $\beta=6/g^2$,
and $a m_{u,d}$, $a m_s$ are the bare masses of the domain-wall sea quarks. The
parameters $a m_b$, $u_{0L}$, $c_3$, and $c_4$ enter in the NRQCD action for the $b$ quarks. The lattice spacings, $a$, were
computed in Ref.~\cite{Meinel:2010pv}. The MD (molecular dynamics) range specifies the range of the gauge-field generation Markov chain \cite{Aoki:2010dy}
for which ``measurements'' are performed. The measurements are separated by the given step size in MD time, and are done for $n_{\rm src}$
different source locations [$(\bs{x'}, t')$ in Eq.~(\ref{eq:2ptA})] on each gauge field configuration.}
\end{table*}

The lattice NRQCD action for the $b$ quarks has the same form as in Ref.~\cite{Meinel:2010pv}. It can be written as
\begin{equation}
S_{\psi}=a^3\sum_{\bs{x},t}\psi^\dagger(\bs{x},t)\big[{\psi}(\bs{x},t)
-K(t) \: {\psi}(\bs{x},t-a) \big], \label{eq:latact}
\end{equation}
where $\psi$ is a two-component spinor, and $K(t)$ is given by \cite{Lepage:1992tx}
\begin{equation}
K(t)=\left(1-\frac{a\:\delta H|_t}{2}\right)
\left(1-\frac{a H_0|_t}{2n} \right)^n U_4^\dag(t-a)\left(1-\frac{a H_0|_{t-a}}{2n} \right)^n
\left(1-\frac{a\:\delta H|_{t-a}}{2}\right),\label{eq:mNRQCD_action_kernel}
\end{equation}
with the leading-order kinetic energy operator,
\begin{equation}
H_0 = -\frac{\Delta^{(2)}}{2 m_b}, \label{eq:H0}
\end{equation}
and the following higher-order relativistic and discretization corrections:
\begin{eqnarray}
\nonumber\delta H&=&-c_1\:\frac{\left(\Delta^{(2)}\right)^2}{8 m_b^3}+c_2\:\frac{ig}{8 m_b^2}\:\Big(\bs{\nabla}\cdot\bs{\widetilde{E}}
-\bs{\widetilde{E}}\cdot\bs{\nabla}\Big)\\
\nonumber&&
\label{eq:adder1}\\
\nonumber&&-c_3\:\frac{g}{8 m_b^2}\:\bs{\sigma}\cdot
\left(\bs{\widetilde{\nabla}}\times\bs{\widetilde{E}}
-\bs{\widetilde{E}}\times\bs{\widetilde{\nabla}} \right)-c_4\:\frac{g}{2 m_b}\:\bs{\sigma}\cdot\bs{\widetilde{B}}\\
\nonumber && + c_5\:\frac{a^2\Delta^{(4)}}{24m_b}
-c_6\:\frac{a\left(\Delta^{(2)}\right)^2}{16n\:m_b^2}\\
&&-c_7\:\frac{g}{8 m_b^3}\Big\{ \Delta^{(2)}, \: \bs{\sigma}\cdot\bs{\widetilde{B}} \Big \}
-c_8\:\frac{3g}{64 m_b^4}\left\{ \Delta^{(2)}, \: 
\bs{\sigma}\cdot \left(\bs{\widetilde{\nabla}}\times\bs{\widetilde{E}}-\bs{\widetilde{E}}\times\bs{\widetilde{\nabla}} \right) \right\}
-c_9\:\frac{i g^2}{8 m_b^3}\:\bs{\sigma}\cdot(\bs{\widetilde{E}}\times\bs{\widetilde{E}}).
\label{eq:dH_full}
\end{eqnarray}
Here, $\bs{E}$ and $\bs{B}$ are the chromoelectric and chromomagnetic components of a lattice gluon
field strength tensor. Unlike in the previous sections, the tilde appearing on some of the quantities in Eq.~(\ref{eq:dH_full}) does not denote smearing; instead it denotes
improvement corrections which reduce discretization errors \cite{Lepage:1992tx}. The action is also tadpole-improved \cite{Lepage:1992xa}, with the values
of the Landau gauge mean link $u_{0L}$ as given in Table \ref{tab:lattices}. The heavy-quark masses in lattice units, $a m_b$, are set to the physical
values as determined for the same gauge field ensembles in Ref.~\cite{Meinel:2010pv}.

In Eq.~(\ref{eq:dH_full}), the terms with matching coefficients $c_1$, $c_2$, $c_3$, and $c_4$ are the relativistic corrections of order $v^4$.
The terms with coefficients $c_5$ and $c_6$ are spatial and temporal discretization improvements for $H_0$. Finally,
the terms with coefficients $c_7$, $c_8$, and $c_9$ are the spin-dependent order-$v^6$ terms. In principle, additional
operators containing four (or more) quark fields are introduced through gluon loops, but these are not included here.

At tree level in the matching of NRQCD to QCD, the coefficients $c_i$ in Eq.~(\ref{eq:dH_full}) are all equal to 1.  Because
the terms in $\delta H$ are suppressed relative to $H_0$ by at least one power of $v^2$, using the tree-level values for $c_i$ already provides accuracy
of order $\alpha_s v^2 \approx 0.02$ for the radial and orbital energy splittings in the $b\bar{b}$ and $bbb$ systems.
However, spin splittings first arise through the operators with coefficients $c_3$ and $c_4$, and therefore these two coefficients are tuned nonperturbatively here.
The tuning condition used here is that, when calculated with the lattice NRQCD action, the following two combinations of bottomonium $1P$ energy levels
agree with experiment:
\begin{eqnarray}
&&-2E(\chi_{b0})-3E(\chi_{b1})+5E(\chi_{b2}), \label{eq:bbarspinorbitsplitting} \\
&&-2E(\chi_{b0})+3E(\chi_{b1})-E(\chi_{b2}). \label{eq:bbartensorsplitting}
\end{eqnarray}
As discussed in Ref.~\cite{Gray:2005ur} and confirmed numerically in Ref.~\cite{Meinel:2010pv}, to a good approximation
the combination (\ref{eq:bbarspinorbitsplitting}) is proportional to $c_3$, while (\ref{eq:bbartensorsplitting}) is proportional to $c_4^2$.
Table VII of Ref.~\cite{Meinel:2010pv} gives numerical results for (\ref{eq:bbarspinorbitsplitting}) and (\ref{eq:bbartensorsplitting}),
computed with $c_i=1$ for the same order-$v^6$ NRQCD action on the same gauge field ensembles.
Using these results, one can then solve for $c_3$ and $c_4$ so that the experimental values \cite{Nakamura:2010zzi}
for (\ref{eq:bbarspinorbitsplitting}) and (\ref{eq:bbartensorsplitting}) are reproduced:
\begin{eqnarray}
\nonumber c_3 &=& \left \{ \begin{array}{ll} 1.196 \pm 0.106, & a\approx0.11\:\:{\rm fm}, \\ 1.175 \pm 0.084, & a\approx0.08\:\:{\rm fm}, \end{array}\right.\\
 c_4 &=& \left \{ \begin{array}{ll} 1.168 \pm 0.081, & a\approx0.11\:\:{\rm fm}, \\ 1.113 \pm 0.053, & a\approx0.08\:\:{\rm fm}. \end{array}\right. \label{eq:c3c4}
\end{eqnarray}
In the present work, the main calculations of the $bbb$ spectrum are performed directly at $c_3$ and $c_4$ set equal to the central values in Eq.~(\ref{eq:c3c4}),
and with $c_1=c_2=c_5=c_6=c_7=c_8=c_9=1$.
The uncertainties in (\ref{eq:c3c4}) are mainly statistical, and the resulting uncertainties in the $bbb$ spectrum will be included in the final results (Sec.~\ref{sec:chiral}).

\section{\label{sec:fitting}Fits of the two-point functions and angular momentum identification}

The two-point functions defined in Eq.~(\ref{eq:2ptA}) are labeled by $\Gamma$ and $\Gamma'$, which determine the baryon interpolating operators
at the sink and source, respectively. The two-point functions vanish when $\Gamma$ and $\Gamma'$ correspond to different irreducible representations (irreps)
of the double-cover octahedral group, or when $\Gamma$ and $\Gamma'$ correspond to different rows of the same irrep. In the remaining
cases of equal irrep and equal row at source and sink, one can average over the different rows.
In the following, we use the notation $\Lambda^{(i)}_r$ for row $r$ of the $i$-th operator in irrep $\Lambda$, according to Table \ref{tab:operators}. Then the row-averaged
two-point functions are defined as
\begin{equation}
 C^{(\Lambda)}_{ij}(t-t') = \frac{1}{\mathrm{dim}(\Lambda)} \sum_{r=1}^{\mathrm{dim}(\Lambda)}\:\:C_{\displaystyle \Lambda^{(i)}_r,\:\Lambda^{(j)}_r}(t-t'). \label{eq:rowav2pt}
\end{equation}
\begin{figure*}[t!]
 \hfill \includegraphics[height=6.5cm]{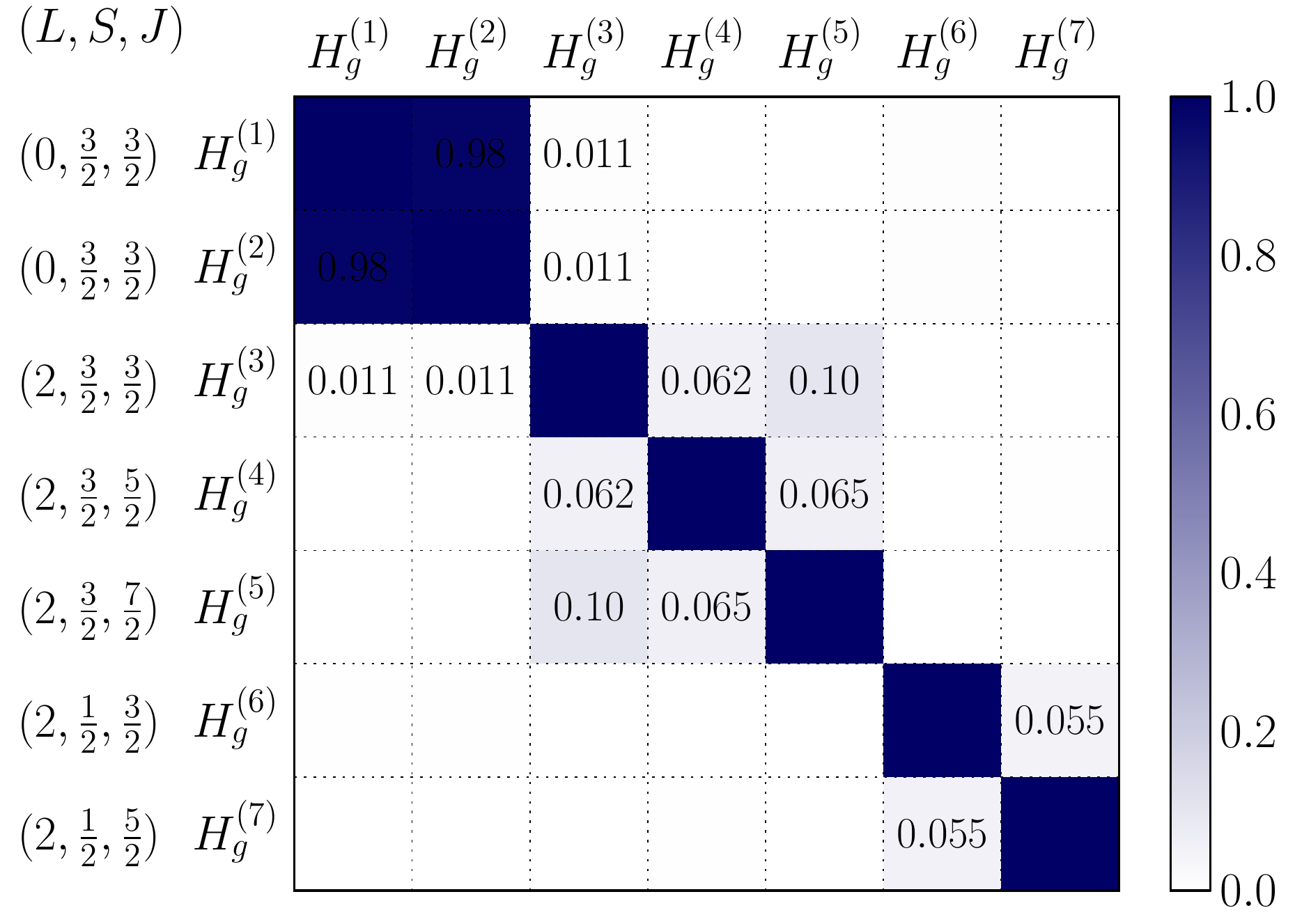} \hfill \includegraphics[height=6.5cm]{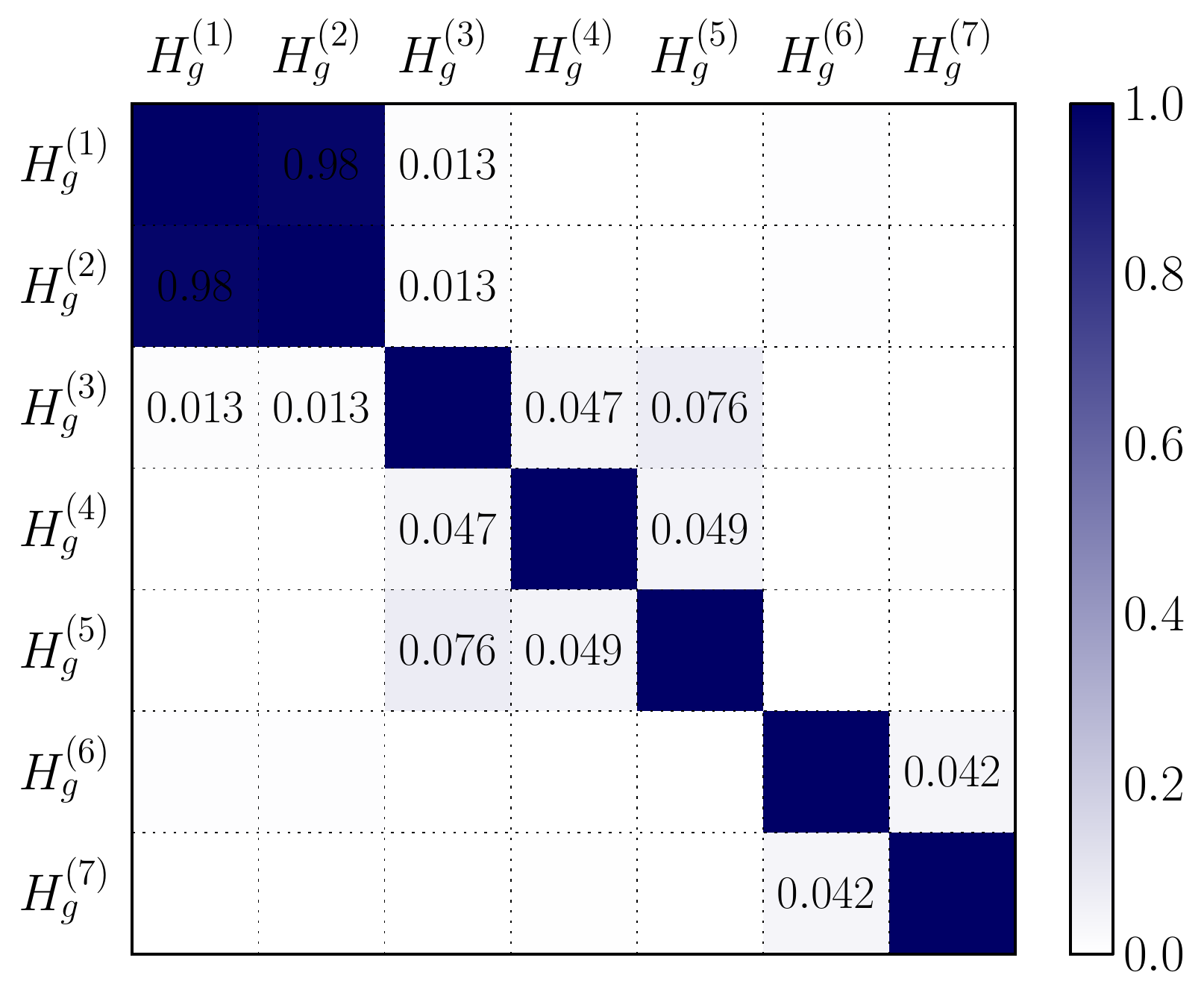}  \hfill \null
\caption{\label{fig:twoptmagnHg} Visualization of rescaled matrix two-point functions $|C_{ij}|/\sqrt{C_{ii} C_{jj}}$
in the $H_g$ irreducible representation, at one time slice. Off-diagonal entries larger than 0.01 are also given numerically (the $i=1$, $j=2$ entry is 0.98).
The values of $L$, $S$, and $J$ from which each operator $H_g^{(i)}$ was subduced are indicated.
Left plot: $a\approx0.11$ fm, $am_{u,d}=0.005$, $(t-t')/a=5$. Right plot: $a\approx0.08$ fm, $am_{u,d}=0.004$, $(t-t')/a=6$. }
\end{figure*}
\begin{figure*}[t!]
 \hfill \includegraphics[height=3.25cm]{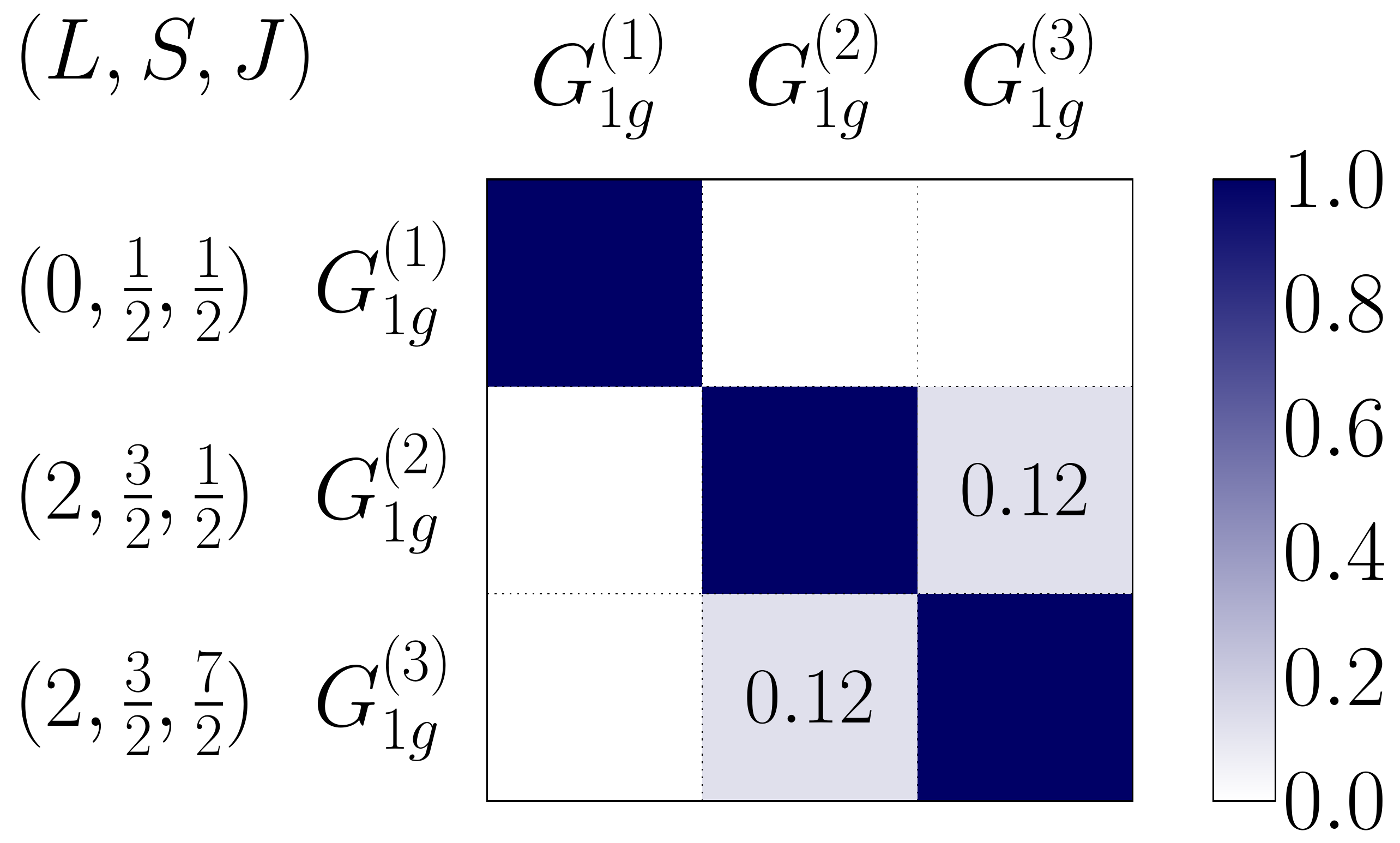} \hfill \includegraphics[height=3.25cm]{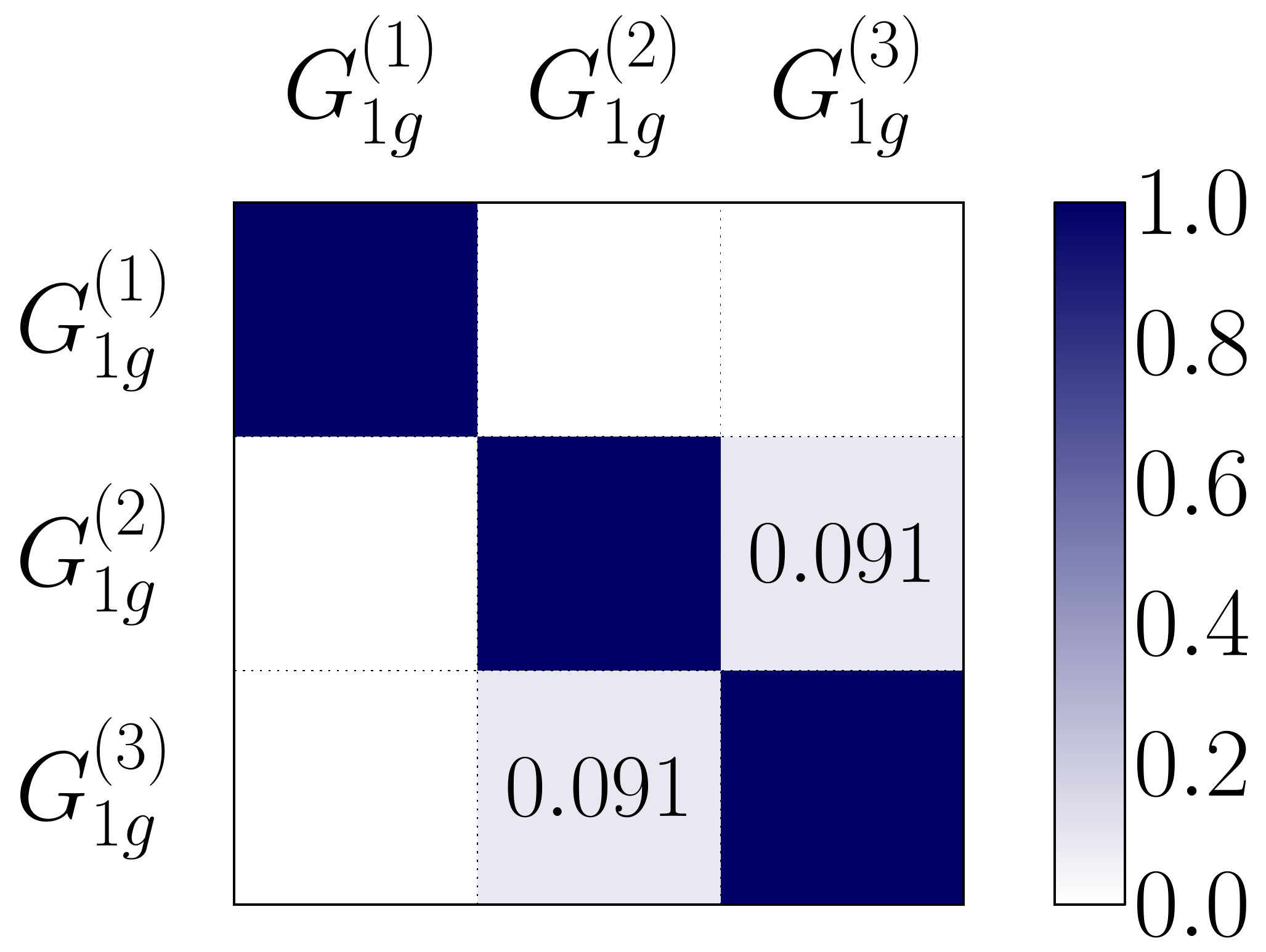}  \hfill \null
\caption{\label{fig:twoptmagnG1g} Like Fig.~\ref{fig:twoptmagnHg}, but for the $G_{1g}$ irreducible representation.}
\end{figure*}
\begin{figure*}[t!]
 \hfill \includegraphics[height=3.25cm]{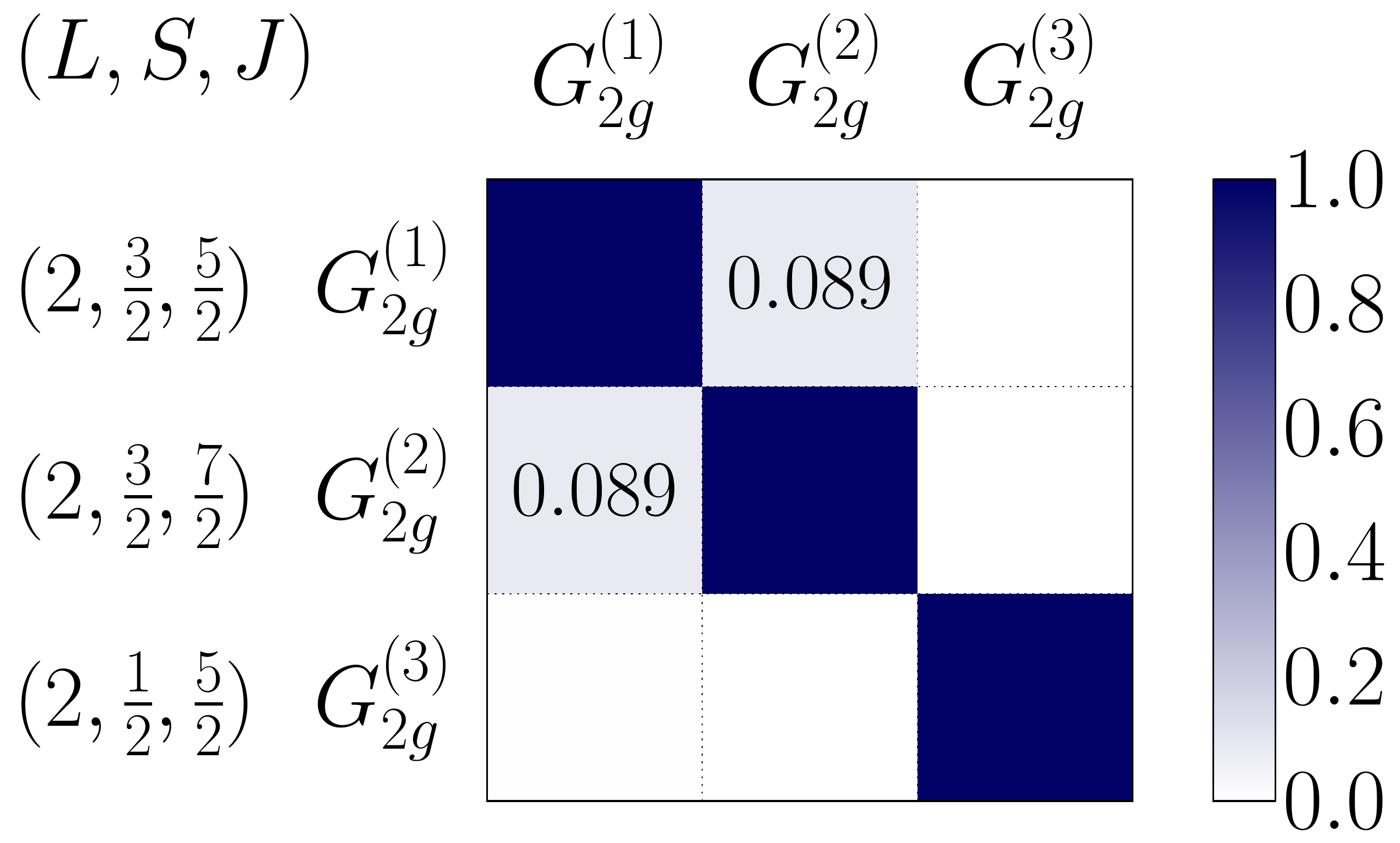} \hfill \includegraphics[height=3.25cm]{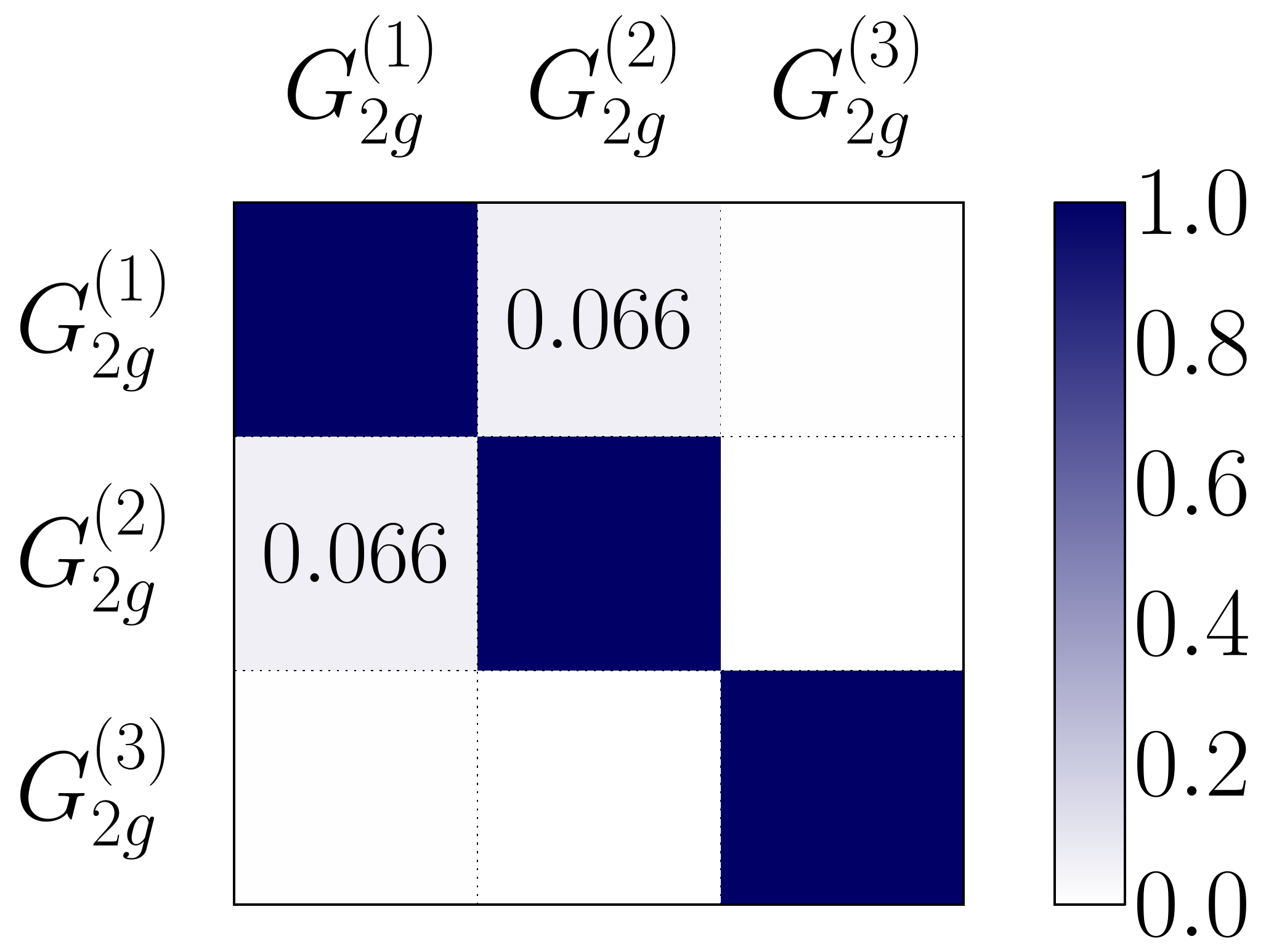}  \hfill \null
\caption{\label{fig:twoptmagnG2g} Like Fig.~\ref{fig:twoptmagnHg}, but for the $G_{2g}$ irreducible representation.}
\end{figure*}
For the operators in Table \ref{tab:operators}, one obtains a $(7\times7)$ matrix of two-point functions
in the $H_g$ irrep, $(3\times3)$ matrices in the $G_{1g}$ and $G_{2g}$ irreps, and $(1\times1)$ ``matrices'' in the $H_u$ and $G_{1u}$ irreps.
The magnitudes of the rescaled two-point functions $|C^{(\Lambda)}_{ij}|/\sqrt{C^{(\Lambda)}_{ii} C^{(\Lambda)}_{jj}}$ at one time slice are shown in Figs.~\ref{fig:twoptmagnHg},
\ref{fig:twoptmagnG1g}, and \ref{fig:twoptmagnG2g} for the $H_g$, $G_{1g}$, and $G_{2g}$ irreps, respectively.
The first important observation is that cross-correlations between operators subduced from continuum operators that differ in \emph{at least one of} the quantum numbers
$L$, $S$, or $J$ are small. Note that $J$ is an exactly conserved quantum number in the continuum, but $L$ and $S$ are not.
The weak coupling between operators subduced from different $J$-values indicates that rotational symmetry breaking by the lattice is small.
This has also been observed in Ref.~\cite{Edwards:2011jj} for light baryons.
On the other hand, the weak coupling between operators subduced from common $J$-values but different $L$- or different $S$-values
is a new feature appearing here.
Because of the large mass of the $b$ quarks, the dynamics is approximately nonrelativistic, and the spin-orbit coupling is
suppressed, so that $L$ and $S$ are approximately conserved. In fact, for the lattice spacings considered here,
the operator overlaps between different $L$- or $S$-values appear to be smaller than that between different $J$-values.
Furthermore, the overlaps between operators subduced from different $J$-values (for example between $H_g^{(3)}$ and $H_g^{(5)}$,
which are subduced from $J=\frac32$ and $J=\frac72$, respectively) appear to be somewhat larger than what was seen for light baryons in Ref.~\cite{Edwards:2011jj}.
This may be a consequence of the much smaller physical extent of the $bbb$ baryons [as modelled
by the initial smearing width of $r_S\approx 0.14$ fm in Eq.~(\ref{eq:smear_op})], which makes
the operators more sensitive to the non-zero lattice spacing.

As can be seen in Fig.~\ref{fig:twoptmagnHg}, there is a strong overlap between the $H_g^{(1)}$ and $H_g^{(2)}$ operators, because both
are subduced from continuum operators with the common quantum numbers $L=0$, $S=\frac32$, $J=\frac32$.
All other cross-correlations, also in the $G_{1g}$ and $G_{2g}$ irreps (Figs.~\ref{fig:twoptmagnG1g} and \ref{fig:twoptmagnG2g})
are small, because there is suppression as a consequence of different $J$, $L$, or $S$.

Further information can be gained by looking at the lattice-spacing dependence of the operator overlaps. In each of the figures,
the left plot shows data from $a\approx0.11$ fm, while the right plot shows data from $a\approx0.08$ fm. It can be seen that the cross-correlations
between operators subduced from different continuum $J$ are smaller at the finer lattice spacing, demonstrating the improvement of rotational symmetry as $a$ is reduced.
On the other hand, the overlaps between $H_g^{(3)}$ and $H_g^{(1)}$, as well as between $H_g^{(3)}$ and $H_g^{(2)}$, are not smaller at the finer lattice spacing. In that
case, the operators are all subduced from the same $J$ ($=\frac32$), and one does not expect the cross-correlations to vanish in the continuum limit.

In this work, the matrix two-point functions in each irrep $\Lambda$ were fitted directly using the form
\begin{equation}
  C^{(\Lambda)}_{ij}(t-t') = \sum_{n=1}^N \:A_{n,i}^{(\Lambda)} \:  A_{n,j}^{(\Lambda)}\: e^{-E_n^{(\Lambda)} (t -t')}.  \label{eq:2ptfit}
\end{equation}
The number of exponentials was chosen to be equal to the dimension of the matrix, i.e.~equal to the number of interpolating operators
for each irrep: $N=7$ for $H_g$, $N=3$ for $G_{1g}$ and $G_{2g}$, and $N=1$ for $H_u$ and $G_{1u}$. Of course, the complete spectral decomposition of the two-point functions
also contains an infinite number of higher-energy exponentials. Therefore, only the data with $t-t' \geq t_{\rm min}$ with sufficiently large $t_{\rm min}$ 
were included in the fit, so that the contributions from these higher states are negligible. The dependence of the results on $t_{\rm min}$ will be discussed later.

The fits performed here fully take into account the statistical correlations between all data points. The dimension of the data correlation matrix 
for an $(N\times N)$ matrix fit is equal to $N_t\: N^2 $, where $N_t$ is the number of time slices included in the fit ($N_t=t_{\rm max}/a-t_{\rm min}/a + 1$).
The definition of $\chi^2$ contains the inverse of this data correlation matrix, and one has to make sure that the number of measurements
used to estimate the data correlation matrix is much larger than its dimension. Because the number of measurements was of order $n_{\rm src} \times n_{\rm cfg} \sim 10^4$ for each ensemble,
these large, fully correlated matrix fits were possible here (for sufficiently small $N_t$). In order to reduce the dimension of the data correlation matrix to $N_t\:N(N+1)/2$ and thereby allow
slightly larger $N_t$, the symmetry of the data in $i$, $j$ (which is exact for infinite statistics) was used. The data for the two-point functions were first symmetrized explicitly measurement by measurement,
and then the fits using Eq.~(\ref{eq:2ptfit}) were performed only for $i\geq j$. 

Within each irrep $\Lambda$, the operators $\Lambda^{(i)}$ in Table \ref{tab:operators} are labeled by $i$ such that they are ordered by the
energy of the state with which they have the strongest overlap (this ordering was not known a priori and was only assigned after some initial fits).
For each irrep $\Lambda$, the amplitudes in Eq.~(\ref{eq:2ptfit}) were then rewritten as follows:
\begin{equation}
  A_{n,i}^{(\Lambda)} = \left\{ \begin{array}{ll} A_{i}^{(\Lambda)}, & \mathrm{for}\:\:n=i, \\ B_{n,i}^{(\Lambda)}\:A_{i}^{(\Lambda)}, & \mathrm{for}\:\:n\neq i, \end{array} \right. \label{eq:Bi}
\end{equation}
using the new parameters $A_{i}^{(\Lambda)}$ and $B_{n,i}^{(\Lambda)}$ instead of $A_{n,i}^{(\Lambda)}$ in the fits.
The parameters $B_{n,i}^{(\Lambda)}$ then describe the overlaps of the operator $\Lambda^{(i)}$ with the other states $n\neq i$,
relative to the state with $n=i$.

Furthermore, the energies $E_n^{(\Lambda)}$ in Eq.~(\ref{eq:2ptfit}) were rewritten for $n>1$ as
\begin{equation}
 E_n^{(\Lambda)} = E_1^{(\Lambda)} + \delta_1^{(\Lambda)} + ... + \delta_{n-1}^{(\Lambda)},\hspace{2ex}\mathrm{with}\hspace{2ex}\delta_n^{(\Lambda)} = E_{n+1}^{(\Lambda)}-E_n^{(\Lambda)}, \label{eq:En}
\end{equation}
using the ground-state energy $E_1^{(\Lambda)}$ and the energy splittings $\delta_1^{(\Lambda)}$, ..., $\delta_{N-1}^{(\Lambda)}$ (all in units of $1/a$) as the actual fit parameters.
When computing $E_n^{(\Lambda)}$ (and other combinations of energy levels) from the fit results for $E_1^{(\Lambda)}$ and $\delta_1^{(\Lambda)}$, ..., $\delta_{N-1}^{(\Lambda)}$,
the uncertainties were added in a fully covariant way, using the parameter covariance matrix obtained from the second derivatives of $\chi^2$.

\begin{figure*}[t!]
\hfill \framebox{\includegraphics[width=0.385\linewidth]{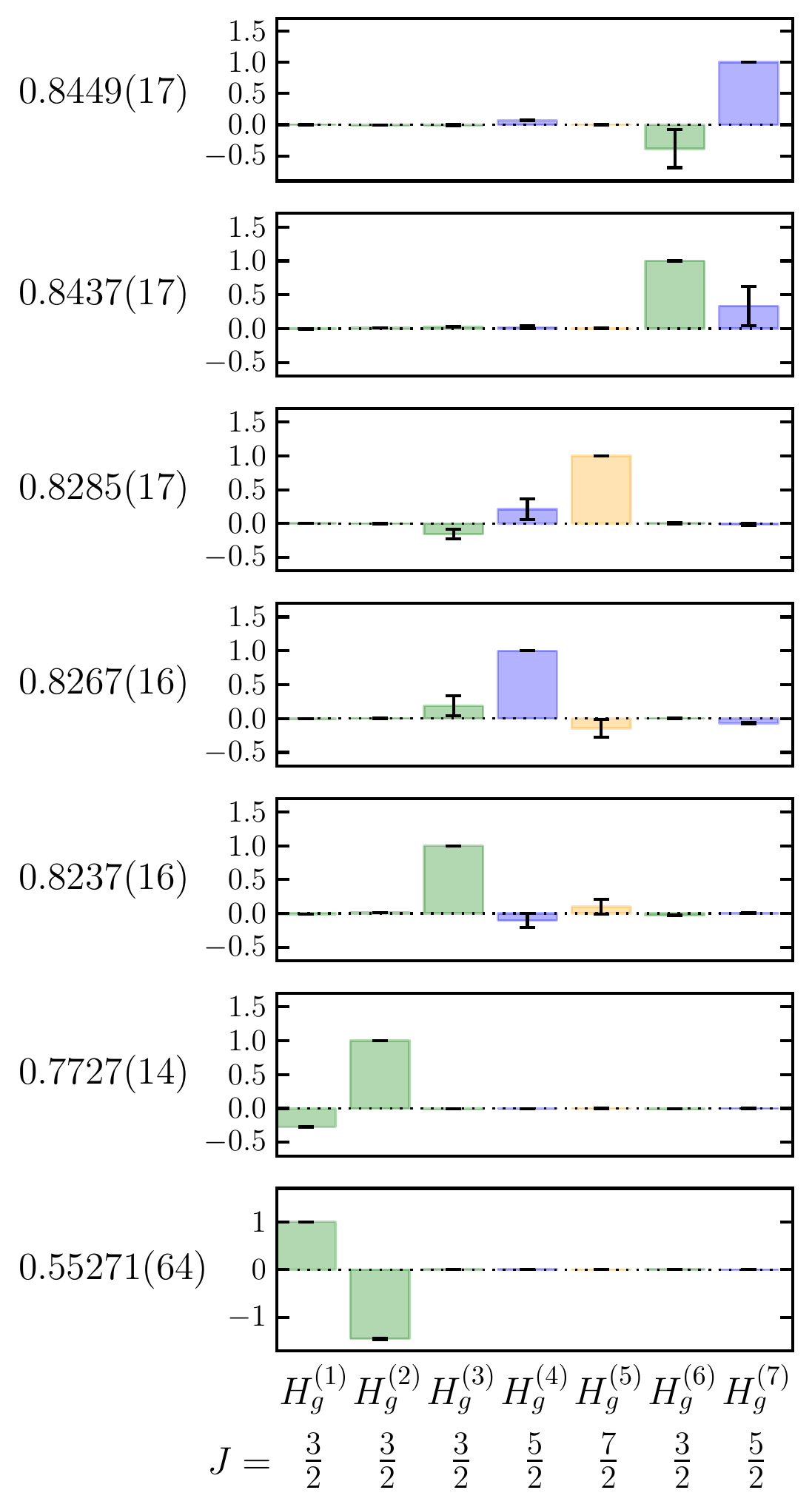}} \hfill \framebox{\includegraphics[width=0.242\linewidth]{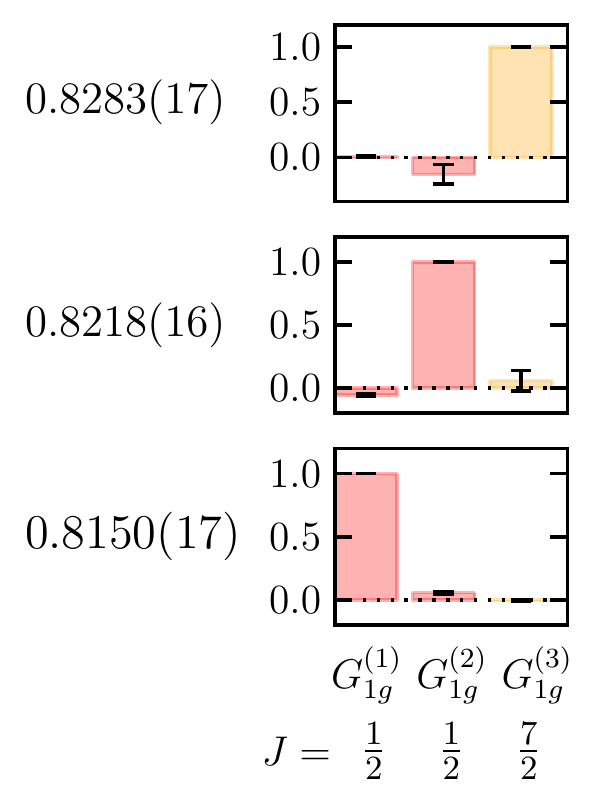}}  \hfill \framebox{\includegraphics[width=0.242\linewidth]{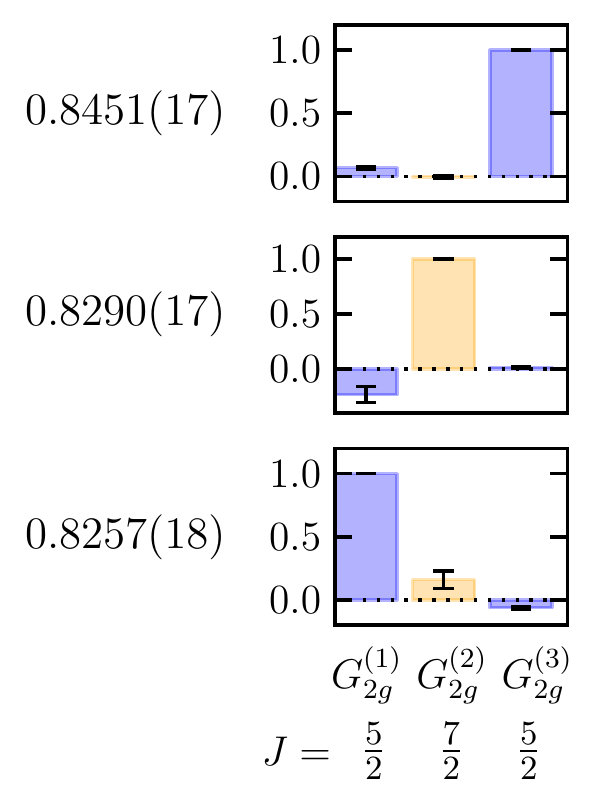}} \hfill \null
\caption{\label{fig:spin_identification_L32_004_Hg} Fitted energies $E_n^{(\Lambda)}$ (in lattice units; from bottom to top: $n=1,...,N$), together
with histograms of the corresponding relative overlap factors $A_{n,i}^{(\Lambda)}/A_{i}^{(\Lambda)}$ [see Eqs.~(\ref{eq:2ptfit}) and (\ref{eq:Bi})].
The fits for the three different irreps were performed independently.
For each $i$, the continuum angular momentum $J$ from which the operator $\Lambda^{(i)}$ was subduced is given at the bottom. These values of $J$ are also
indicated by the colors in the histograms (red: $J=\frac12$, green: $J=\frac32$, blue: $J=\frac52$, orange: $J=\frac72$). The data shown
here are from the ensemble with $a\approx 0.08$ fm and $a m_{u,d}=0.004$; the fits have $t_{\rm min}/a=6$.}
\end{figure*}

Following Ref.~\cite{Edwards:2011jj}, the spectral overlaps $A_{n,i}^{(\Lambda)}$ are used here to assign values of the continuum angular momentum $J$ to each
energy level $E_n^{(\Lambda)}$. Examples of fitted energies $E_n^{(\Lambda)}$, together with the relative overlap factors $A_{n,i}^{(\Lambda)}/A_{i}^{(\Lambda)}$,
are shown in Fig.~\ref{fig:spin_identification_L32_004_Hg} for the $H_g$, $G_{1g}$, and $G_{2g}$ irreps (in the cases of the $G_{1u}$ and $H_u$ irreps, there
is only one operator each, subduced trivially from $J=\frac12$ and $J=\frac32$, respectively). The angular momentum identification proceeds as follows: for
each energy level $E_n^{(\Lambda)}$, the operator $\Lambda^{(i)}$ with the largest relative overlap factor $A_{n,i}^{(\Lambda)}/A_{i}^{(\Lambda)}$ is determined.
The value of $J$ from which this operator was subduced is then assigned to this energy level. As can be seen in Fig.~\ref{fig:spin_identification_L32_004_Hg}, no ambiguity arises here.
Notice that the two $J=\frac52$ levels appearing in the $H_g$ irrep also show up in the $G_{2g}$ irrep, with nearly identical energies. Similarly, the $J=\frac72$ level
appears in all three irreps $H_g$, $G_{1g}$, and $G_{2g}$, again with nearly identical energies. For these levels, the absolute overlap factors were
also found to be consistent across the different irreps, confirming the assignment of $J$.

Because of the strong statistical correlations across irreps, the tiny splittings of the $J=\frac52$ and $J=\frac72$ levels
into the different lattice irreps, which are caused by rotational symmetry breaking, can be computed with smaller uncertainties than the individual energies of these levels.
To this end, simultaneous fits of the two-point functions in the $H_g$, $G_{1g}$ and $G_{2g}$ irreps were performed, where a global correlated
$\chi^2$ was formed but all fit parameters remained independent for each irrep. The results for the rotational-symmetry-breaking-induced energy splittings,
converted to MeV, are given in Table \ref{tab:irrep_splittings} for two gauge field ensembles.
Up to some statistical fluctuations, the splittings are smaller at $a\approx 0.08$ fm compared to $a\approx0.11$ fm, consistent with the 
discretization errors proportional to $\alpha_s\: a^2$ that are expected for the improved lattice NRQCD action used here.
Along with the behavior of the off-diagonal matrix elements that was discussed at the beginning of this section, the results shown
in Table \ref{tab:irrep_splittings} provide another demonstration of the improvement of rotational symmetry when
the lattice spacing $a$ is reduced.

\begin{table}
\begin{tabular}{cccccccc}
\hline\hline
Continuum $J^P$  & \hspace{2ex} &  Splitting  & \hspace{2ex} & $a\approx 0.11$ fm  & \hspace{2ex} &  $a \approx 0.08$ fm \\
\hline
\\[-2.5ex]
$\frac52^+$ &&  $E_4^{(H_{g})}-E_1^{(G_{2g})}$    && $5.8(2.0)\nb$    &&  $2.5(2.0)\nb$ \\
\\[-2.5ex]
$\frac52^+$ &&  $E_3^{(G_{2g})}-E_7^{(H_{g})}$    && $0.70(44)\wdt$   &&  $0.44(64)\wdt$ \\
\\[-2.5ex]
$\frac72^+$ &&  $E_2^{(G_{2g})}-E_3^{(G_{1g})}$   && $2.1(1.1)\nb$    &&  $1.6(1.4)\nb$ \\
\\[-2.5ex]
$\frac72^+$ &&  $E_5^{(H_{g})}-E_3^{(G_{1g})}$    && $1.49(78)\wdt$   &&  $0.38(79)\wdt$ \\
\\[-2.5ex]
$\frac72^+$ &&  $E_2^{(G_{2g})}-E_5^{(H_{g})}$    && $0.59(45)\wdt$   &&  $1.24(72)\wdt$ \\
\\[-2.5ex]
\hline\hline
\end{tabular}
\caption{\label{tab:irrep_splittings}Splitting of continuum energy levels with $J>\frac32$ into different irreducible representations of the double-cover octahedral group. All results in MeV.
The data at $a\approx 0.11$ fm are from the ensemble with $a m_{u,d}=0.005$, while the data at $a\approx0.08$ fm are from the ensemble with $a m_{u,d}=0.004$.}
\end{table}

\begin{figure*}[ht!]
 \framebox{\includegraphics[width=0.67\linewidth]{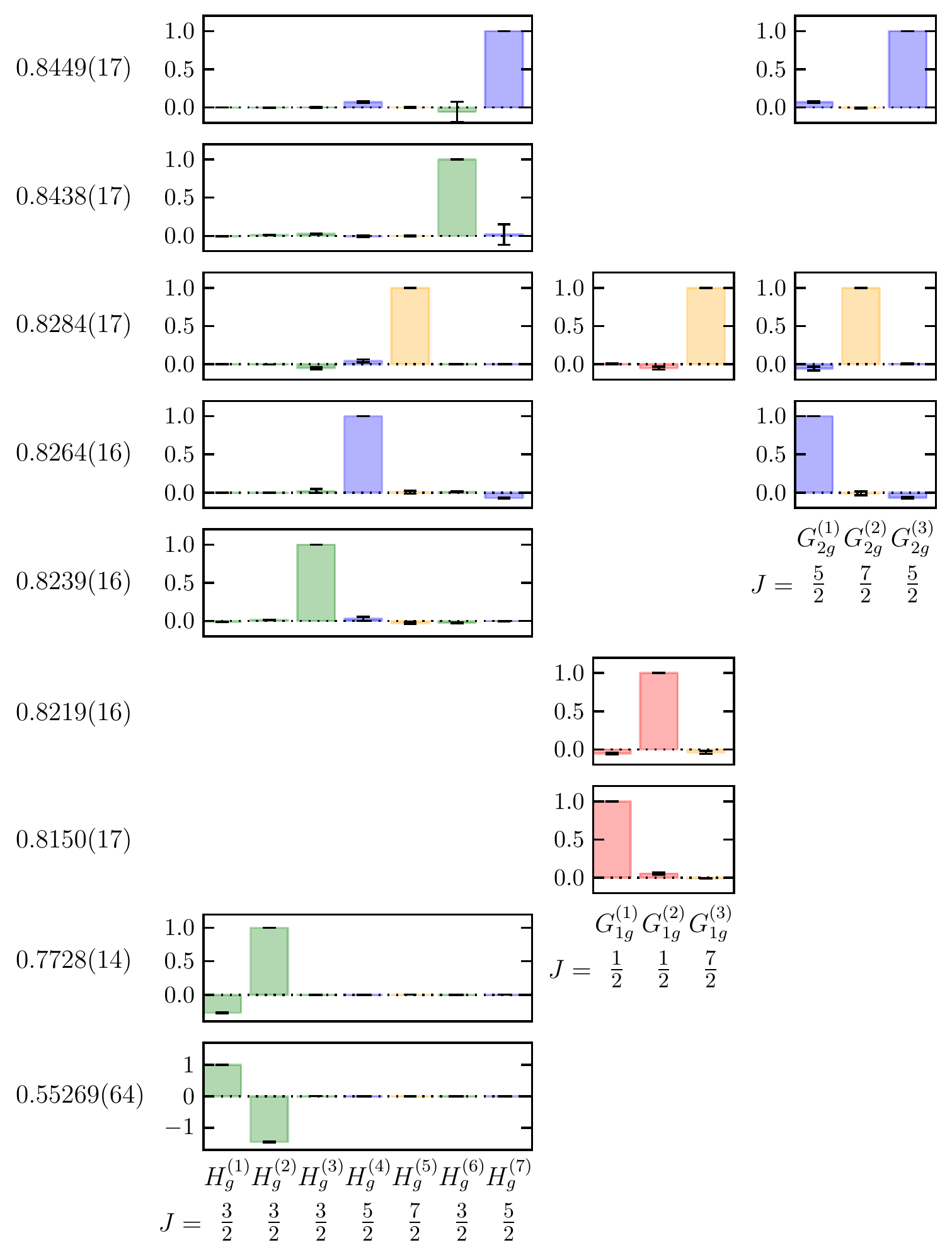}}
\caption{\label{fig:spin_identification_L32_004_simultaneous}Like Fig.~\ref{fig:spin_identification_L32_004_Hg}, but for a coupled fit containing the $H_g$, $G_{1g}$, and $G_{2g}$ irreps,
where the equalities of common $J=\frac52$ and $J=\frac72$ energy levels are enforced: $E_4^{(H_g)}=E_1^{(G_{2g})}$, $E_7^{(H_g)}=E_3^{(G_{2g})}$, and
$E_5^{(H_g)}=E_3^{(G_{1g})}=E_2^{(G_{2g})}$.}
\end{figure*}

Finally, to get the best possible estimates of the continuum energy levels, new simultaneous fits of the two-point functions
in the $H_g$, $G_{1g}$ and $G_{2g}$ irreps were performed, in which the fitted energies for the matching $J=\frac52$ and $J=\frac72$ levels
in different irreps were forced to be equal:
\begin{eqnarray}
\nonumber && E_4^{(H_g)}=E_1^{(G_{2g})}, \\
\nonumber && E_7^{(H_g)}=E_3^{(G_{2g})}, \\
&& E_5^{(H_g)}=E_3^{(G_{1g})}=E_2^{(G_{2g})}. \label{eq:energyequalities}
\end{eqnarray}
This was implemented by augmenting the $\chi^2$ function of the simultaneous fit in the following way:
\begin{eqnarray}
\nonumber \chi^2 &\rightarrow& \chi^2 + \left[ E_4^{(H_g)} - E_1^{(G_{2g})} \right]^2  / \sigma^2 + \left[ E_7^{(H_g)} - E_3^{(G_{2g})} \right]^2  / \sigma^2 \\
&& \phantom{\chi^2} + \left[ E_5^{(H_g)} - E_3^{(G_{1g})} \right]^2  / \sigma^2 + \left[ E_3^{(G_{1g})} - E_2^{(G_{2g})} \right]^2  / \sigma^2, \label{eq:augmchisqr}
\end{eqnarray}
where the energies $E_n^{(\Lambda)}$ are expressed in terms of the actual fit parameters as $E_n^{(\Lambda)} = E_1^{(\Lambda)} + \delta_1^{(\Lambda)} + ... + \delta_{n-1}^{(\Lambda)}$.
The width $\sigma$ in Eq.~(\ref{eq:augmchisqr}) was chosen about two orders of magnitude smaller than the typical statistical uncertainty in the energies. By minimizing the augmented
$\chi^2$, fit parameters are returned that satisfy the conditions (\ref{eq:energyequalities}) up to the input width $\sigma$. These
new fits still had $\chi^2/{\rm d.o.f}\approx 1$, because of the smallness of the energy splittings between the different irreps. Performing
the simultaneous fit with the enforced relations (\ref{eq:energyequalities}) also stabilizes the extraction of the very close energy levels (such as $E_6^{(H_g)}$
and $E_7^{(H_g)}$), and makes the spectral overlap factors more sharply peaked, as can be seen in Fig.~\ref{fig:spin_identification_L32_004_simultaneous}.
Note that in this work no further constraints beyond that of Eq.~(\ref{eq:augmchisqr}) were imposed on any of the fit parameters.

These simultaneous fits, along with simple one-exponential fits in the $H_u$ and $G_{1u}$ irreps, yield 11 different $bbb$ energy levels. Having
performed the angular momentum identification, these levels can now be labeled by $J^P$ and a new subscript counting the states in each $J^P$
channel by increasing energy:
\begin{eqnarray}
\nonumber && E_1(\sfrac12^+),\: E_2(\sfrac12^+), \\
\nonumber && E_1(\sfrac32^+),\: E_2(\sfrac32^+),\: E_3(\sfrac32^+),\: E_4(\sfrac32^+), \\
\nonumber && E_1(\sfrac52^+),\: E_2(\sfrac52^+), \\
\nonumber && E_1(\sfrac72^+), \\
\nonumber && E_1(\sfrac12^-), \\
          && E_1(\sfrac32^-).
\end{eqnarray}
Because NRQCD is used in this work, the extracted energies do not include the rest masses of the three $b$ quarks, i.e. they
are all shifted by a common amount that is not known with sufficient precision. Therefore, only energy \emph{differences} are considered
in the following.

\begin{figure*}
 \centerline{\includegraphics[width=0.9\linewidth]{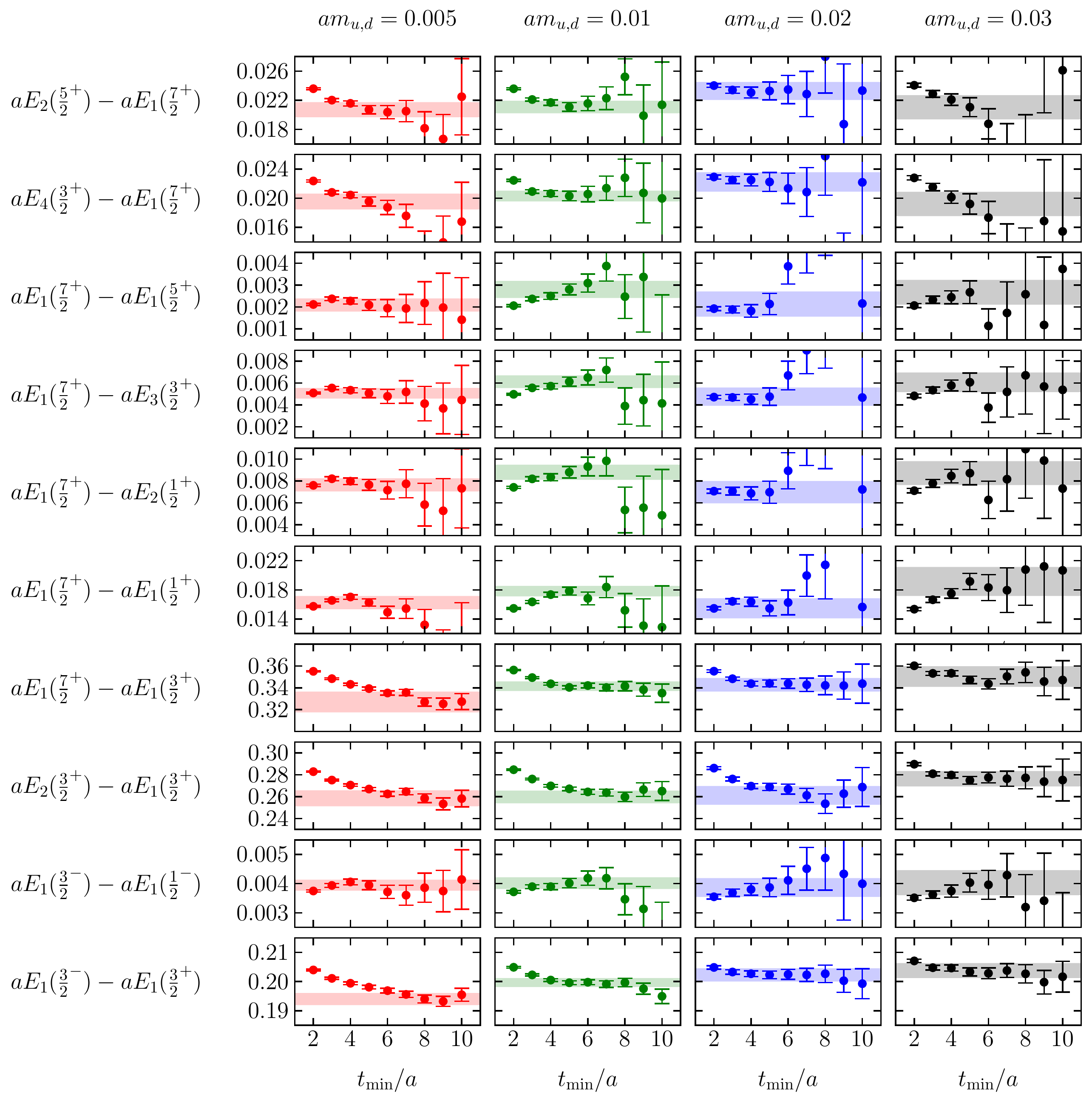}}
\caption{\label{fig:E_vs_tmin_L24}Dependence of the results for the $bbb$ energy splittings on the start time slice $t_{\rm min}$ that is used in the fit. The data shown here are for the
ensembles with $a\approx0.11$ fm, with the light quark masses of $a m_{u,d}=0.005,\: 0.01,\: 0.02,\: 0.03$ (from left to right). The shaded bands
indicate the best possible estimates of the energy splittings.}
\end{figure*}

\begin{figure*}
 \centerline{\includegraphics[width=0.78\linewidth]{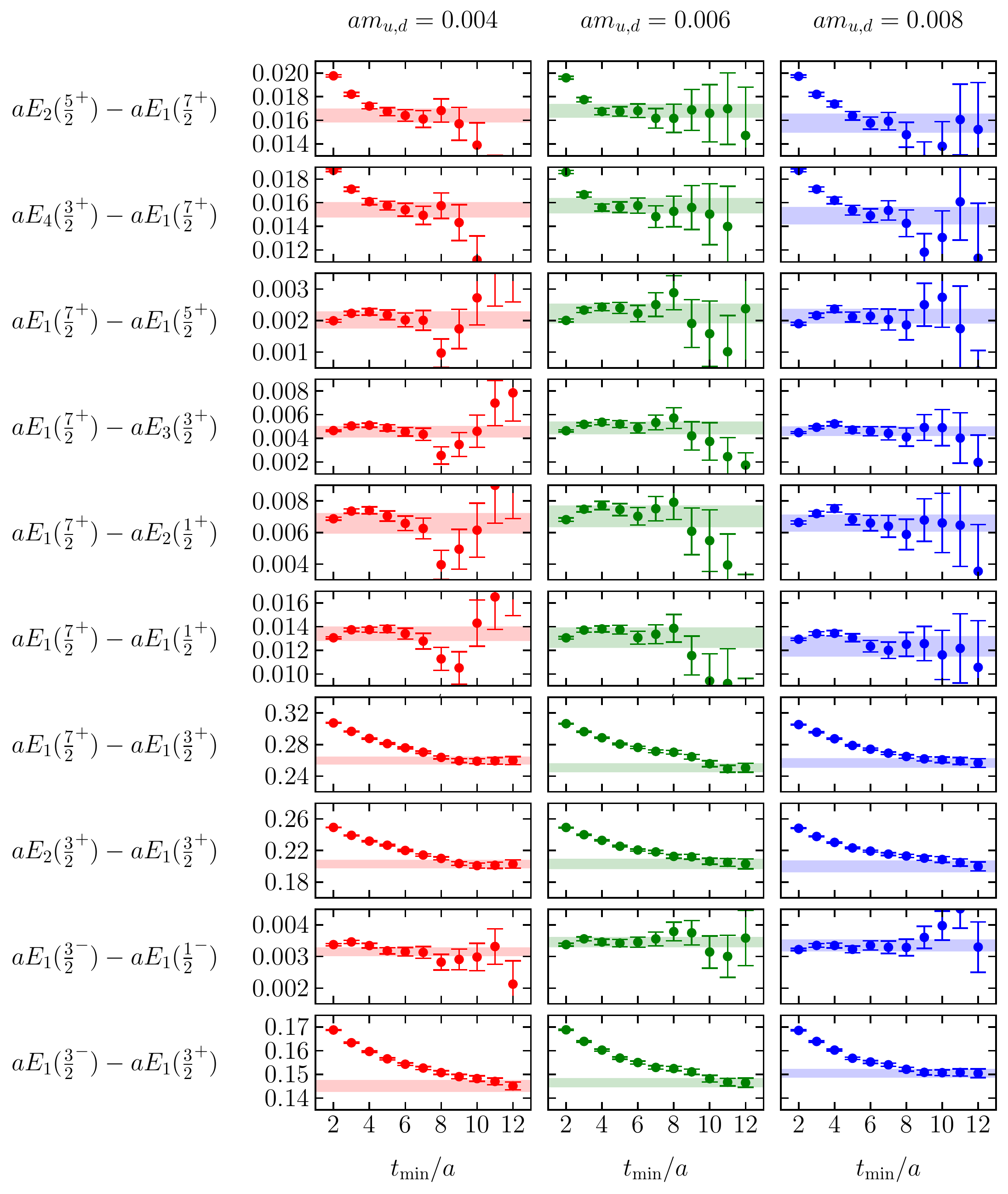} }
\caption{\label{fig:E_vs_tmin_L32}Dependence of the results for the $bbb$ energy splittings on the start time slice $t_{\rm min}$ that is used in the fit. The data shown here are for the
ensembles with $a\approx0.08$ fm, with the light quark masses of $a m_{u,d}=0.004,\: 0.006,\: 0.008$ (from left to right). The shaded bands
indicate the best possible estimates of the energy splittings.}
\end{figure*}

The remaining point to be discussed in this section is the choice of $t_{\rm min}$, the starting time slice from which the fits are performed.
This parameter has to be chosen large enough such that the contamination from higher-excited states, which decay exponentially with $t$, is negligible.
However, $t_{\rm min}$ must not be made too large either, as the statistical uncertainties increase with $t_{\rm min}$ and the fits eventually become unstable.
Figures \ref{fig:E_vs_tmin_L24} and \ref{fig:E_vs_tmin_L32} show the $t_{\rm min}$-dependence of the set of ten independent energy splittings chosen here.
For the matrix two-point functions in the $H_g$, $G_{1g}$, and $G_{2g}$ irreps, the total number of time slices included in the fit,
$N_t=t_{\rm max}/a-t_{\rm min}/a + 1$, was held constant as $t_{\rm min}$ was varied, to keep the dimension of the data correlation matrix fixed at a manageable size
($N_t=5,8,8$ for the $H_g$, $G_{1g}$, $G_{2g}$ irreps, respectively).

As can be seen in Figs.~\ref{fig:E_vs_tmin_L24} and \ref{fig:E_vs_tmin_L32}, for the energy splittings $aE_1(\frac32^-)-aE_1(\frac32^+)$, $aE_2(\frac32^+)-aE_1(\frac32^+)$, and $aE_1(\frac72^+)-aE_1(\frac32^+)$,
which are large energy differences between $bbb$ states of rather different spatial structure, the plateaus set in later than for the other, smaller splittings,
which mainly constitute the fine- and hyperfine structure. To extract the best possible estimates for the further analysis,
at the coarse lattice spacing the three large energy splittings were taken from the fits with $t_{\rm min}/a=8$ or $t_{\rm min}/a=7$, while the other splittings were taken from $t_{\rm min}/a=5$.
At the fine lattice spacing, $t_{\rm min}/a=12$ was selected for the three large splittings, and $t_{\rm min}/a=6$ for all other splittings.
Possible remaining systematic uncertainties resulting from the choice of $t_{\rm min}/a$ were estimated by computing the shift in the energy splittings when
reducing $t_{\rm min}/a$ from the selected values by one unit. These shifts were added in quadrature to the original statistical uncertainties,
and the resulting total fitting uncertainties are indicated by the shaded bands in Figs.~\ref{fig:E_vs_tmin_L24} and \ref{fig:E_vs_tmin_L32}.

\clearpage

\section{\label{sec:chiral}Final results for the $\mathbi{bbb}$ spectrum}

In the previous section, ten $bbb$ energy splittings were computed for each of the seven different ensembles of gauge fields.
These results are given by the horizontal bands in Figs.~\ref{fig:E_vs_tmin_L24} and \ref{fig:E_vs_tmin_L32}. The values of the
light sea-quark masses used in the generation of the gauge field ensembles correspond to pion masses that are larger than physical (see Table \ref{tab:lattices}).
The final step of the analysis is to perform extrapolations of the $bbb$ spectrum to the physical value of the pion mass.
These extrapolations are done here using the same method that was used for the bottomonium spectrum in Ref.~\cite{Meinel:2010pv}.
The light quarks influence the $bbb$ spectrum only through their vacuum-polarization effects, and the dependence
on $m_{u,d}$ is weak. Therefore, it is sufficient to perform the extrapolations linearly in $m_{u,d}$, and hence linearly in $m_\pi^2$.

The $bbb$ energy splittings were first converted to MeV using the values of the lattice spacings as given in Table \ref{tab:lattices}.
Then, coupled fits to the data for the two different values of the gauge coupling, $\beta_1=2.25$ and $\beta_2=2.13$, were performed using
\begin{eqnarray}
\nonumber E(m_\pi^2,\: \beta_1) &=& E(0, \beta_1)+A \:m_\pi^2,\\
E(m_\pi^2,\: \beta_2) &=& E(0, \beta_2)+A \:m_\pi^2, \label{eq:sim_chiral_extrap}
\end{eqnarray}
where $E(m_\pi^2, \beta)$ denotes a generic $bbb$ energy splitting. The ensembles with $\beta=\beta_1$ have $a\approx0.08$ fm, while
the ensembles with $\beta=\beta_2$ have $a\approx0.11$ fm. The free fit parameters in Eq.~(\ref{eq:sim_chiral_extrap}) are
$E(0, \beta_1)$, $E(0, \beta_2)$, and $A$. No continuum extrapolation is performed here, because lattice NRQCD is an effective field
theory that requires a cut-off $a^{-1} \lesssim m_b$. The only assumption made here is that higher-order effects proportional
to terms like $a^2 m_\pi^2$ are negligible, so that the same parameter $A$ can be used for both values of $\beta$.

\begin{figure*}[ht!]
 \includegraphics[width=0.45\linewidth]{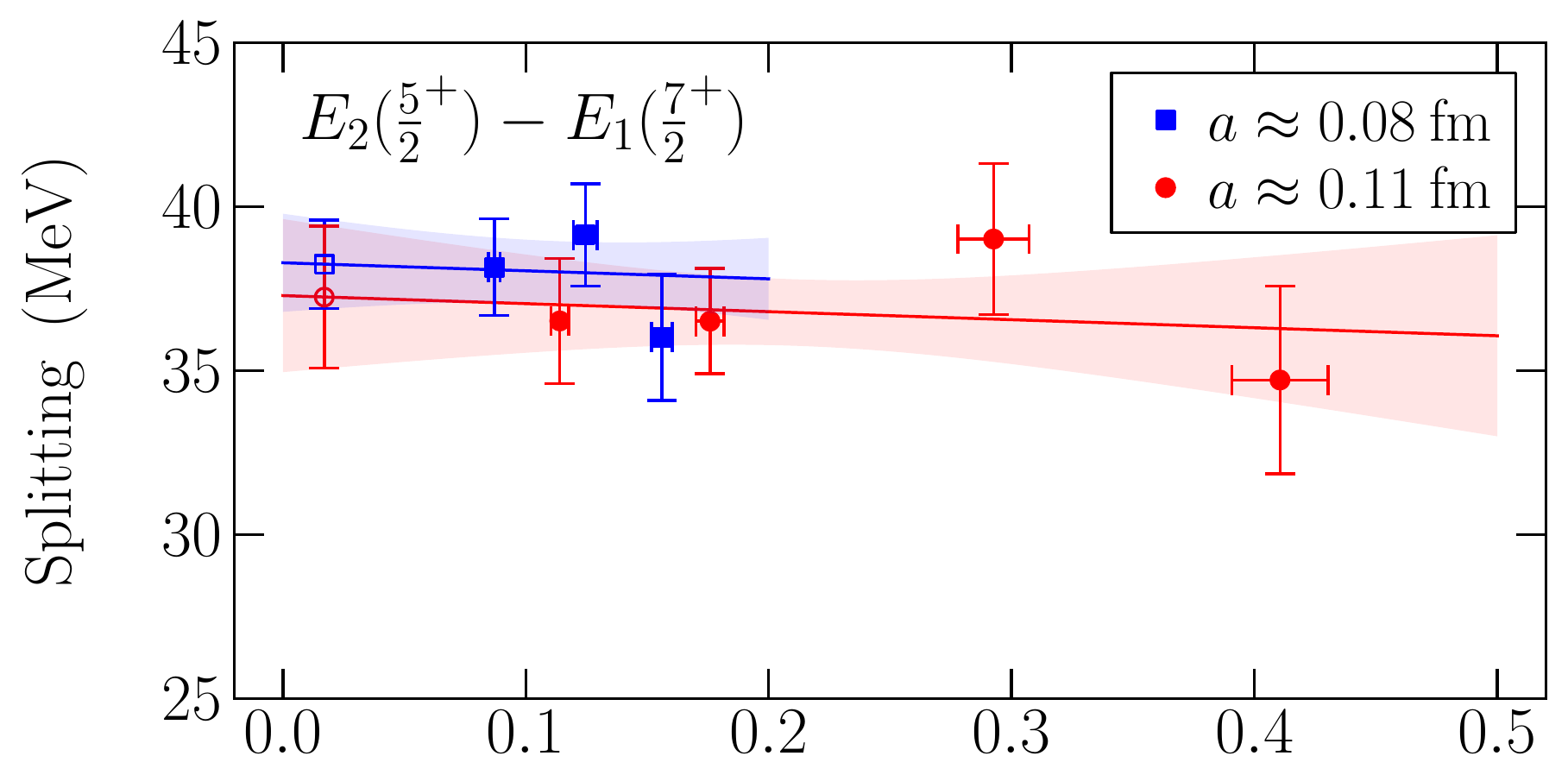}   \hfill \includegraphics[width=0.45\linewidth]{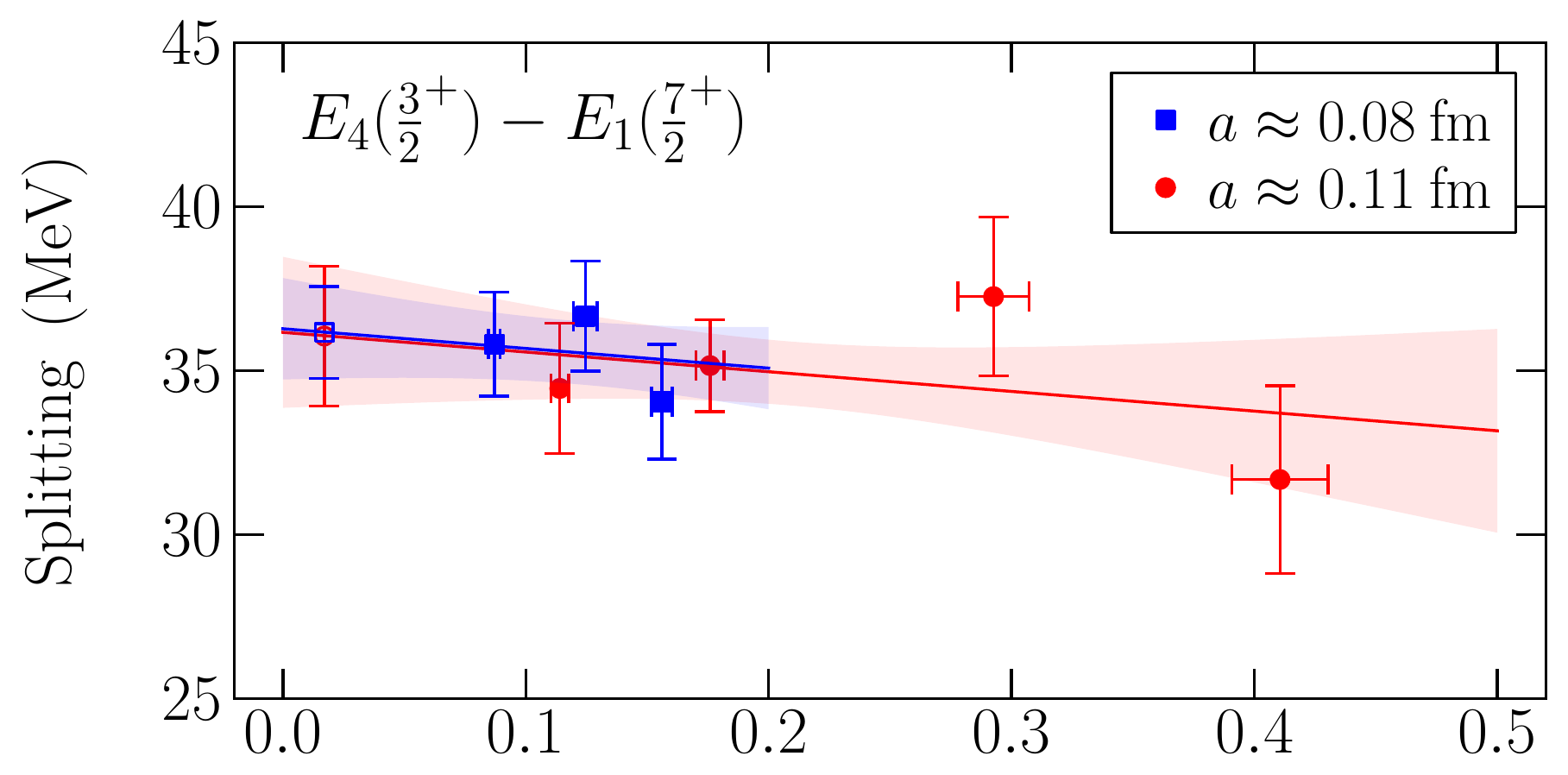}
 \includegraphics[width=0.45\linewidth]{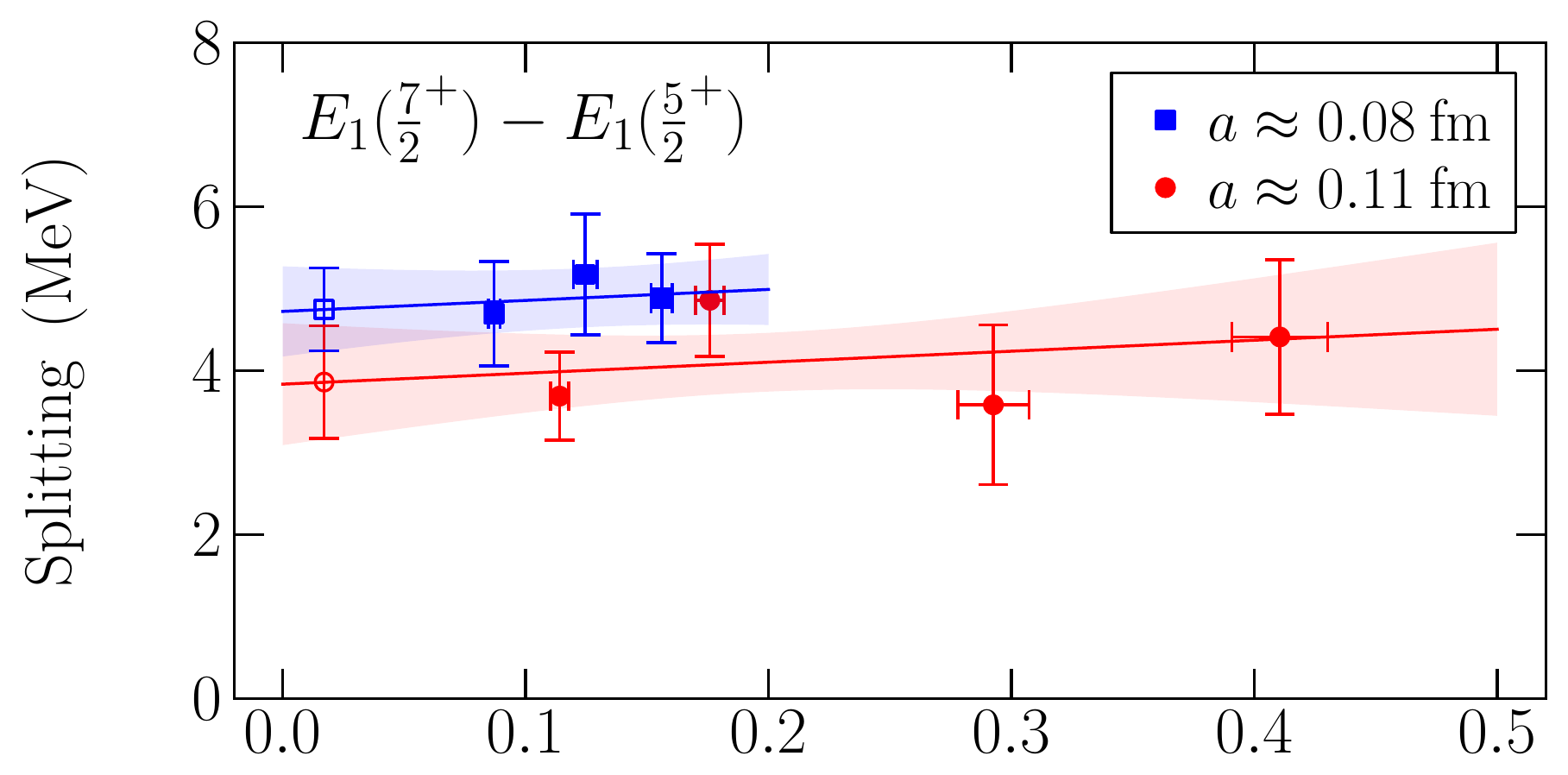}   \hfill \includegraphics[width=0.45\linewidth]{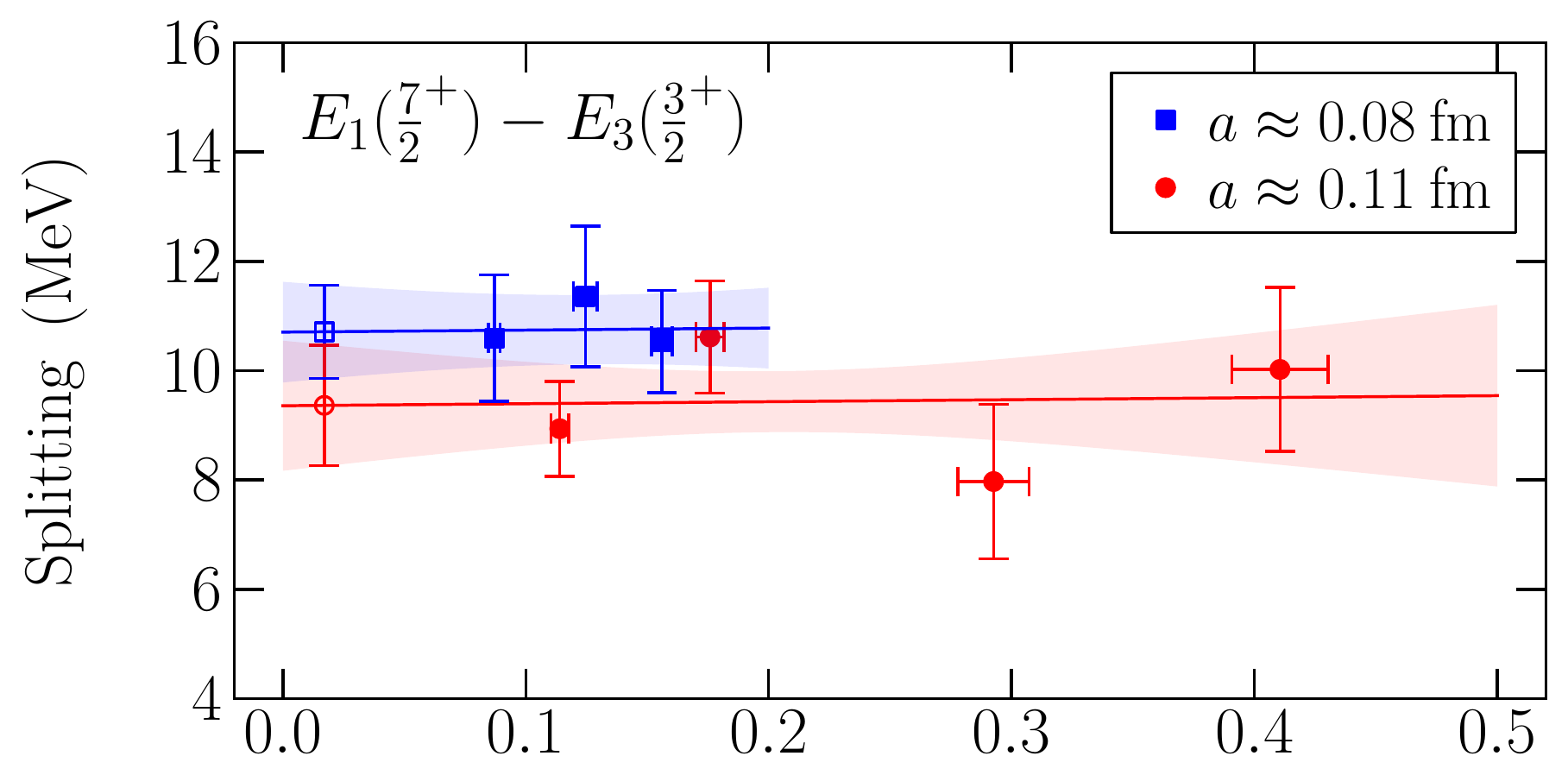}
 \includegraphics[width=0.45\linewidth]{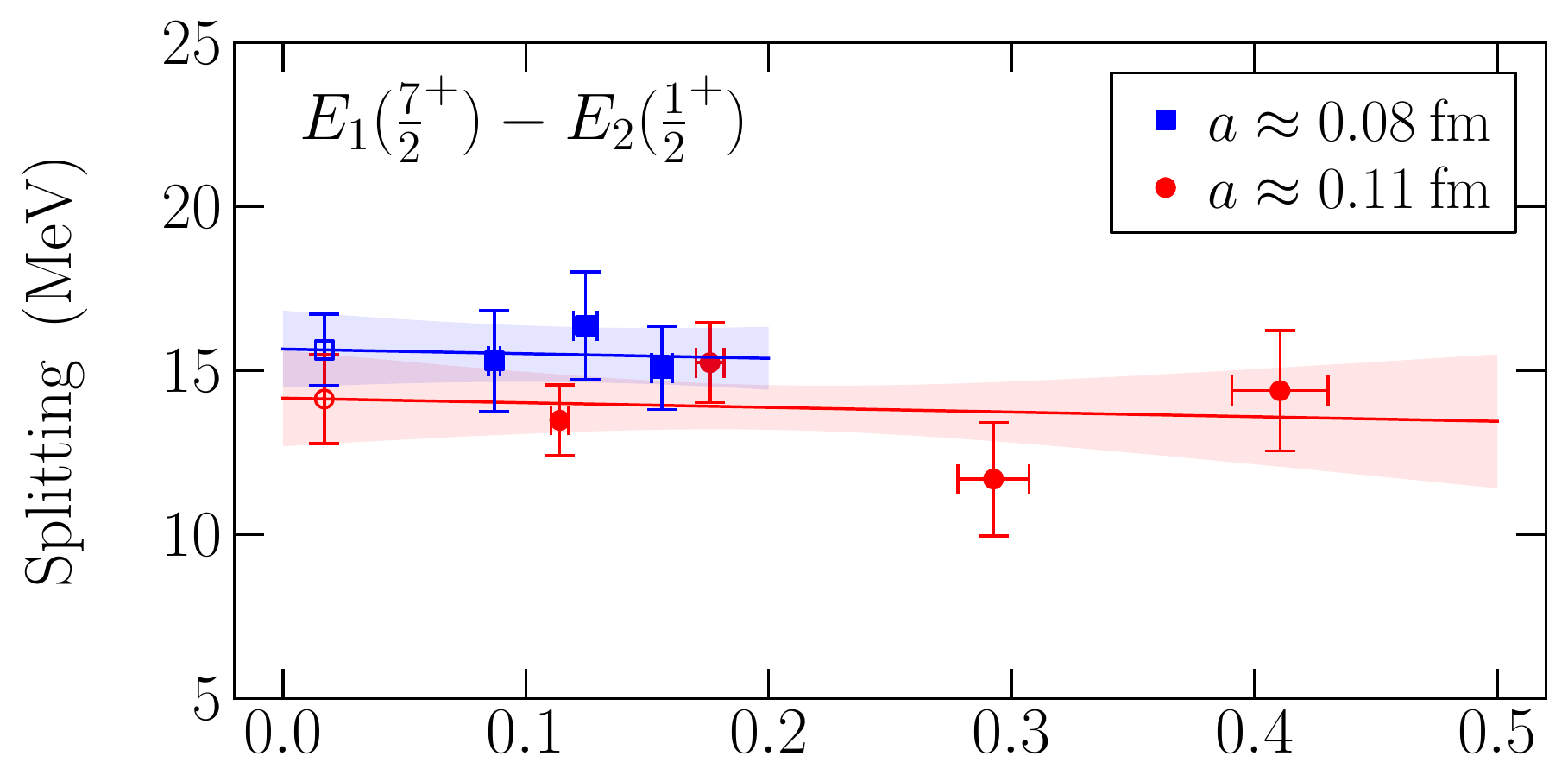}   \hfill \includegraphics[width=0.45\linewidth]{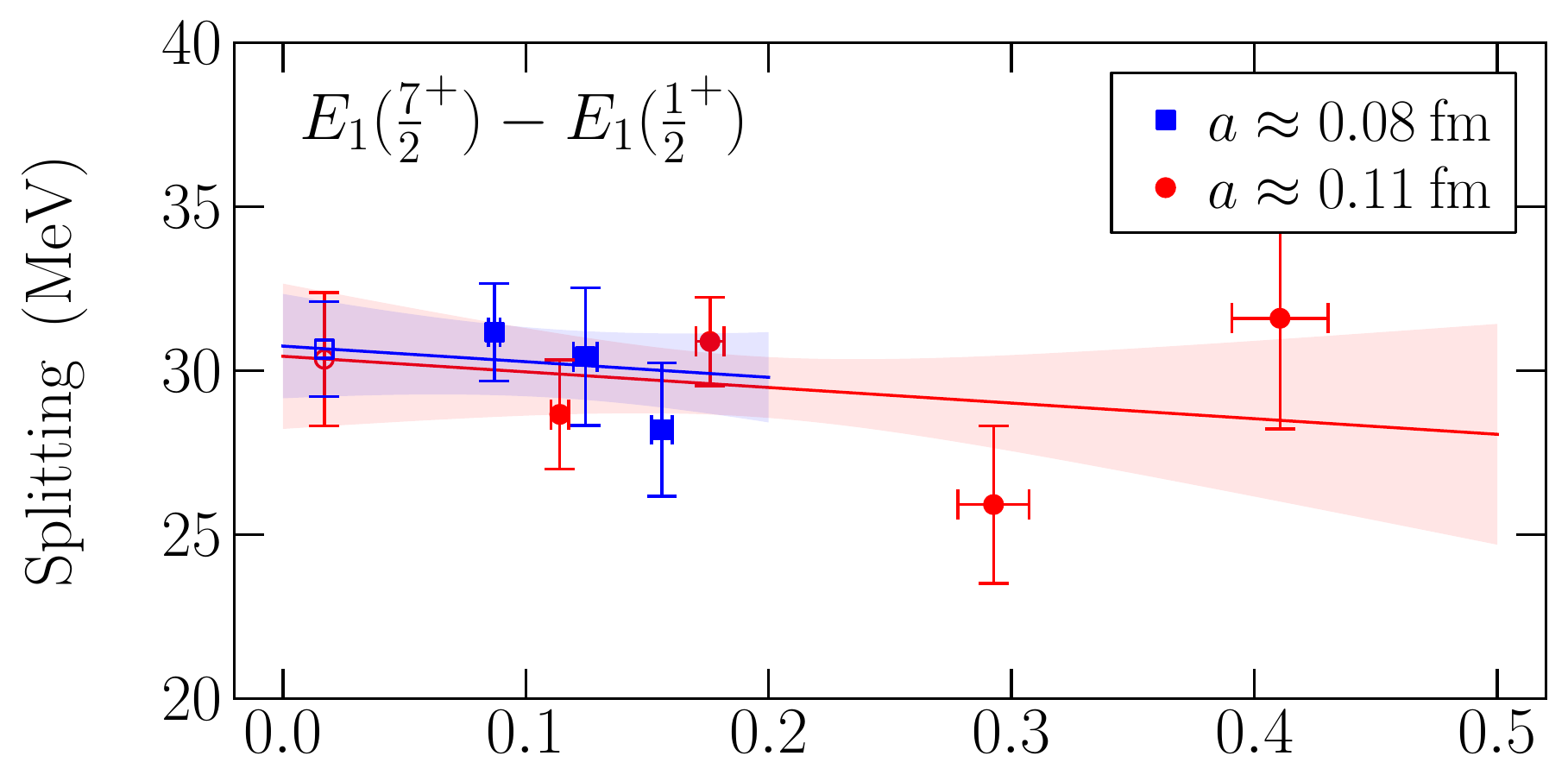}
 \includegraphics[width=0.45\linewidth]{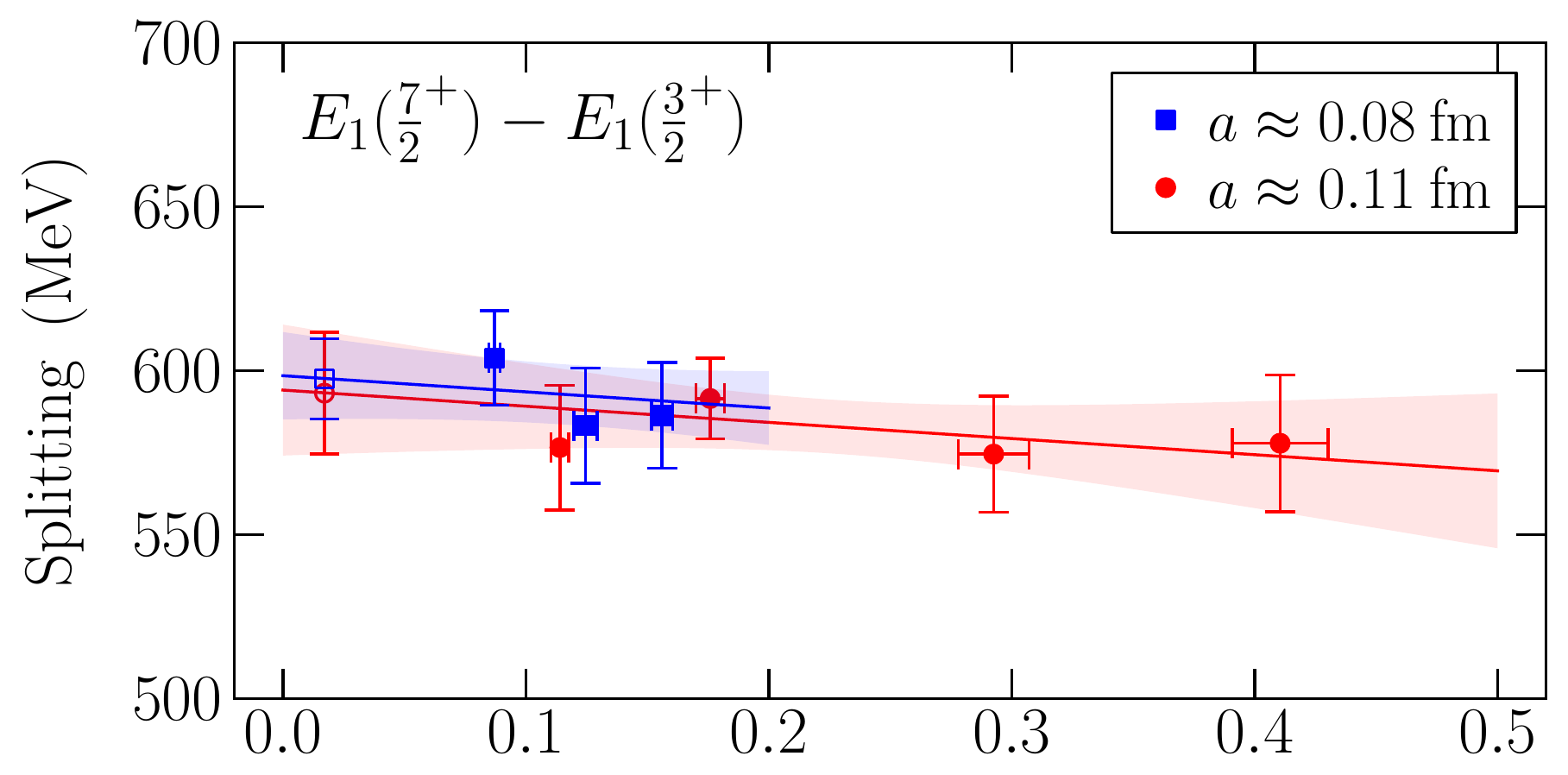}   \hfill \includegraphics[width=0.45\linewidth]{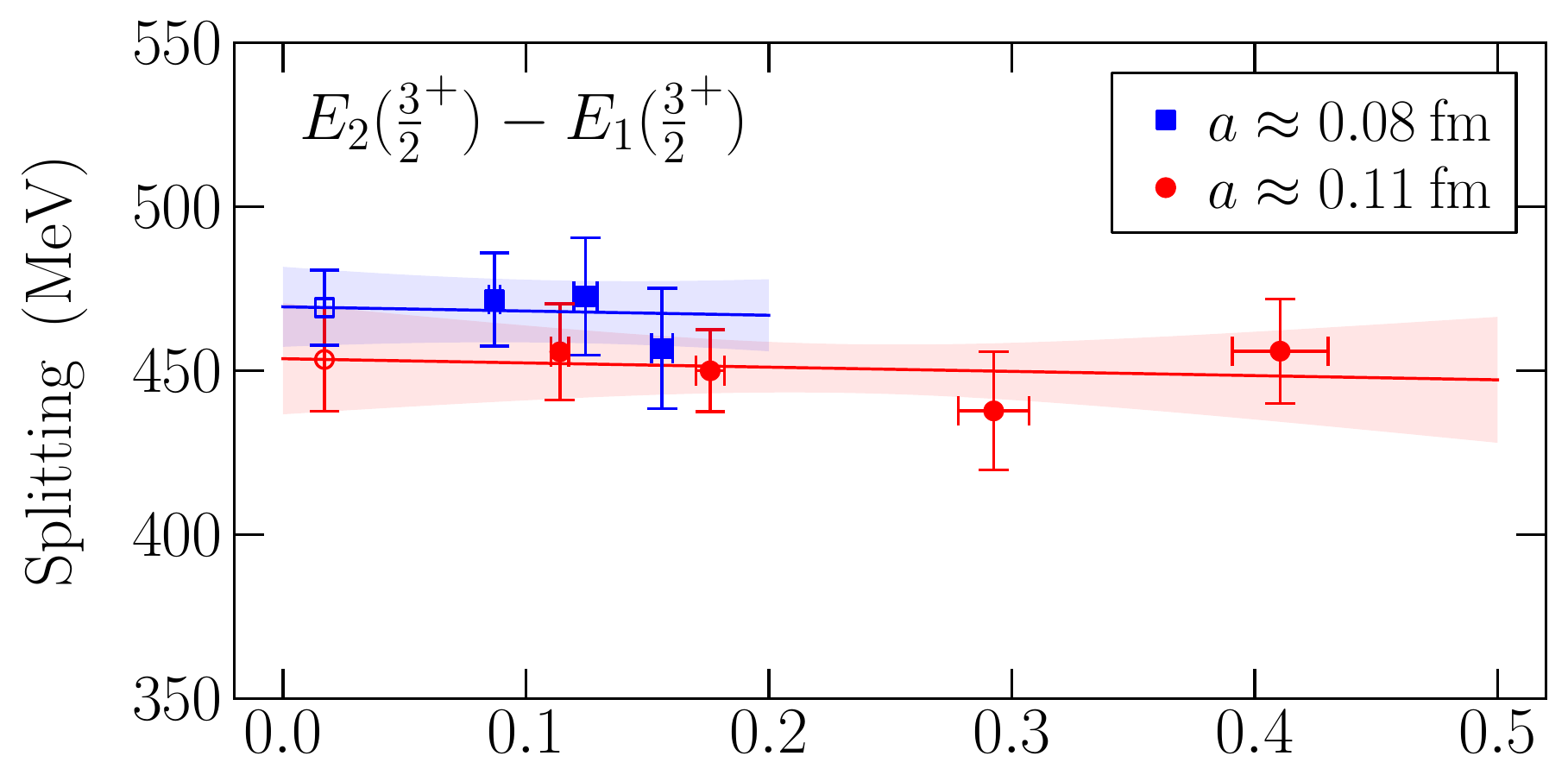}
 \includegraphics[width=0.45\linewidth]{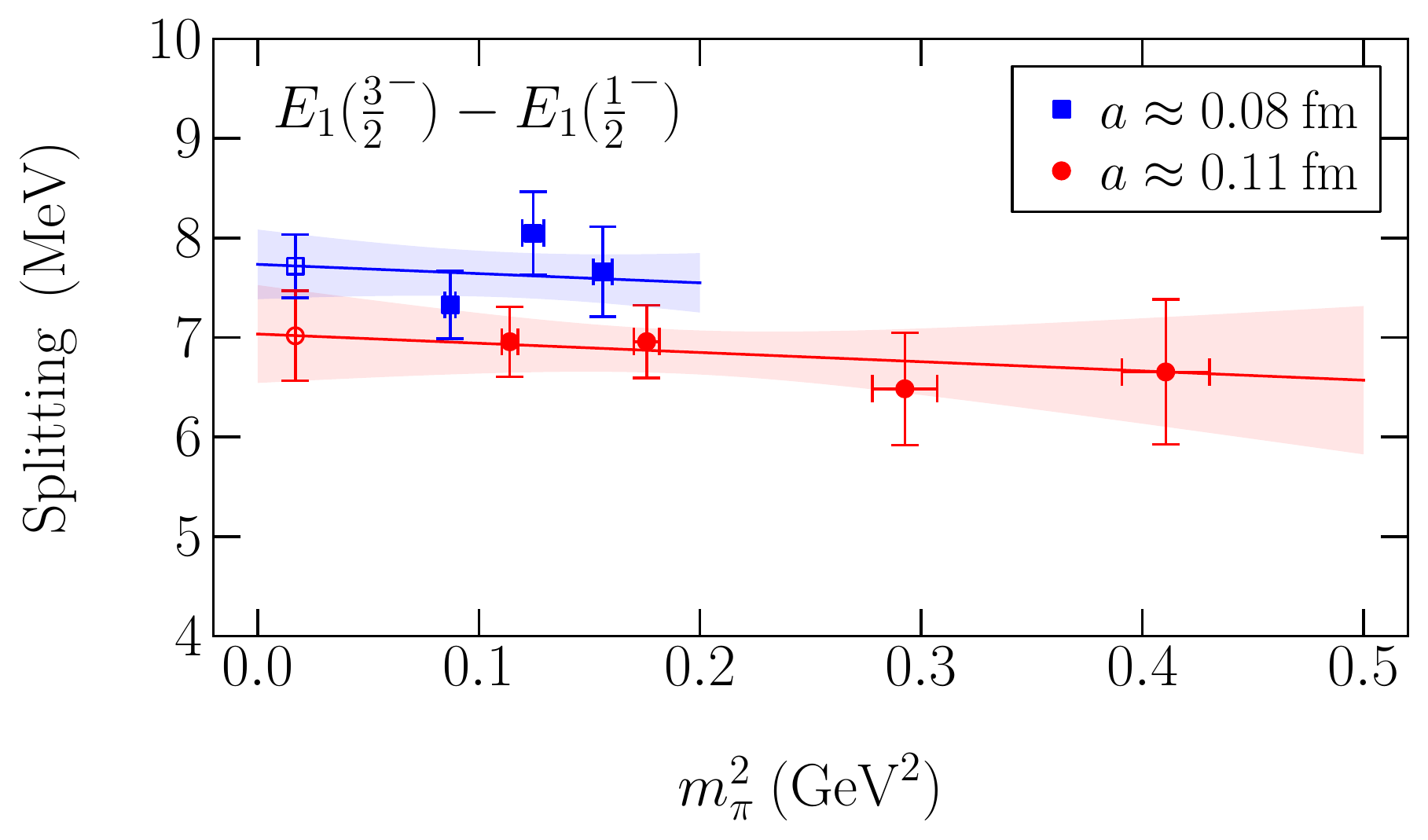}   \hfill \includegraphics[width=0.45\linewidth]{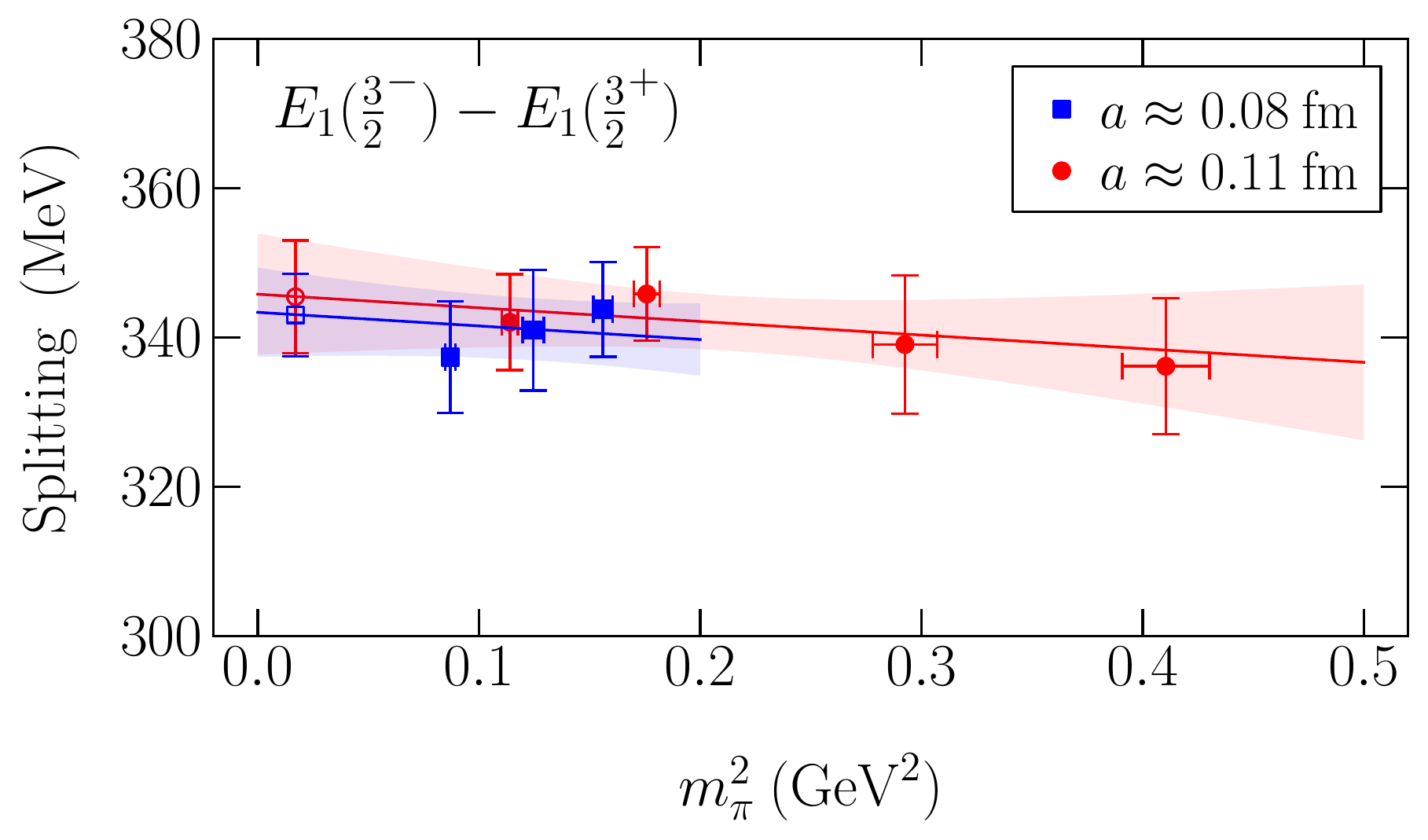}
\caption{\label{fig:chiral_extrap}Extrapolation of the $bbb$ energy splittings to the physical pion mass. The fits are linear in $m_\pi^2$, and were done simultaneously for the data
at the two different lattice spacings. The data are plotted with closed symbols, and the extrapolated results at $m_\pi=138$ MeV are plotted with open symbols. The fitted
functions and their 1-sigma uncertainty are given by the lines and the shaded regions.}
\end{figure*}

The fits to the data for the ten $bbb$ energy splittings using Eq.~(\ref{eq:sim_chiral_extrap}) are visualized in Fig.~\ref{fig:chiral_extrap}.
Evaluating the fitted functions for $m_\pi=138$ MeV leads to the results given in Table \ref{tab:results}. In addition to
the ten independent energy splittings discussed so far, the Table also gives some further combinations for convenience, in particular
the energy differences of all ten excited states to the ground state $E_1(\frac32^+)$, and a result for the very small
splitting $E_4(\frac32^+)-E_2(\frac52^+)$ that, as a consequence of the strong correlations, has smaller absolute uncertainties
than the other splittings involving the same levels.

As can be seen in Fig.~\ref{fig:chiral_extrap} and Table \ref{tab:results}, the results for the $bbb$ spectrum show only a weak dependence on the
lattice spacing, which in most cases is not statistically significant. The results at $a\approx 0.08$ fm and $m_\pi=138$ MeV
can be quoted as the predicted values for the continuum $bbb$ spectrum, once the remaining systematic uncertainties have been estimated.
These estimates can be made using information from Sec.~\ref{sec:coeff_dep} about the dependence of the $bbb$ energy splittings
on the couplings $c_i$ in the NRQCD action [see Eq.~(\ref{eq:dH_full})]. The systematic uncertainty is computed individually
for each energy splitting $E$, using the formula
\begin{equation}
\sigma_E^{(\rm syst)} = \left[  \left(\frac{\partial E}{\partial c_3}\right)^{\!2}\!\!\sigma_{c_3}^2 +  \left(\frac{\partial E}{\partial c_4}\right)^{\!2}\!\!\sigma_{c_4}^2 
+ \Big(0.02\: E_{\rm SI}\Big)^2 + \Big(0.07\: (E - E_{\rm SI}) \Big)^2 \right]^{1/2}, \label{eq:systerrs}
\end{equation}
which takes into account the varying contributions from spin-dependent and spin-independent NRQCD interactions.

The first two terms
in Eq.~(\ref{eq:systerrs}) correspond to the uncertainty in $E$ that results from the uncertainty in the tuning of the NRQCD
coefficients $c_3$, and $c_4$ [see Eq.~(\ref{eq:c3c4})]. The derivatives with respect to $c_3$ and $c_4$
are approximated using discrete difference quotients formed from the results in the last three columns of Table \ref{tab:coeff_dep}.
To save computer time, the results in Table \ref{tab:coeff_dep} were obtained at the coarser lattice spacing $a\approx 0.11$ fm.
However, for the purpose of estimating $\sigma_E^{(\rm syst)}$, it is sufficient to approximate the derivatives with respect to $c_3$ and $c_4$
at $a\approx 0.08$ fm as being equal to those at $a\approx 0.11$ fm, and then setting $\sigma_{c_3}=0.084$ and $\sigma_{c_4}=0.053$ according
to Eq.~(\ref{eq:c3c4}) for $a\approx0.08$ fm.

The third term in Eq.~(\ref{eq:systerrs}) describes the systematic uncertainty in the spin-\emph{independent} contribution to the energy splitting.
This contribution, $E_{\rm SI}$, is obtained by setting $c_3=c_4=c_7=c_8=c_9=0$ in the NRQCD action. Given the weak $a$-dependence
of the spectrum, $E_{\rm SI}$ can be taken from the second column of Table \ref{tab:coeff_dep}. However, the estimate of a 2\% systematic uncertainty is specific
to $a\approx 0.08$ fm. It includes the radiative, discretization, and relativistic errors, and is based on the discussion of radial and
orbital energy splittings for the same lattice spacing in bottomonium \cite{Meinel:2010pv}.
The estimates of uncertainties for bottomonium are also valid for triply-bottom baryons, since the energy- and momentum scales involved are the same (indeed, the
results of Sec.~\ref{sec:coeff_dep} confirm that the $v^2$-expansion converges at a similar rate for the $bbb$ system as for bottomonium).

The last term in Eq.~(\ref{eq:systerrs}) describes the systematic uncertainty in the spin-\emph{dependent} contribution to the energy splitting.
This contribution can be isolated by computing the difference $(E - E_{\rm SI})$, where $E$ is the result from the full NRQCD action.
Because the leading spin-dependent couplings $c_3$ and $c_4$ have been tuned nonperturbatively (and their
tuning uncertainty is already taken into account), and because the spin-dependent order-$v^6$ terms have been included in the NRQCD action at tree-level,
the dominant remaining sources of error for the spin splittings are discretization errors and the missing radiative corrections
in the $v^6$-terms. Following the discussion of the bottomonium fine- and hyperfine splittings in Ref.~\cite{Meinel:2010pv},
a systematic uncertainty of 7\% is assigned here to the spin-dependent contributions at $a\approx 0.08$ fm. Again, the values of
$(E - E_{\rm SI})$ can be taken from Table \ref{tab:coeff_dep} (the differences of the results from columns six and two),
because the spectrum has a weak $a$-dependence.

\begin{table}[t!]
\begin{tabular}{ccllllll}
\hline\hline
     & \hspace{2ex} & $a\approx 0.11$ fm  & \hspace{2ex} &  $a\approx 0.08$ fm & \hspace{2ex} &  \hspace{2ex} Final result \\
\hline
\\[-2.5ex]
$E_1(\frac12^+)-E_1(\frac32^+)$ &&  $\wm563(21)$     &&  $\wm567(14)$    &&  $\wm567 \pm 14 \pm 12$    \\
$E_2(\frac12^+)-E_1(\frac32^+)$ &&  $\wm579(20)$     &&  $\wm582(13)$    &&  $\wm582 \pm 13 \pm 13$    \\
$E_2(\frac32^+)-E_1(\frac32^+)$ &&  $\wm453(16)$     &&  $\wm469(11)$    &&  $\wm469 \pm 11 \pm 9$    \\
$E_3(\frac32^+)-E_1(\frac32^+)$ &&  $\wm584(20)$     &&  $\wm587(13)$    &&  $\wm587 \pm 13 \pm 12$    \\
$E_4(\frac32^+)-E_1(\frac32^+)$ &&  $\wm629(21)$     &&  $\wm634(14)$    &&  $\wm634 \pm 14 \pm 13$    \\
$E_1(\frac52^+)-E_1(\frac32^+)$ &&  $\wm589(19)$     &&  $\wm593(13)$    &&  $\wm593 \pm 13 \pm 12$    \\
$E_2(\frac52^+)-E_1(\frac32^+)$ &&  $\wm630(21)$     &&  $\wm636(14)$    &&  $\wm636 \pm 14 \pm 13$    \\
$E_1(\frac72^+)-E_1(\frac32^+)$ &&  $\wm593(19)$     &&  $\wm598(12)$    &&  $\wm598 \pm 12 \pm 12$    \\
$E_1(\frac12^-)-E_1(\frac32^+)$ &&  $\wm338.4(8.0)$  &&  $\wm335.3(5.8)$ &&  $\wm335.3 \pm 5.8 \pm 7.4$ \\
$E_1(\frac32^-)-E_1(\frac32^+)$ &&  $\wm345.5(7.5)$  &&  $\wm343.0(5.5)$ &&  $\wm343.0 \pm 5.5 \pm 7.2$ \\
\\[-1ex]
$E_1(\frac12^+)-E_1(\frac72^+)$ &&  $-30.3(2.0)$     &&  $-30.7(1.4)$    &&  $-30.7 \pm 1.4 \pm 0.8$    \\
$E_2(\frac12^+)-E_1(\frac72^+)$ &&  $-14.1(1.4)$     &&  $-15.6(1.1)$    &&  $-15.6 \pm 1.1 \pm 1.6$    \\
$E_3(\frac32^+)-E_1(\frac72^+)$ &&  $-9.4(1.1)$      &&  $-10.71(85)$    &&  $-10.7 \pm 0.9 \pm 1.2$    \\
$E_4(\frac32^+)-E_1(\frac72^+)$ &&  $\wm36.1(2.1)$   &&  $\wm36.2(1.4)$  &&  $\wm36.2 \pm 1.4 \pm 1.4$    \\
$E_1(\frac52^+)-E_1(\frac72^+)$ &&  $-3.86(69)$      &&  $-4.75(50)$     &&  $-4.75 \pm 0.50 \pm 0.55$     \\
$E_2(\frac52^+)-E_1(\frac72^+)$ &&  $\wm37.2(2.2)$   &&  $\wm38.2(1.4)$  &&  $\wm38.2 \pm 1.4 \pm 1.1$    \\
\\[-1ex]
$E_1(\frac12^-)-E_1(\frac32^-)$ &&  $-7.02(45)$      &&  $-7.72(32)$     &&  $-7.72 \pm 0.32 \pm 0.90$   \\
\\[-1ex]
$E_4(\frac32^+)-E_2(\frac52^+)$ &&  $-1.63(62)$      &&  $-2.06(48)$     &&  $-2.06 \pm 0.48 \pm 0.59$   \\
\\[-2.5ex]
\hline\hline
\end{tabular}
\caption{\label{tab:results}Energy splittings in MeV between various $bbb$ states, extrapolated to the physical pion mass. In the final results (last column),
the central values and statistical/fitting/scale setting uncertainties are taken from $a\approx 0.08$ fm, and estimates of the total systematic uncertainties
computed using Eq.~(\ref{eq:systerrs}) are given. The ground-state mass is equal to $E_1(\frac32^+)= 14371 \pm 4 \pm 11$ MeV \cite{Meinel:2010pw}.}
\end{table}

\begin{figure*}[ht!]
 \includegraphics[width=0.7\linewidth]{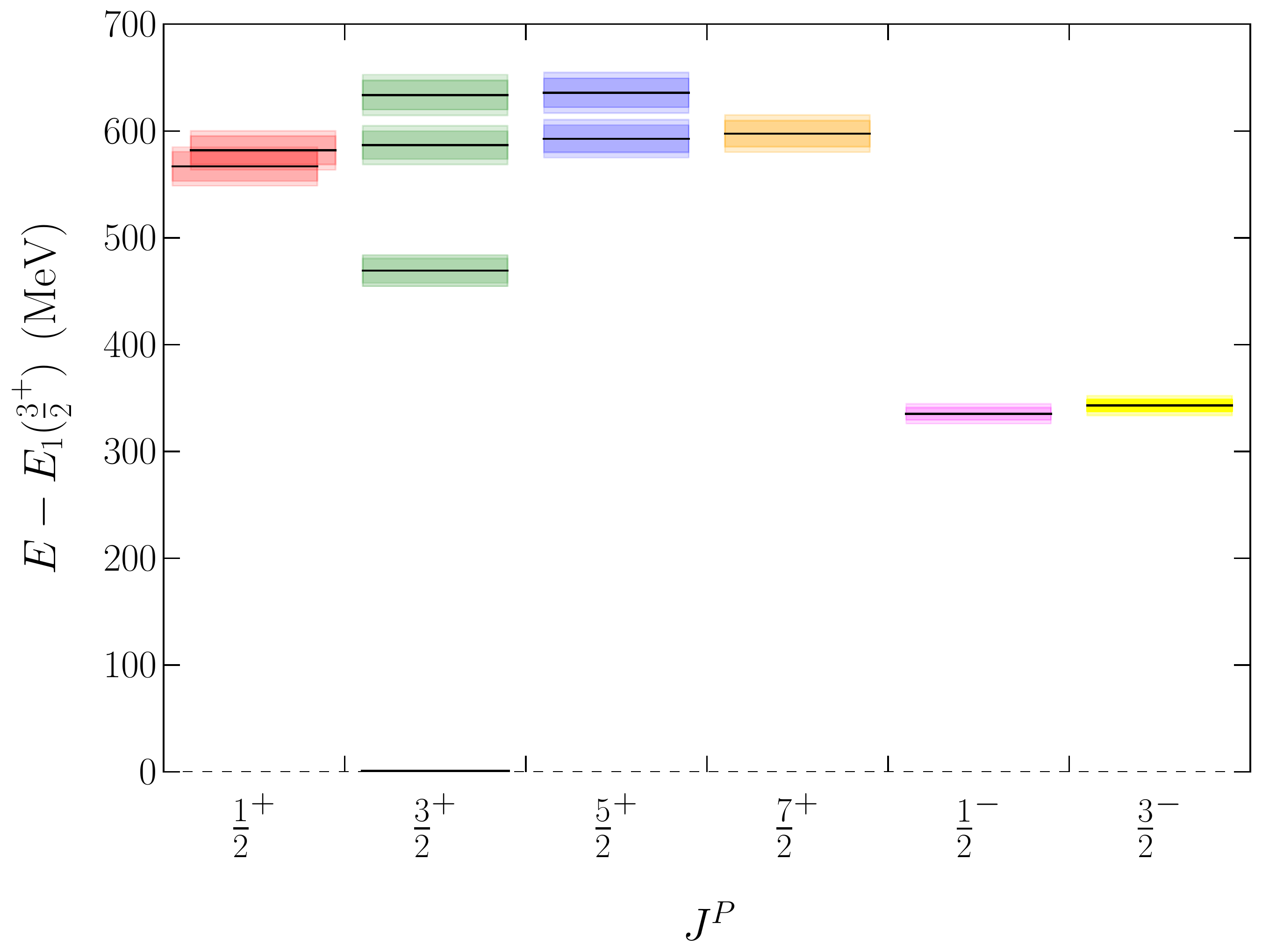} 
\caption{\label{fig:bbb_spectrum_all_L32}Final results for the $bbb$ spectrum relative to ground state $E_1(\frac32^+)$ (see the last column of
Table \ref{tab:results} for the numerical values).
The superimposed shaded regions show the statistical/fitting/scale setting uncertainty and the total (including systematic) uncertainty, respectively.
The results are highly correlated, and the uncertainties for energy differences between nearby states are in fact much smaller
than suggested by this plot. See Fig.~\ref{fig:bbb_spectrum_wrtJ72_L32} for close-ups of the spectra near $E_1(\frac72^+)$ and $E_1(\frac32^-)$,
where advantage of the correlations is taken by computing the energy differences relative these levels.}
\end{figure*}

\begin{figure*}[ht!]
 \includegraphics[height=58ex]{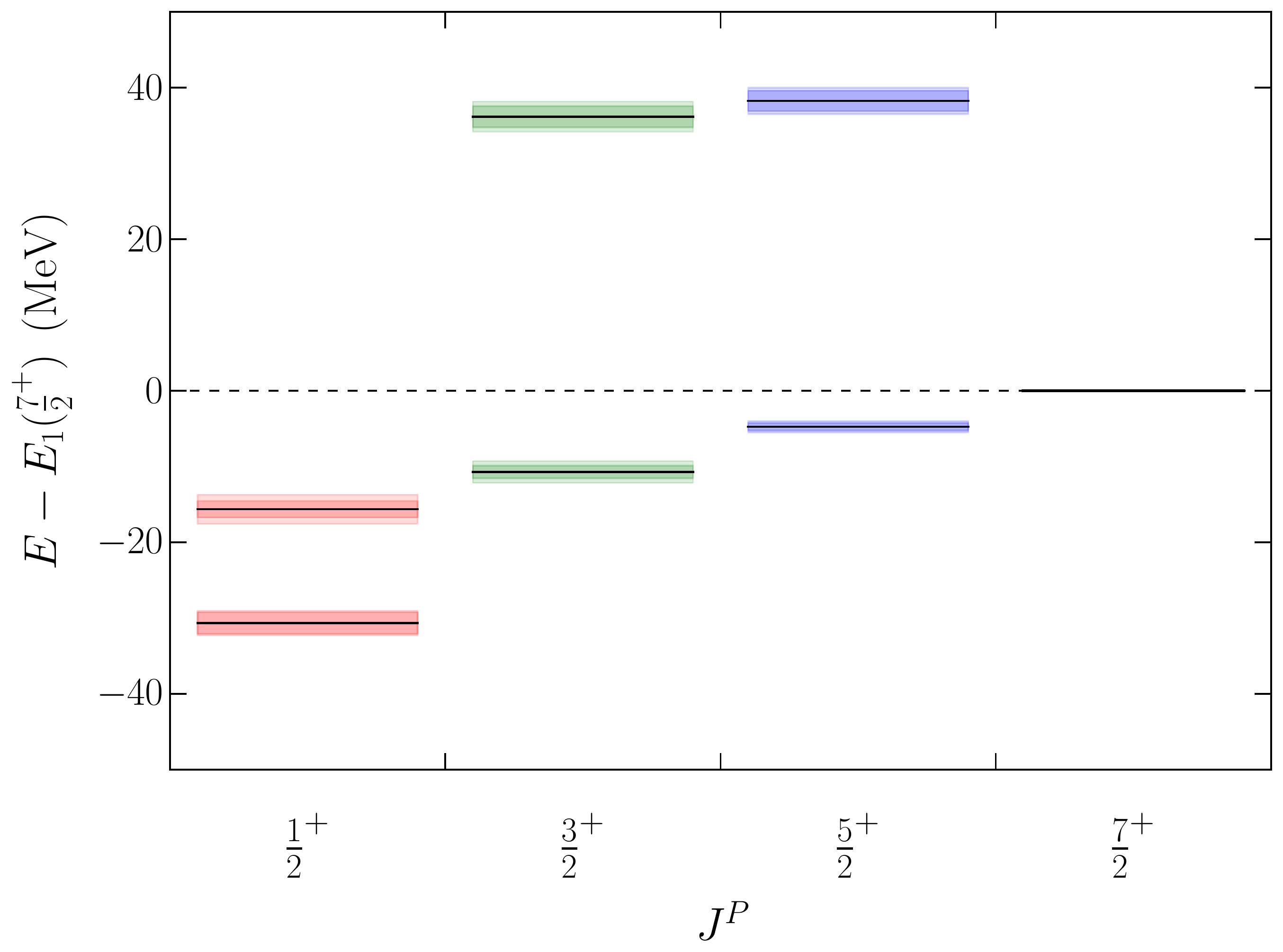} \hspace{4ex} \includegraphics[height=58ex]{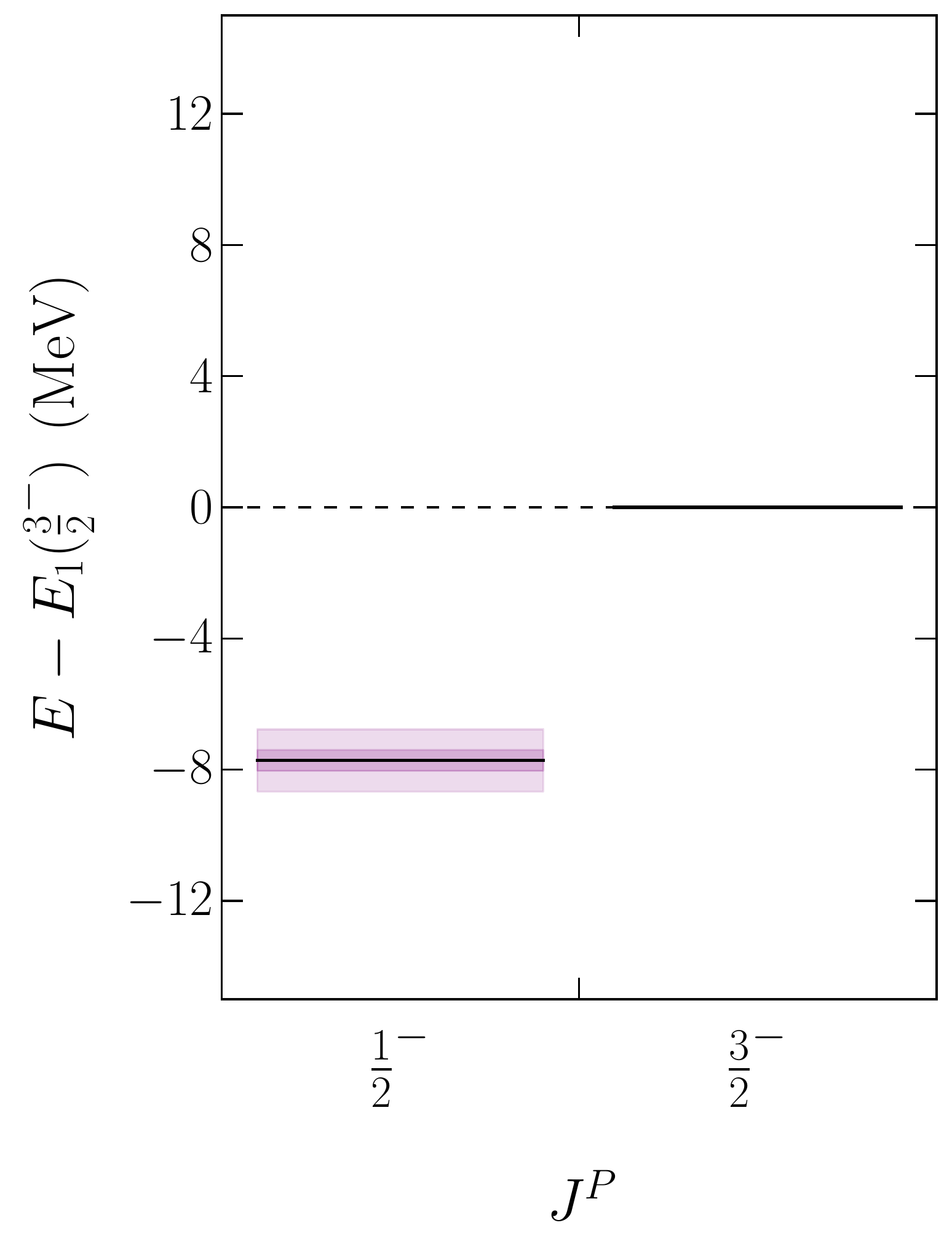}
\caption{\label{fig:bbb_spectrum_wrtJ72_L32}Final results for the $bbb$ spectrum relative to $E_1(\frac72^+)$ (left panel) and $E_1(\frac32^-)$ (right panel),
showing only the states in the vicinity of these levels. The superimposed shaded regions show the statistical/fitting/scale setting uncertainty and the total
(including systematic) uncertainty, respectively. See the last column of Table \ref{tab:results} for the numerical values.}
\end{figure*}

The final results for the $bbb$ spectrum, with systematic uncertainties computed using Eq.~(\ref{eq:systerrs}), are given in the last column of Table
\ref{tab:results}. The energy differences of the ten excited states to the ground state $\Omega_{bbb}$ are plotted in Fig.~\ref{fig:bbb_spectrum_all_L32}.
The results for the different energy levels are highly correlated, and the small splittings between nearby states
can in fact be computed with much smaller absolute uncertainties. These smaller energy splittings are
given in the lower part of Table \ref{tab:results}, and are plotted in Fig.~\ref{fig:bbb_spectrum_wrtJ72_L32}.

It is interesting to compare the QCD results obtained here to the potential-model calculation of Ref.~\cite{SilvestreBrac:1996bg} (see Fig.~5 therein). The numbers
of states in the considered energy region are in agreement, and the energy differences to the ground state predicted by Ref.~\cite{SilvestreBrac:1996bg}
are found to be within 10\% of the QCD results. However, the potentials used in Ref.~\cite{SilvestreBrac:1996bg} did not
include any spin-orbit or tensor interactions, so that the results obtained there have the exact degeneracies
$E_2(\frac12^+)=E_3(\frac32^+)=E_1(\frac52^+)=E_1(\frac72^+)$, $E_4(\frac32^+)=E_2(\frac52^+)$, and $E_1(\frac12^-)=E_1(\frac32^-)$.
As can be seen in Fig.~\ref{fig:bbb_spectrum_wrtJ72_L32}, the QCD calculation performed here is so precise that
the spin-dependent effects that lift these degeneracies are clearly resolved. These effects will be discussed further in Sec.~\ref{sec:coeff_dep}.

Reference \cite{SilvestreBrac:1996bg} also calculated the higher-lying $bbb$ spectrum, and these additional states
were all found to be separated by energy gaps of order 300 MeV from the states considered here. Along with the plateaus observed in
Figs.~\ref{fig:E_vs_tmin_L24} and \ref{fig:E_vs_tmin_L32}, the large energy gaps found in Ref.~\cite{SilvestreBrac:1996bg}
provides further confidence that the contamination from higher states in the fits of Sec.~\ref{sec:fitting} is negligible.

Remarkably, the three energy splittings $E_2(\frac12^+)-E_1(\frac32^+)$, $E_1(\frac12^-)-E_1(\frac32^+)$, and $E_2(\frac32^+)-E_1(\frac32^+)$
that were computed in the early bag-model calculation of Ref.~\cite{Hasenfratz:1980ka} also agree with the results obtained
here to within 10\%. On the other hand, the energy splittings calculated recently using a quark model in Ref.~\cite{Roberts:2007ni} (see Table 19 therein)
are in dramatic disagreement with the QCD results obtained here: by about a factor of two for the larger splittings and by about
a factor of 10 for the smaller splittings.

\clearpage

\section{\label{sec:coeff_dep}Dependence of the spectrum on the coefficients in the NRQCD action}

In Sec.~\ref{sec:chiral}, the $bbb$ spectrum was computed with coefficients $c_i$ in the
lattice NRQCD action tuned such that the effective field theory reproduces relativistic QCD.
Table \ref{tab:results} and Figs.~\ref{fig:bbb_spectrum_all_L32}, \ref{fig:bbb_spectrum_wrtJ72_L32}
give the best possible results obtained here for the $bbb$ energy levels in the real world.
However, with lattice NRQCD, one can perform simulations for arbitrary values of the coefficients
$c_i$. The ability to selectively turn on and off the different terms in the NRQCD action and compute
the effect on the $bbb$ energy levels can be exploited to gain deeper insight into the interactions between three
heavy quarks.

The numerical results of this section are summarized in Table \ref{tab:coeff_dep}. Shown there
are the values of the $bbb$ energy splittings computed for eight different choices of the coefficients
in the NRQCD action. The various terms in the NRQCD action were already discussed in Sec.~\ref{sec:parameters}, and
their coefficients $c_i$ were defined in Eq.~(\ref{eq:dH_full}). The calculations in this
section were done for a single gauge field ensemble only ($a\approx 0.11$ fm, $a m_{u,d}=0.005$), to save computer time.
As shown in Sec.~\ref{sec:chiral}, the dependence of the $bbb$ spectrum on $a$ and $m_{u,d}$ is weak, and therefore
a single ensemble is sufficient for the purpose of studying the $c_i$-dependence.
In all cases, the $b$ quark mass and the Symanzik-improvement coefficients in the NRQCD action remained unchanged
($a m_b=2.487$, $c_5=c_6=1$).
The following discussion focuses on the energy regions near $E_1(\frac72^+)$ and $E_1(\frac32^-)$, as this is
where all the spin-dependent level splittings are found.

\begin{figure*}[ht!]
 \includegraphics[height=58ex]{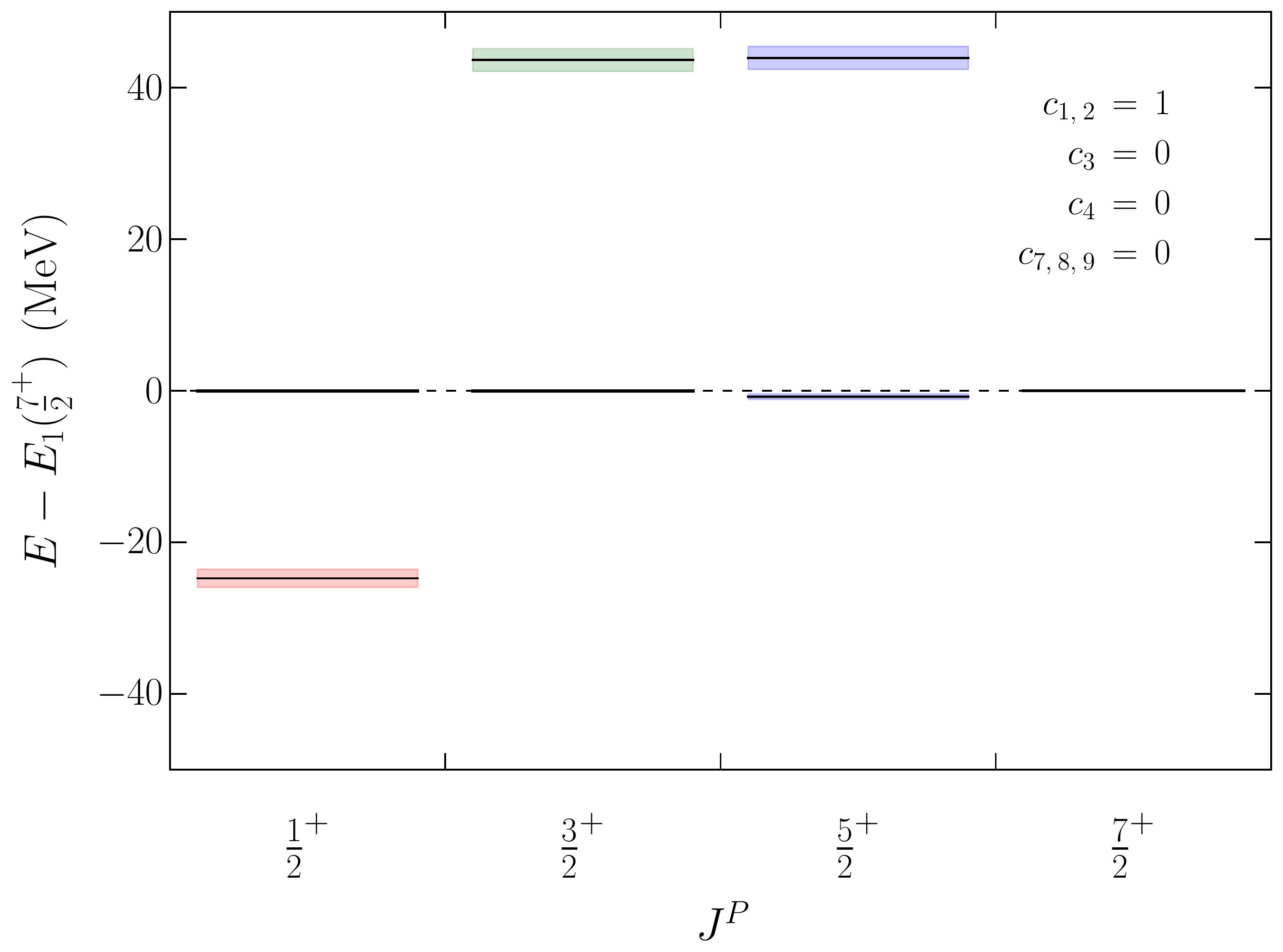} \hspace{4ex} \includegraphics[height=58ex]{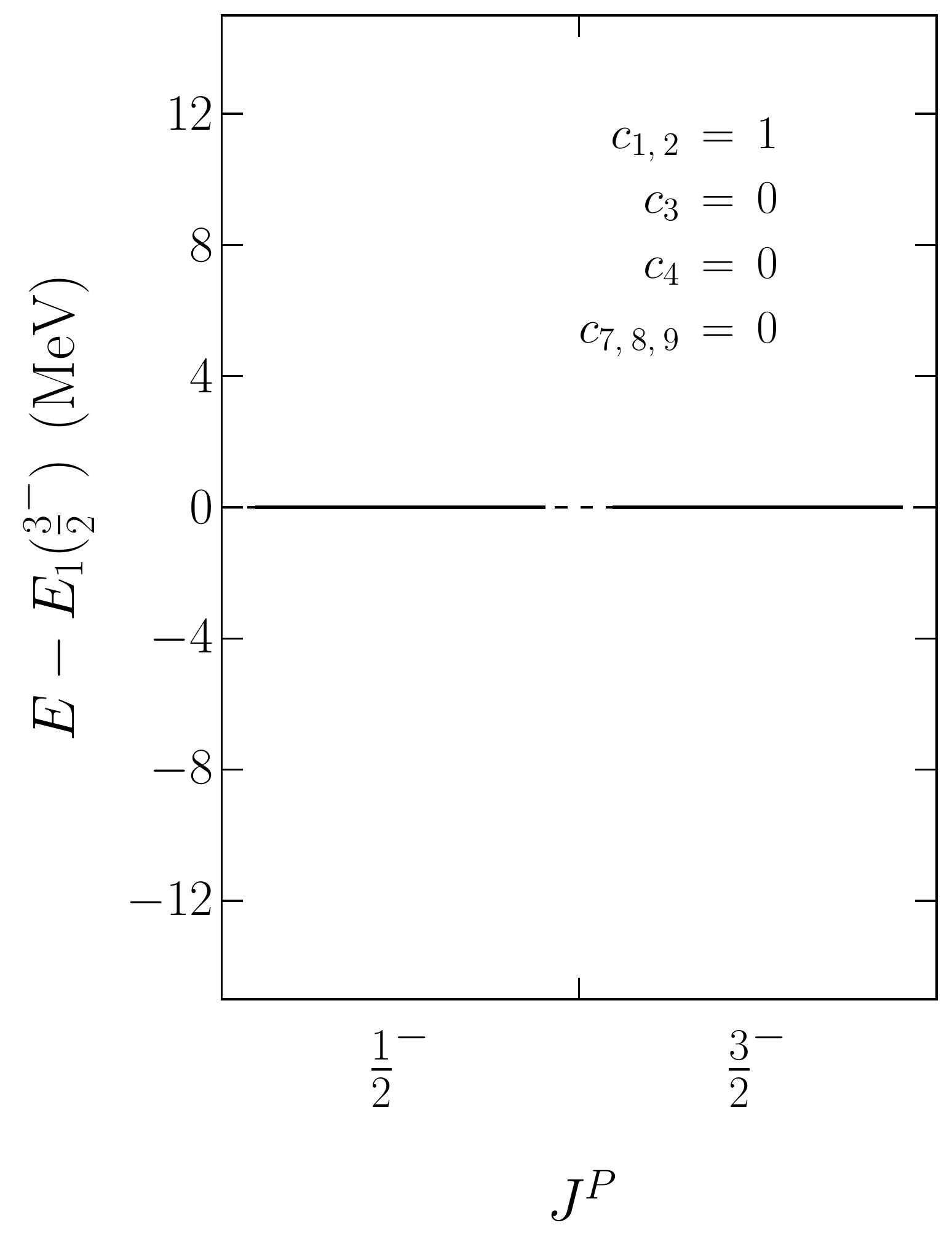}
\caption{\label{fig:coeff_dependence_SI}Dependence of the spectrum near $E_1(\frac72^+)$ and $E_1(\frac32^-)$ on the coefficients $c_i$ in the NRQCD action (at $a\approx 0.11$ fm, $am_{u,d}=0.005$).
Shown here is the case of the spin-independent order-$v^4$ NRQCD action, obtained by setting $c_3=c_4=c_7=c_8=c_9=0$.
In the absence of rotational symmetry breaking, this leads to the exact degeneracies $E_2(\frac12^+)=E_3(\frac32^+)=E_1(\frac52^+)=E_1(\frac72^+)$, $E_4(\frac32^+)=E_2(\frac52^+)$, and $E_1(\frac12^-)=E_1(\frac32^-)$.
On the lattice, the relations $E_1(\frac12^-)=E_1(\frac32^-)$ and $E_2(\frac12^+)=E_3(\frac32^+)$ are still exact,
but the degeneracies with $J>\frac32$ levels are only approximate.}
\end{figure*}

The energy splittings in the first column of Table \ref{tab:coeff_dep} were computed with
the order-$v^2$ NRQCD action, which contains only $H_0 = -\frac{1}{2 m_b}\Delta^{(2)}$ (and the associated lattice discretization improvement terms with $c_5$ and $c_6$).
Turning on also the spin-\emph{independent} order-$v^4$ terms, $-c_1\frac{1}{8 m_b^3}\left(\Delta^{(2)}\right)^2$ and $c_2\frac{ig}{8 m_b^2}\:\big(\bs{\nabla}\cdot\bs{\widetilde{E}}
-\bs{\widetilde{E}}\cdot\bs{\nabla}\big)$, gives the results in the second column
of Table \ref{tab:coeff_dep}. These results are plotted in Fig.~\ref{fig:coeff_dependence_SI}.
In both cases, the action does not depend on the heavy-quark spin, so that $L$ and $S$
become separately conserved quantum numbers, up to the small effects of rotational symmetry breaking introduced by the lattice.
In the absence of rotational symmetry breaking, one would then have the exact level degeneracies
$E_2(\frac12^+)=E_3(\frac32^+)=E_1(\frac52^+)=E_1(\frac72^+)$, $E_4(\frac32^+)=E_2(\frac52^+)$, and $E_1(\frac12^-)=E_1(\frac32^-)$.
The relations $E_1(\frac12^-)=E_1(\frac32^-)$ and $E_2(\frac12^+)=E_3(\frac32^+)$ actually remain exact on the lattice,
an observation that can be related to the trivial subduction of these two $J$ values into lattice irreps (cf.~Sec.~\ref{sec:subduction}).
The degeneracies with $J>\frac32$ are only approximate, but the splittings remain very small. Note that the energies quoted here
for the higher-$J$ levels were obtained by averaging over the different irreps into which a continuum level splits
[see the discussion around Eq.~(\ref{eq:augmchisqr}); also see Table \ref{tab:irrep_splittings} for the size of the original splittings
between the different irreps].

\begin{figure*}[ht!]
\null
\vspace{14ex}
 \includegraphics[height=58ex]{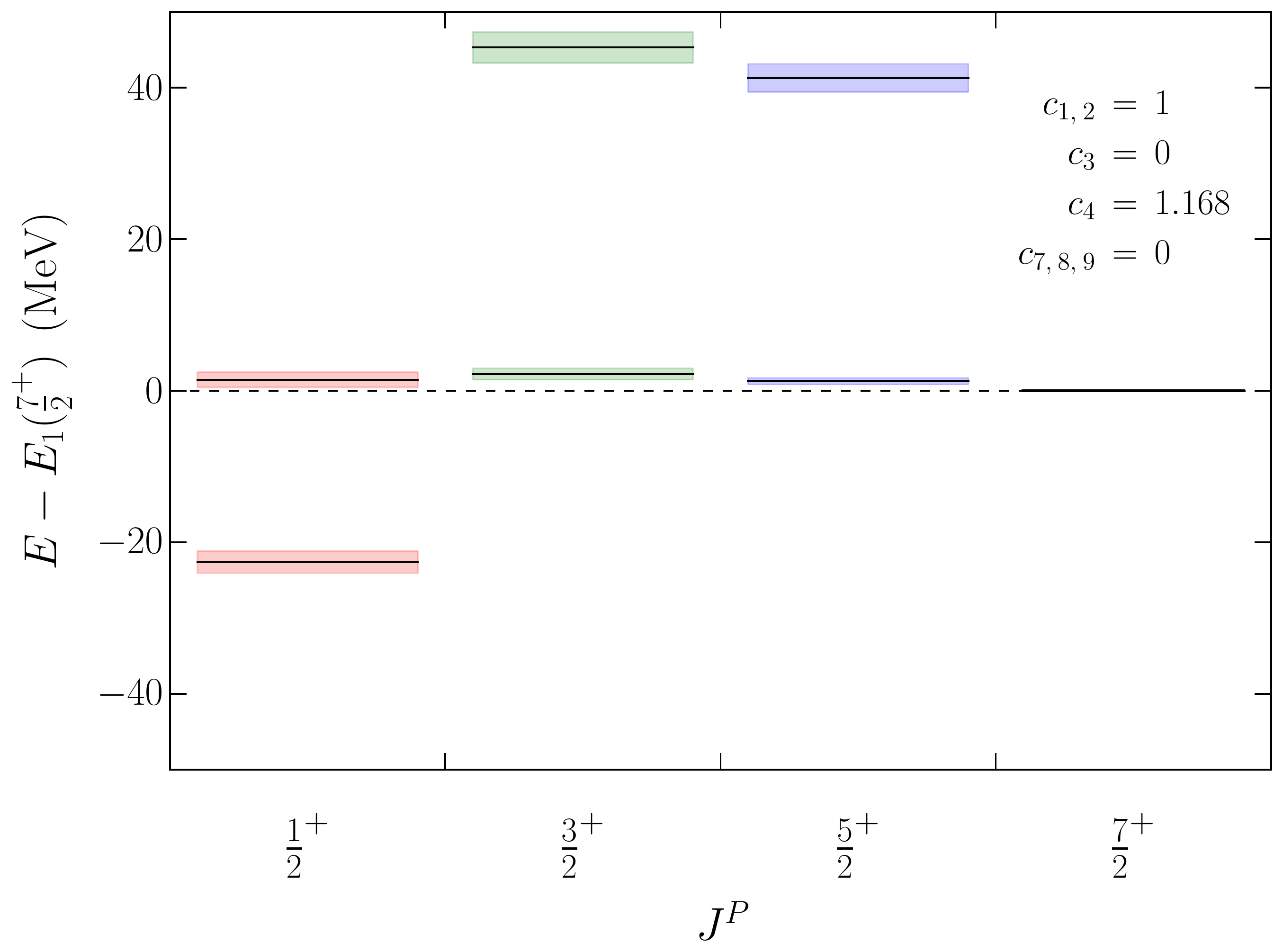} \hspace{4ex} \includegraphics[height=58ex]{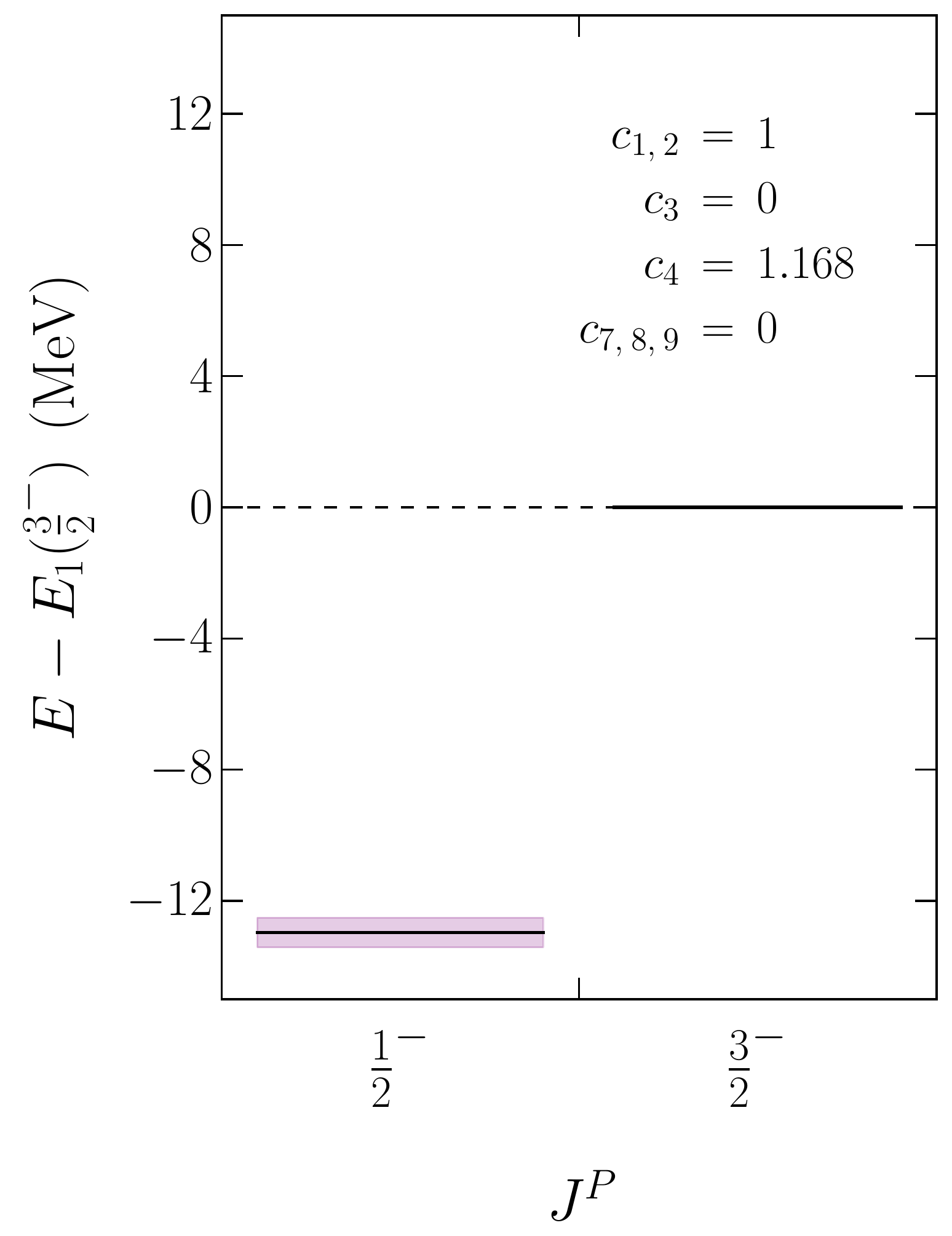}
\caption{\label{fig:coeff_dependence_c3_0}Dependence of the spectrum near $E_1(\frac72^+)$ and $E_1(\frac32^-)$ on the coefficients $c_i$ in the NRQCD action (at $a\approx 0.11$ fm, $am_{u,d}=0.005$).
Shown here is the case of the order-$v^4$ NRQCD action, but with the coefficient of the operator \mbox{$\bs{\sigma}\cdot\left(\bs{\widetilde{\nabla}}\times\bs{\widetilde{E}}-\bs{\widetilde{E}}\times\bs{\widetilde{\nabla}} \right)$} set to zero, so that
the only remaining spin-dependent interaction is \mbox{$-c_4\:\displaystyle\frac{g}{2 m_b}\:\bs{\sigma}\cdot\bs{\widetilde{B}}$}.}
\end{figure*}

Next, Figure \ref{fig:coeff_dependence_c3_0} shows the spectrum after additionally turning on the leading interaction with the
chromomagnetic moment of the heavy quark:
\begin{equation}
 -c_4\:\frac{g}{2 m_b}\:\bs{\sigma}\cdot\bs{\widetilde{B}}. \label{eq:sigmadotB}
\end{equation}
This interaction causes small positive splittings $\big[E_2(\frac12^+)-E_1(\frac72^+)\big]_{\rm subtr.}=1.5(1.0)$ MeV, $\big[E_3(\frac32^+)-E_1(\frac72^+)\big]_{\rm subtr.}=2.23(74)$ MeV,
$\big[E_1(\frac52^+)-E_1(\frac72^+)\big]_{\rm subtr.}=2.05(56)$ MeV, and $\big[E_4(\frac32^+)-E_2(\frac52^+)\big]_{\rm subtr.}=4.28(49)$ MeV, where
the rotational-symmetry-breaking-induced splittings seen at $c_4=0$ (second column of Table \ref{tab:coeff_dep}) have been subtracted.
The operator (\ref{eq:sigmadotB}) also introduces a very significant splitting of
the two odd-parity levels considered here: $E_1(\frac12^-)-E_1(\frac32^-)=-12.97(45)$ MeV. For heavy quarkonium,
the operator (\ref{eq:sigmadotB}) is mainly associated with spin-spin and tensor interactions. However, simple potential models for baryons that
include only spin-spin and tensor interactions predict $E_1(\frac12^-)-E_1(\frac32^-)=0$ \cite{Isgur:1978xj, Chao:1980em, Gromes:1982ze}. Thus, one
can conclude that the operator (\ref{eq:sigmadotB}) also plays an important role in the generation of spin-orbit interactions. This
can indeed be seen in the derivation of spin-dependent potentials using pNRQCD \cite{Pineda:2000sz}.

\begin{figure*}[ht!]
 \includegraphics[height=58ex]{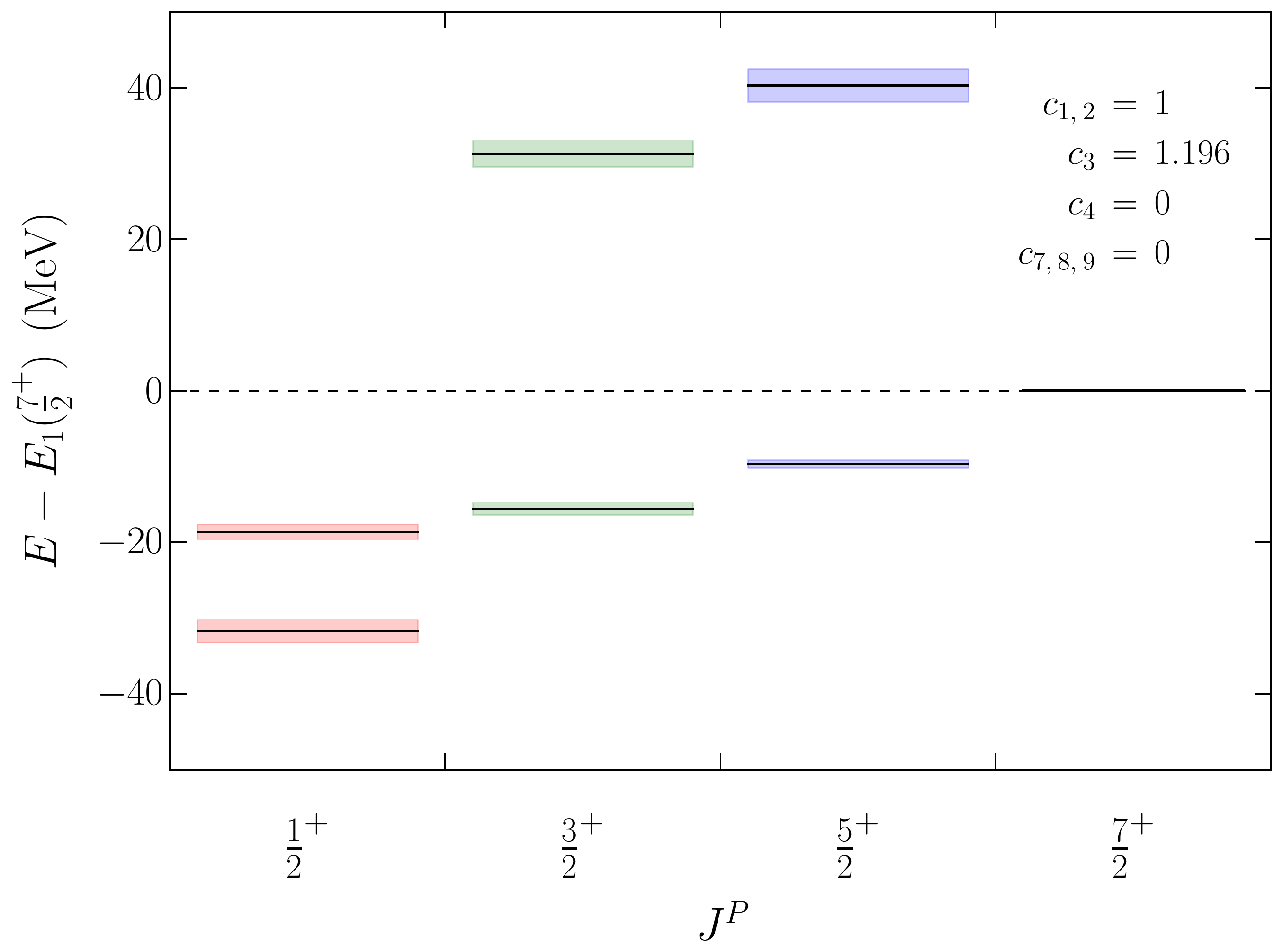} \hspace{4ex} \includegraphics[height=58ex]{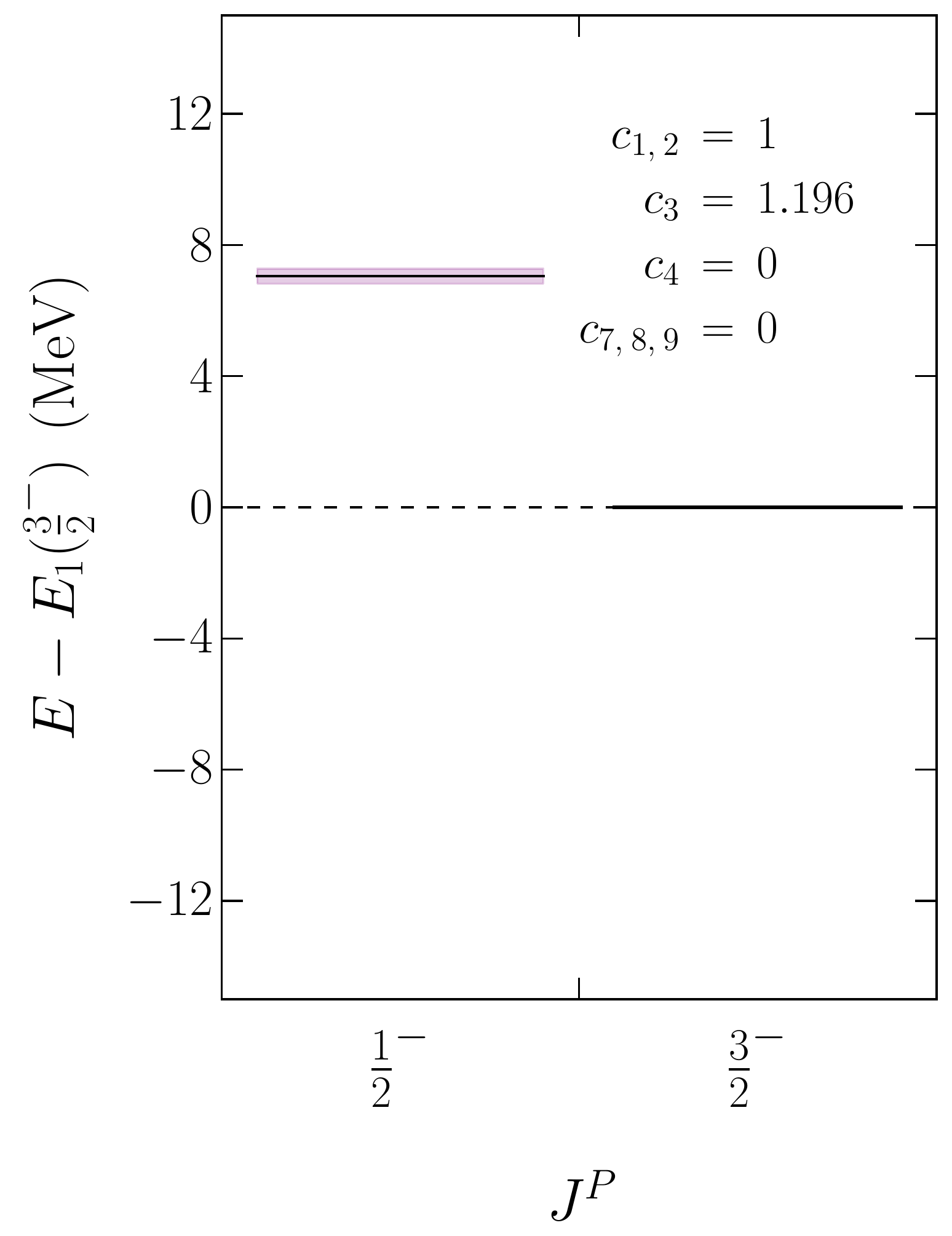}
\caption{\label{fig:coeff_dependence_c4_0}Dependence of the spectrum near $E_1(\frac72^+)$ and $E_1(\frac32^-)$ on the coefficients $c_i$ in the NRQCD action (at $a\approx 0.11$ fm, $am_{u,d}=0.005$).
Shown here is the case of the order-$v^4$ NRQCD action, but with the coefficient of the operator \mbox{$\bs{\sigma}\cdot\bs{\widetilde{B}}$} set to zero, so that
the only remaining spin-dependent interaction is \mbox{$-c_3\:\displaystyle\frac{g}{8 m_b^2}\:\bs{\sigma}\cdot\left(\bs{\widetilde{\nabla}}\times\bs{\widetilde{E}}-\bs{\widetilde{E}}\times\bs{\widetilde{\nabla}} \right)$}.}
\end{figure*}

The other spin-dependent interaction of order $v^4$ is given by
\begin{equation}
 -c_3\:\displaystyle\frac{g}{8 m_b^2}\:\bs{\sigma}\cdot\left(\bs{\widetilde{\nabla}}\times\bs{\widetilde{E}}-\bs{\widetilde{E}}\times\bs{\widetilde{\nabla}} \right). \label{eq:sigmadotDxE}
\end{equation}
Setting $c_4=0$ again, and turning on the interaction (\ref{eq:sigmadotDxE}) instead, produces the results shown in Fig.~\ref{fig:coeff_dependence_c4_0}.
For the $bbb$ levels considered here, the operator (\ref{eq:sigmadotDxE}) results in spin splittings with the opposite
sign compared to those introduced by (\ref{eq:sigmadotB}): $\big[E_2(\frac12^+)-E_1(\frac72^+)]_{\rm subtr.}=-18.63(99)$ MeV, $\big[E_3(\frac32^+)-E_1(\frac72^+)]_{\rm subtr.}=-15.58(84)$ MeV,
$\big[E_1(\frac52^+)-E_1(\frac72^+)]_{\rm subtr.}=-8.89(64)$ MeV, $E_4\big[(\frac32^+)-E_2(\frac52^+)]_{\rm subtr.}=-8.74(53)$ MeV,
and $E_1(\frac12^-)-E_1(\frac32^-)=7.05(23)$ MeV. Notice in particular that for the
$bbb$ levels with approximate structure $L=2$, $S=\frac32$, the effect of (\ref{eq:sigmadotDxE}) is an order of magnitude larger than
the effect of (\ref{eq:sigmadotB}). Furthermore, the shifts introduced for these levels by the operator (\ref{eq:sigmadotDxE})
are approximately proportional to $2\,\bs{L}\cdot\bs{S}=J(J+1)-L(L+1)-S(S+1)$. This is what is expected for a spin-orbit interaction
in baryons levels with totally symmetric spatial wavefunctions \cite{Gromes:1976cr}.

\begin{table*}[ht!]
\begin{ruledtabular}
\begin{tabular}{lllllllllll}
Coefficient(s) \\
\hline
\\[-2.5ex]
$c_1$, $c_2$                    & \hspace{3ex} 0  & \hspace{3ex} 1  & \hspace{3ex} 1     & \hspace{3ex} 1     & \hspace{3ex} 1     & \hspace{3ex} 1     & \hspace{3ex} 1     & \hspace{3ex} 1     \\
$c_3$                           & \hspace{3ex} 0  & \hspace{3ex} 0  & \hspace{3ex} 0     & \hspace{3ex} 1.196 & \hspace{3ex} 1.196 & \hspace{3ex} 1.196 & \hspace{3ex} 1.196 & \hspace{3ex} 1     \\
$c_4$                           & \hspace{3ex} 0  & \hspace{3ex} 0  & \hspace{3ex} 1.168 & \hspace{3ex} 0     & \hspace{3ex} 1.168 & \hspace{3ex} 1.168 & \hspace{3ex} 1     & \hspace{3ex} 1.168 \\
$c_7$, $c_8$, $c_9$             & \hspace{3ex} 0  & \hspace{3ex} 0  & \hspace{3ex} 0     & \hspace{3ex} 0     & \hspace{3ex} 0     & \hspace{3ex} 1     & \hspace{3ex} 1     & \hspace{3ex} 1     \\
\\[-2.5ex]
\hline
\\[-2.5ex]
Splitting \\
\hline
\\[-2.5ex]
$E_1(\frac12^+)-E_1(\frac32^+)$ & $\wm592(12)$    & $\wm582(11)$    & $\wm546(15)$    & $\wm559(15)$    & $\wm545(15)$    & $\wm548(21)$     & $\wm551(15)$    & $\wm548(15)$    \\
$E_2(\frac12^+)-E_1(\frac32^+)$ & $\wm617(11)$    & $\wm607(10)$    & $\wm570(14)$    & $\wm572(15)$    & $\wm557(15)$    & $\wm563(20)$     & $\wm566(14)$    & $\wm565(14)$    \\
$E_2(\frac32^+)-E_1(\frac32^+)$ & $\wm467(11)$    & $\wm457(10)$    & $\wm454(12)$    & $\wm458.9(9.9)$ & $\wm454(12)$    & $\wm456(15)$     & $\wm457(12)$    & $\wm456(12)$    \\
$E_3(\frac32^+)-E_1(\frac32^+)$ & $\wm617(11)$    & $\wm607(10)$    & $\wm571(14)$    & $\wm576(15)$    & $\wm563(14)$    & $\wm568(20)$     & $\wm570(14)$    & $\wm569(14)$    \\
$E_4(\frac32^+)-E_1(\frac32^+)$ & $\wm661(12)$    & $\wm650(12)$    & $\wm614(15)$    & $\wm622(16)$    & $\wm606(15)$    & $\wm611(21)$     & $\wm614(15)$    & $\wm612(15)$    \\
$E_1(\frac52^+)-E_1(\frac32^+)$ & $\wm617(11)$    & $\wm606(11)$    & $\wm570(14)$    & $\wm581(14)$    & $\wm570(14)$    & $\wm573(20)$     & $\wm576(14)$    & $\wm573(14)$    \\
$E_2(\frac52^+)-E_1(\frac32^+)$ & $\wm662(12)$    & $\wm651(12)$    & $\wm610(15)$    & $\wm631(16)$    & $\wm610(15)$    & $\wm613(21)$     & $\wm617(15)$    & $\wm613(15)$    \\
$E_1(\frac72^+)-E_1(\frac32^+)$ & $\wm617(11)$    & $\wm607(10)$    & $\wm568(13)$    & $\wm591(14)$    & $\wm575(13)$    & $\wm577(19)$     & $\wm580(13)$    & $\wm576(13)$    \\
$E_1(\frac12^-)-E_1(\frac32^+)$ & $\wm358.6(6.8)$ & $\wm356.1(6.0)$ & $\wm330.3(6.0)$ & $\wm356.0(6.4)$ & $\wm333.7(6.0)$ & $\wm335.1(6.1)$  & $\wm339.3(6.2)$ & $\wm334.5(6.1)$ \\
$E_1(\frac32^-)-E_1(\frac32^+)$ & $\wm358.6(6.8)$ & $\wm356.1(6.0)$ & $\wm343.3(6.4)$ & $\wm348.9(6.6)$ & $\wm339.4(6.3)$ & $\wm342.0(6.4)$  & $\wm344.5(6.5)$ & $\wm342.7(6.4)$ \\
\\[-1ex]
$E_1(\frac12^+)-E_1(\frac72^+)$ & $-25.6(1.3)$    & $-24.8(1.2)$    & $-22.6(1.5)$    & $-31.7(1.5)$    & $-29.8(1.8)$    & $-28.7(1.7)$     & $-29.1(1.6)$    & $-27.5(1.6)$    \\
$E_2(\frac12^+)-E_1(\frac72^+)$ & $-0.023(17)$    & $-0.017(16)$    & $\wm1.4(1.0)$   & $-18.64(99)$    & $-17.2(1.3)$    & $-13.5(1.1)$     & $-13.63(95)$    & $-10.51(98)$    \\
$E_3(\frac32^+)-E_1(\frac72^+)$ & $-0.023(17)$    & $-0.017(16)$    & $\wm2.21(74)$   & $-15.60(84)$    & $-11.7(1.1)$    & $-8.94(87)$      & $-9.51(77)$     & $-6.67(78)$     \\
$E_4(\frac32^+)-E_1(\frac72^+)$ & $\wm44.0(1.6)$  & $\wm43.6(1.5)$  & $\wm45.3(2.1)$  & $\wm31.3(1.8)$  & $\wm31.5(1.8)$  & $\wm34.5(2.0)$   & $\wm34.5(2.0)$  & $\wm36.6(2.1)$  \\
$E_1(\frac52^+)-E_1(\frac72^+)$ & $-0.80(37)$     & $-0.77(35)$     & $\wm1.28(44)$   & $-9.66(54)$     & $-4.94(66)$     & $-3.69(54)$      & $-4.30(48)$     & $-2.59(48)$     \\
$E_2(\frac52^+)-E_1(\frac72^+)$ & $\wm44.3(1.6)$  & $\wm43.9(1.5)$  & $\wm41.3(1.9)$  & $\wm40.3(2.2)$  & $\wm35.0(1.8)$  & $\wm36.5(1.9)$   & $\wm37.1(2.0)$  & $\wm37.4(1.9)$  \\
\\[-1ex]
$E_1(\frac12^-)-E_1(\frac32^-)$ & $\wm0$          & $\wm0$          & $-12.97(45)$    & $\wm7.05(23)$   & $-5.70(35)$     & $-6.96(35)$      & $-5.19(28)$     & $-8.14(37)$   \\
\\[-1ex]
$E_4(\frac32^+)-E_2(\frac52^+)$ & $-0.28(15)$     & $-0.26(14)$     & $\wm4.02(47)$   & $-9.00(51)$     & $-3.50(50)$     & $-2.06(46)$      & $-2.62(39)$     & $-0.78(45)$   \\
\\[-2.5ex]
\end{tabular}
\end{ruledtabular}
\caption{\label{tab:coeff_dep}Dependence of the $bbb$ spectrum on the coefficients $c_i$ in the NRQCD action [see Eq.~(\ref{eq:dH_full})].
All results are given in MeV. The data are from the ensemble with $a\approx0.11$ fm and $a m_{u,d}=0.005$.}
\end{table*}

\vspace{16ex}

\begin{figure*}[ht!]
\null
\vspace{14ex}
 \includegraphics[height=58ex]{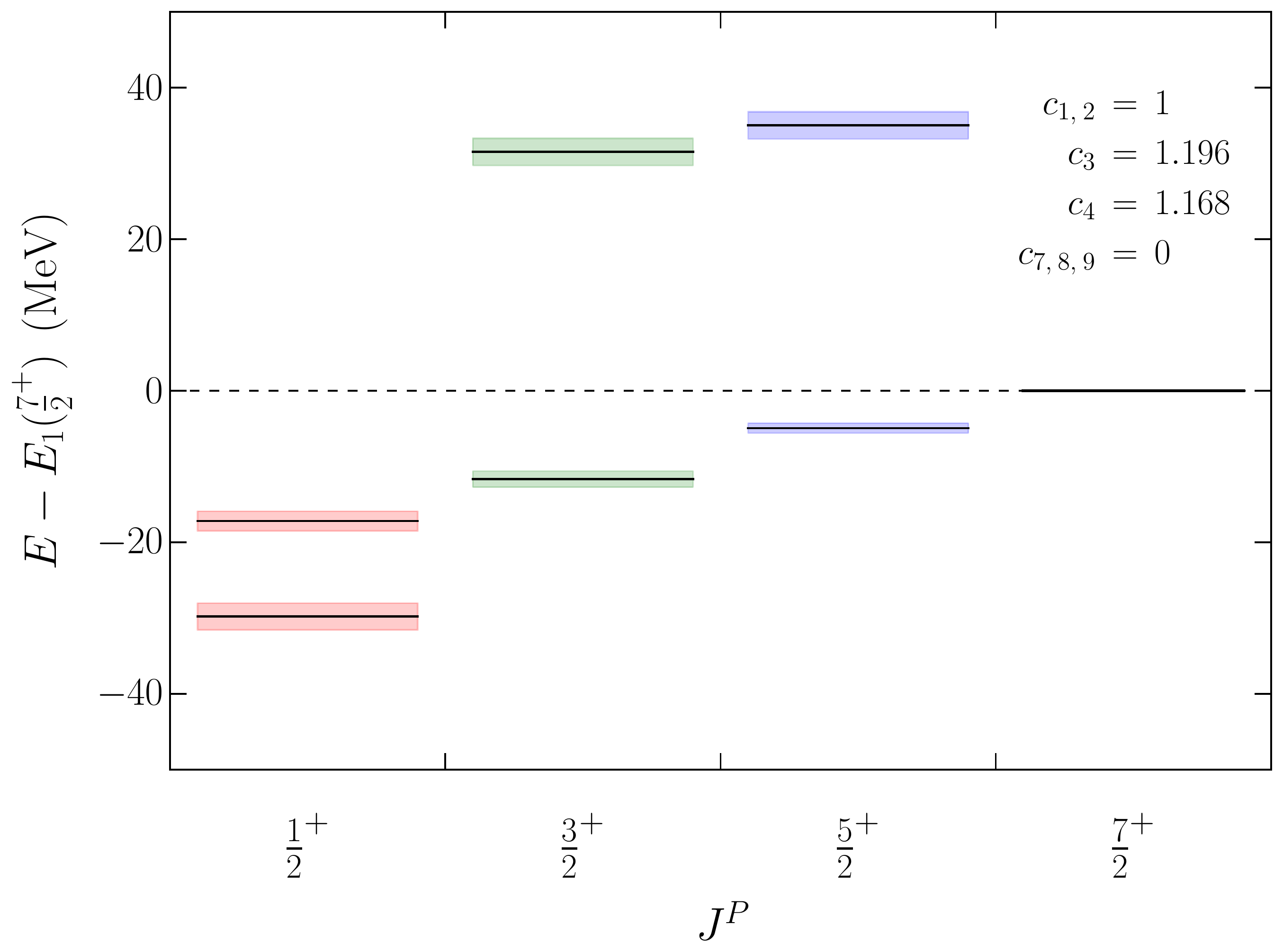} \hspace{4ex} \includegraphics[height=58ex]{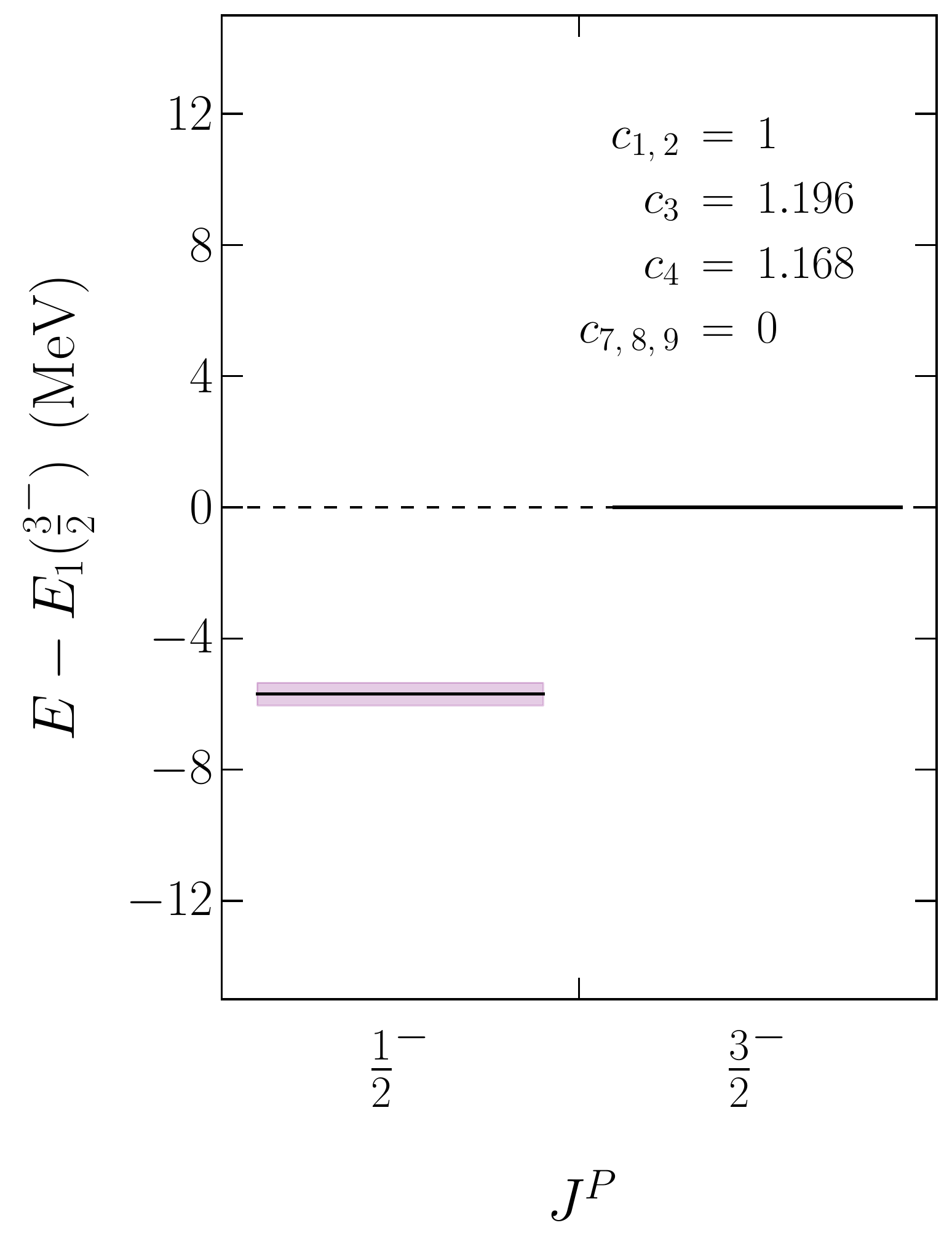}
\caption{\label{fig:coeff_dependence_v4}Dependence of the spectrum near $E_1(\frac72^+)$ and $E_1(\frac32^-)$ on the coefficients $c_i$ in the NRQCD action (at $a\approx 0.11$ fm, $am_{u,d}=0.005$).
Shown here is the case of the complete order-$v^4$ NRQCD action.}
\end{figure*}

Next, Fig.~\ref{fig:coeff_dependence_v4} shows the $bbb$ spectrum with both (\ref{eq:sigmadotB}) and (\ref{eq:sigmadotDxE}) turned on
(fifth column of Table \ref{tab:coeff_dep}). For $\big[E_2(\frac12^+)-E_1(\frac72^+)]_{\rm subtr.}$ and $E_1(\frac12^-)-E_1(\frac32^-)$,
the new results are consistent with the sums of the results from separately turning on (\ref{eq:sigmadotB}) and (\ref{eq:sigmadotDxE}),
but there is some evidence for nonlinear behavior in the other spin splittings. For example, the splitting $\big[E_1(\frac52^+)-E_1(\frac72^+)\big]_{\rm subtr.}$
is equal to $-4.17(74)$ MeV now, while the sum of the splittings obtained from separately activating (\ref{eq:sigmadotB}) and (\ref{eq:sigmadotDxE}) is $-6.85(85)$ MeV.
Of course there is no reason to expect linearity here: the lattice calculation is fully nonperturbative.

\begin{figure*}[ht!]
 \includegraphics[height=58ex]{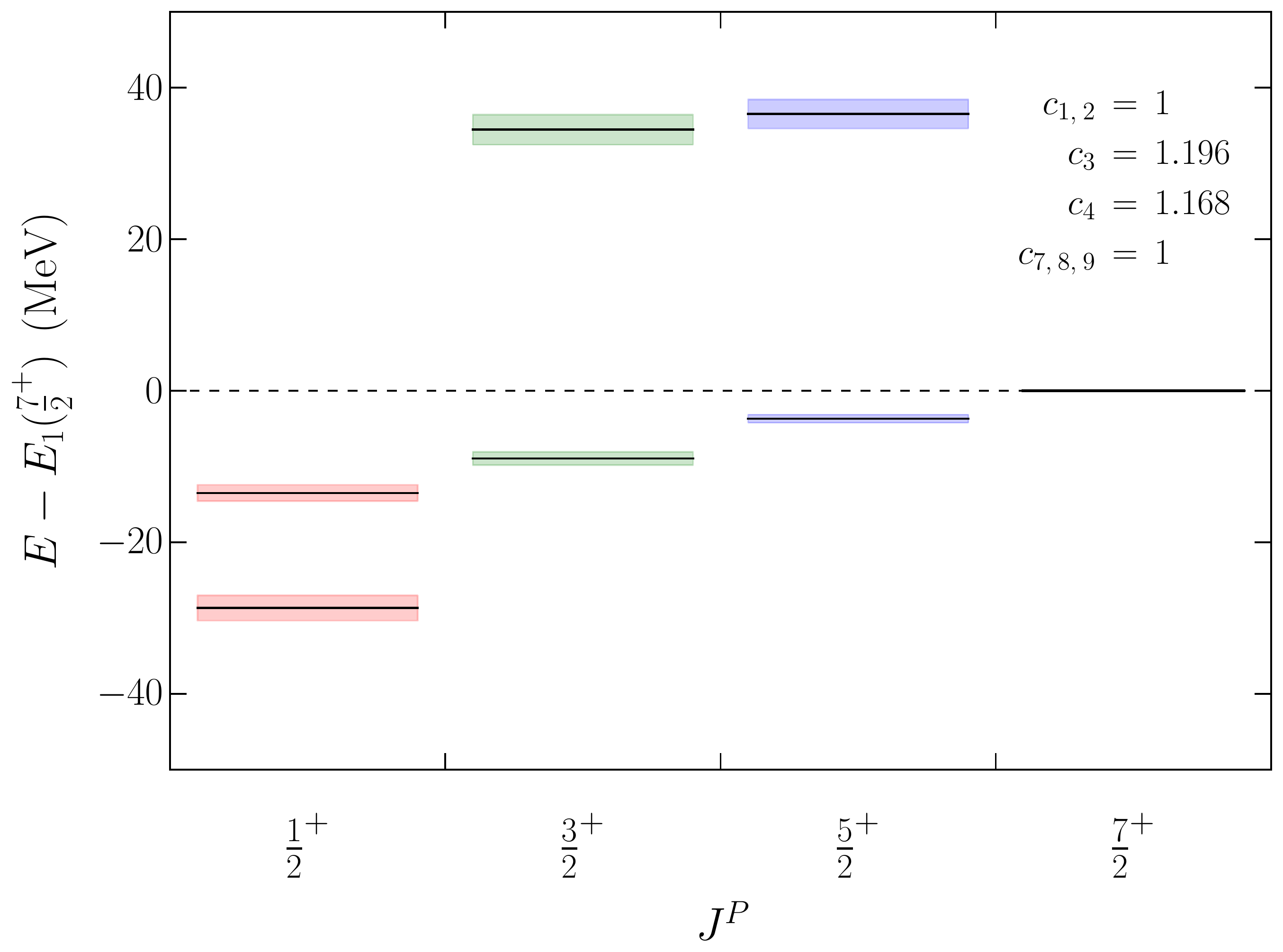} \hspace{4ex} \includegraphics[height=58ex]{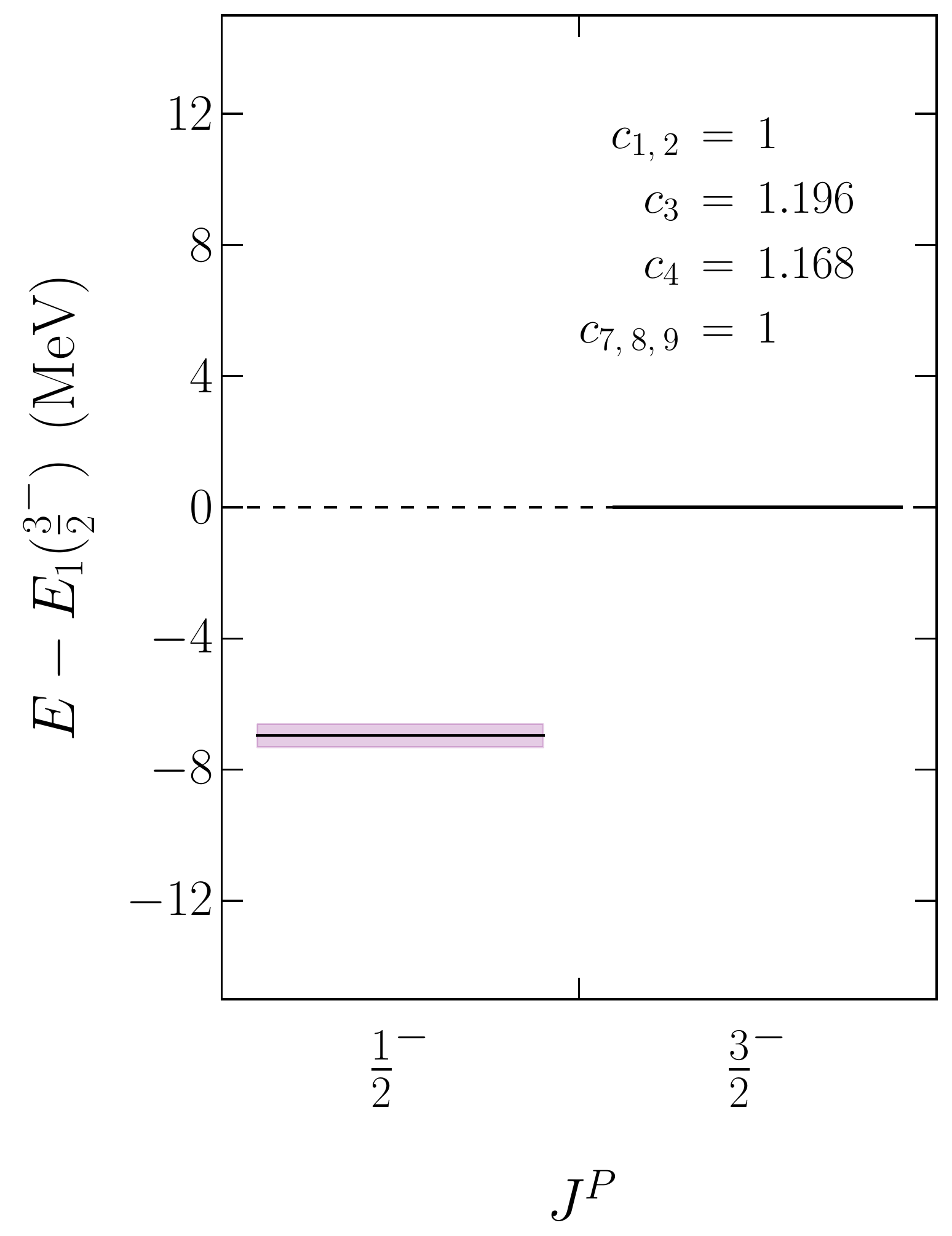}
\caption{\label{fig:coeff_dependence_v6}Dependence of the spectrum near $E_1(\frac72^+)$ and $E_1(\frac32^-)$ on the coefficients $c_i$ in the NRQCD action (at $a\approx 0.11$ fm, $am_{u,d}=0.005$).
Shown here is the case of the complete NRQCD action as used in the main calculations of this work, including all terms of order $v^4$ as well as the spin-dependent order-$v^6$ terms.}
\end{figure*}

Having included both (\ref{eq:sigmadotB}) and (\ref{eq:sigmadotDxE}), the action is now complete through order $v^4$. As can be
seen by comparing the results in the first and the fifth columns of Table \ref{tab:coeff_dep}, the radial and orbital $bbb$ energy splittings
obtained with the order-$v^2$ and order-$v^4$ NRQCD actions differ by $\lesssim$ 10\%, demonstrating the convergence
of the NRQCD expansion with $v^2\approx0.1$ as in bottomonium. Finally, turning on additionally
the spin-dependent order-$v^6$ terms by setting $c_7=c_8=c_9=1$ gives the results in the sixth column of Table \ref{tab:coeff_dep},
which are plotted in Fig.~\ref{fig:coeff_dependence_v6}. The order-$v^6$ terms affect some of the $bbb$ spin splittings by as much as 30\%,
showing that including these terms is essential to obtain precise results. Most of $bbb$ spin splittings considered here
decrease in magnitude when the order-$v^6$ terms are included in the NRQCD action, as is familiar from bottomonium \cite{Meinel:2010pv}. However,
one notable exception to this rule is found here: the order-$v^6$ corrections \emph{increase} the magnitude of $E_1(\frac12^-)-E_1(\frac32^-)$.

\vspace{16ex}

\vspace{30ex}

\section{Conclusions}

In this work, the first nonperturbative QCD calculation of the baryonic analogue of the bottomonium spectrum was performed.
By combining improved lattice NRQCD \cite{Lepage:1992tx} with other powerful techniques that have been developed more recently,
the energies of ten $bbb$ excited states were computed with high precision. The calculations include 2+1 dynamical flavors of light quarks,
and the $bbb$ spectrum was extrapolated to the physical pion mass. The main results
are given in Table \ref{tab:results} and are plotted in Figs.~\ref{fig:bbb_spectrum_all_L32} and \ref{fig:bbb_spectrum_wrtJ72_L32}.

The reliable identification of triply-bottom baryon states with angular momentum up to $J=\frac72$ was
greatly simplified by using interpolating operators constructed with the subduction method of Ref.~\cite{Edwards:2011jj}.
As already observed in Ref.~\cite{Edwards:2011jj} for light baryons, the cross-correlations between interpolating
operators subduced from different values of $J$ are small. In the present work, it was additionally shown that these overlaps
decrease when the lattice spacing is reduced. Furthermore, it was possible to resolve the small energy splittings
of continuum $bbb$ levels with $J>\frac32$ into the different irreducible representations of the double-cover octahedral group.
It was shown that these splittings also decrease when the lattice spacing is reduced (see Table \ref{tab:irrep_splittings}),
providing another demonstration of rotational symmetry restoration.
While the suppression of mixing between different $J$-values is a general consequence of the approximate rotational symmetry,
additional suppressions were observed here for the triply-heavy baryon two-point functions
between operators constructed using different values of $L$ or $S$. This feature is likely to be a consequence
of the large $b$ quark mass, resulting in a suppression of the spin-orbit coupling and hence an approximate
individual conservation of $L$ and $S$ (the total orbital angular momentum and total quark spin).

To implement the $b$ quarks on the lattice, an NRQCD action including the spin-dependent order-$v^6$ terms was used here,
and the coefficients of the spin-dependent order-$v^4$ terms were tuned nonperturbatively. Together with the high statistics,
this allowed the calculation of the $bbb$ spin splittings with $\sim$1 MeV total uncertainty.
To learn more about the forces between three heavy quarks, additional simulations were performed on one ensemble
for several ``unphysical'' choices of coefficients in the NRQCD action, thereby disentangling the contributions of
different NRQCD operators to the $bbb$ energy splittings. These additional simulations also clearly demonstrated
the convergence of the velocity expansion for $bbb$ baryons, and facilitated the estimates of the systematic uncertainties
given in Table \ref{tab:results}.

The lattice QCD results obtained here for the triply-bottom baryon spectrum provide a unique opportunity to test quark
models for baryons in the regime were the description using potentials is expected to work best. Most of the past potential-model calculations
of baryon excited states have focused on \emph{light} baryons, for which some experimental data are
available. However, quark-model descriptions are bound to remain poor approximations for these complicated systems.
Now that precise lattice QCD results for the much cleaner $bbb$ spectrum are available for comparison,
it is desirable to perform new continuum-based calculations for triply-heavy baryons, using for example the quark model of Ref.~\cite{Capstick:1986bm}, or
the modern pNRQCD approach \cite{Brambilla:2005yk, LlanesEstrada:2011kc}.

\begin{acknowledgments}
I thank William Detmold, Robert Edwards, and Kostas Orginos for useful discussions, and the RBC/UKQCD collaboration for making their gauge field ensembles available.
This work was supported by the U.S.~Department of Energy under grant number {D}{E}-{S}{C00}01{784}.
The computations were performed using resources at the National Energy Research Scientific Computing Center and the National Institute
for Computational Sciences (XSEDE grant number TG-PHY080014N).
\end{acknowledgments}

\end{document}